\documentclass[fleqn,usenatbib]{mnras}
\usepackage{newtxtext,newtxmath}
\usepackage[T1]{fontenc}

\DeclareRobustCommand{\VAN}[3]{#2}
\let\VANthebibliography\thebibliography
\def\thebibliography{\DeclareRobustCommand{\VAN}[3]{##3}\VANthebibliography}

\usepackage{graphicx}	

\title[\mosel\ survey: EoR analogues at $z\sim 3$]{\mosel\ Survey: Extremely weak outflows in EoR analogues at $z=3-4$}
\author[A. Gupta et al.]{
Anshu Gupta$^{1, 2}$\thanks{E-mail: anshu.gupta@curtin.edu.au},
Kim-Vy Tran$^{2, 3}$,
Trevor Mendel$^{2,4}$,
Anishya Harshan$^{2,3}$,
Ben Forrest$^{5,6}$,
R. L. Davies$^{2,7}$,
\newauthor{
Emily Wisnioski$^{2,4}$, 
Themiya Nanayakkara$^{2,7}$,
Glenn G. Kacprzak$^{2,7}$, and
Lisa J. Kewley$^{2,4}$
}
\\
$^{1}$International Centre for Radio Astronomy Research (ICRAR), Curtin University, Bentley WA, Australia\\
$^{2}$ARC Centre of Excellence for All Sky Astrophysics in 3 Dimensions (ASTRO 3D), Australia\\
$^{3}$School of Physics, University of New South Wales, Sydney, NSW 2052, Australia\\
$^4$Research School of Astronomy and Astrophysics, The Australian National University, Cotter Road, Weston Creek, ACT 2611, Australia\\
$^5$Department of Physics \& Astronomy, University of California, Riverside, 900 University Avenue, Riverside, CA 92521, USA\\
$^6$Department of Physics \& Astronomy, University of California, Davis, One Shields Avenue, Davis, CA 95616, USA\\
$^7$Centre for Astrophysics and Supercomputing, Swinburne University of Technology, Hawthorn, VIC 3122, Australia\\
}

\newcommand{\mosel}{{\tt MOSEL}}

\newcommand{\zfire}{{\tt ZFIRE}}

\newcommand{\zfourge}{{\tt ZFOURGE}}

\newcommand{\oiii}{[\hbox{{\rm O}\kern 0.1em{\sc iii}}]\,5007}
\newcommand{\nii}{[\hbox{{\rm N}\kern 0.1em{\sc ii}}]\,6583}
\newcommand{\ciii}{[\hbox{{\rm C}\kern 0.1em{\sc iii}}]\,1907,1909}
\newcommand{\civ}{\hbox{{\rm C}\kern 0.1em{\sc iv}}\,1550}
\newcommand{\oiiiuv}{[\hbox{{\rm O}\kern 0.1em{\sc iii}}]\,1660,1666}
\newcommand{\oiiite}{[\hbox{{\rm O}\kern 0.1em{\sc iii}}]\,4363}
\newcommand{\heii}{\hbox{{\rm He}\kern 0.1em{\sc ii}}\,1640}
\newcommand{\hii}{\hbox{{\rm H}\kern 0.1em{\sc ii}}}

\newcommand{\hb}{\hbox{{\rm H}\kern 0.1em{\sc $\beta$}}}
\newcommand{\ciiradio}{[\hbox{{\rm C}\kern 0.1em{\sc ii}}]$_{\rm 158 \mu m}$}

\newcommand{\SiII}{[\hbox{{\rm Si}\kern 0.1em{\sc ii}}]}

\newcommand{\OI}{[\hbox{{\rm O}\kern 0.1em{\sc i}}]}

\newcommand{\oiiiradio}{[\hbox{{\rm O}\kern 0.1em{\sc iii}}]$_{\rm 88 \mu m}$}

\newcommand{\msun}{${\rm M_{\odot}}$}
\newcommand{\halpha}{${\rm H\alpha}$}

\newcommand{\lya}{\hbox{{\rm Lyman-}\kern 0.1em{\sc $\alpha$}}}
\newcommand{\logmstar}{$\log(M_*/{\rm M}_{\odot})$}

\newcommand{\siion}{$\xi_{\rm ion}$}

\date{Accepted 2022 November 15. Received 2022 November 3; in original form 2022 March 14}

\pubyear{2022}

\begin{document}
\label{firstpage}
\pagerange{\pageref{firstpage}--\pageref{lastpage}}
\maketitle

\begin{abstract}

This paper presents deep K-band spectroscopic observations of galaxies at $z=3-4$ with composite photometric rest-frame \hb+\oiii\ equivalent widths (EW$_0$)$>600$\,\AA, comparable to the EW of galaxies observed during the epoch of reionisation (EoR, $z>6$). The typical spectroscopic \oiii\ EW$_0$ and stellar mass of our targets is  $\sim 700$\,\AA\ and \logmstar$=8.98$. By stacking the \oiii\ emission profiles, we find evidence of a weak broad component with $\rm F_{broad}/F_{narrow} \sim 0.2$ and velocity width   $\sigma_{\rm broad} \sim 170$\,km/s. The strength and velocity width of the broad component does not change significantly with stellar mass and \oiii\ EW$_0$ of the stacked sample. Assuming similar broad component profiles for \oiii\ and \halpha\ emission, we estimate a mass loading factor $\sim 0.2$, similar to low stellar mass galaxies at $z>1$ even if the star formation rates of our sample is 10 times higher. We hypothesize that either the multi-phase nature of supernovae driven outflows or the suppression of winds in the extreme star-forming regime is responsible for the weak signature of outflows in the EoR analogues.  
\end{abstract}



\section{Introduction}

In the last few years, deep photometric surveys have advanced our understanding of galaxies at high redshift  \citep[$z>6$, ][]{Stark2016a, Livermore2017, DeBarros2019, Endsley2020}. Direct observations find that galaxies in the EoR have rest-frame optical emission line (\oiii+\hb) equivalent width (EW$_0$)  $>1000$\AA\ \citep{Labbe2013, Roberts-Borsani2016, DeBarros2019}. In contrast, the  \oiii+\hb\ EW$_0$ for a typical star-forming galaxy at $z\sim3$ is around 200\AA\ \citep{Reddy2018}. The  high rest-frame optical EWs suggest galaxies at $z>6$ are undergoing a starburst phase with stellar populations younger than $<100$\,Myr, have a stellar mass around $10^8-10^9$\msun\  and low metallicity \citep[$Z<0.2 Z_{\odot}$,][]{Cohn2018, Endsley2020}. The brightness of UV emission lines such as \civ\ and \ciii\ (ten times the typical $z=1-3$ galaxy) is also indicative of the harder ionising spectrum of $z>5$ galaxies \citep{Stark2017, Hutchison2019, Witstok2021}.

With the launch of the {\it James Webb Space Telescope (JWST)}, our understanding of nebular emission lines in EoR galaxies will be greatly advanced in the next decade. ALMA observations have already given a preview of galaxy properties at $z>6$. \cite{Fujimoto2019b} used the Alpine survey \citep{Fevre2019, Bethermin2020} to detect extended \ciiradio\ ($T\sim 60-200$\,K) halos with sizes almost twice that of the rest-frame UV continuum. The radial profile of the \ciiradio\ halo matches closely with the \lya\ emission, suggesting a link between the physical origin of the two lines \citep{Faisst2020}. \cite{Smit2018} even detect two  disk-dominated galaxies at $z=6.8$ with ordered rotation curves using the \ciiradio\ emission. The prevalence  of \ciiradio\ halos at $z=4-6$ suggests the early onset of enrichment of circum-galactic and interstellar medium of galaxies \citep{Fevre2019, Yan2020}.

The high opacity of the intergalactic medium makes direct estimation of escape fraction and/or production efficiency of ionizing photons in EoR galaxies  extremely challenging. An alternate approach that has been successful is to study detailed physical properties of extreme emission-line galaxies at lower redshifts assuming they are analogous to EoR galaxies. The intense \oiii\ emitters (EW$_0$ $>650$\,\AA)  at $1.4<z<2.6$ have almost 1\, dex higher hydrogen ionising production efficiency (\siion) than galaxies with \oiii\ EW$_0$ $< 400$\,\AA\ \citep{Tang2018}. Similarly,  galaxies at cosmic noon ($z\sim 2$) with \oiii\ EW$_0$ $>1000$\,\AA\ have higher \lya\ emission flux and brighter restframe UV emission lines \citep{Du2019, Tang2020}, similar to galaxies at $z>6$. Extremely young (2-3 Myr) and metal-poor stellar populations (0.06-0.08 Z$_{\odot}$) are revealed in four gravitationally lensed strong \ciii\ emitters at $z\sim 2$ by detailed photoionisation modelling, which is only possible because of their  relatively nearby nature and  the lensing magnification \citep{Mainali2020}.

\cite{Onodera2020} use excess flux in the $K$-band to select 19 bright \oiii\ emitters (\oiii\ EW$_0$ = $130-2051$\,\AA) at $z\sim 3.3$, i.e., galaxies analogous to the Epoch of reionisation (EoR). By combining spectroscopic and photometric observations, they find that EoR analogues have harder ionising spectra than normal star-forming galaxies, probably due to their low metallicity ($\rm \log(O/H)=7.5-8.5$) and a factor of 10 higher ionisation  parameter.  These studies show the importance of analogues to  learn about the physical conditions in the interstellar medium (ISM) of ``first galaxies".  

Very few studies have analysed galactic-scale outflows in the extreme emission-line galaxies. Galactic-scale gas outflows set the foundation of baryonic physics by regulating the star formation efficiency \citep{Krumholz2018} and the enrichment history of galaxies \citep{Tremonti2004}. Early models predict significant mass loss in the low mass galaxies due to their shallow potential well \citep{MacLow1999} assuming thermal pressure by supernovae as the primary source of outflows \citep{Larson1974, Dekel1986}. In contrast, detailed hydrodynamical models of starburst in dwarf galaxies revealed that supernovae feedback alone is efficient at transporting energy but might be extremely inefficient at transporting mass out of the system \citep{Strickland2000, Marcolini2005}.

Cosmological hydrodynamical simulations rely on either the energy-driven \citep{Murray2005} or momentum-driven prescription of the supernovae-driven wind \citep{Chevalier1985}.  In the illustrisTNG simulations, stellar feedback is modelled as the kinetic energy driven wind, where the mass loading factor in the outflow decreases with stellar mass at the injection site \citep{Nelson2019, Pillepich2019}. In the EAGLE simulations, particles are explicitly kicked relative to the circular velocity and the rate of the particle loss is assumed to scale negatively with the circular velocity of the system \citep{Mitchell2019}. In EAGLE the mass lost in outflow scales negatively with the halo mass until $10^{12}$\msun\, above which the blackhole feedback dominates driving stronger the outflows in massive halos \citep{Mitchell2019}. In illustrisTNG, low mass galaxies above the star-forming main-sequence have a higher mass-loading factor \citep{Nelson2019} suggesting that a large number of supernovae going off simultaneously might boost the average mass outflow rate of galaxies above the main-sequence.

 By stacking the \ciiradio\ profile from the ALPINE survey \citep{Fevre2019, Bethermin2020}, \cite{Ginolfi2019a} estimates an average mass loading factor close to unity for $z=4-6$  galaxies with \logmstar$\sim 10$.  The high spatial resolution ($\sim2$\,kpc) observations of a  \ciiradio\ emitter on the star-forming main-sequence at $z\sim 5.5$ reveal that local mass loading factors can be $6-10$ times higher than the global mass loading factor \citep{Herrera-Camus2021}. Although, a three-dimensional dynamical analysis of the \ciiradio\ emission around a $z\sim4.5$ galaxy find a non-turbulent uniformly rotating disk with non-existent signs of outflowing gas \citep{Rizzo2021}. The spectroscopic observations of low mass galaxies at $z>1$ ($\sim\,10^9$\,\msun) do not reveal significant statistical evidence of the asymmetry in the emission line profile, which could either be due to the lack of depth or spectral resolution \citep{Maseda2014}. 
 
In this work, we present a sample of EoR analogues observed as part of the Multi-Object Spectroscopic of Emission Line (\mosel) survey \citep{Tran2020, Gupta2020}. The \mosel\ survey targets  a range of star-forming galaxies  with composite  \hb+\oiii\ equivalent widths (EW$_0$) $>230$\,\AA\ at $3.0<z<3.8$  \citep{Forrest2017, Forrest2018}, selected from the ZFOURGE survey \citep{Straatman2016}.  This paper extends the \mosel\ sample presented in \cite{Tran2020} by specifically targeting galaxies with composite  \hb+\oiii\  EW$_0>600$\,\AA\ with the K-band Multi-Object Spectrograph \citep[KMOS,][]{Sharples2012, Sharples2013} on the Very Large Telescope (VLT).  

This paper shows that highly star-forming low mass  (\logmstar\,$\sim 9.0$, SFR $= 10-50$\,\msun) extreme \oiii\ emitters  exhibit extremely weak signatures of outflows. In Section \ref{sec:observations}, we  describe the sample selection, observation and data reduction for the  \mosel\ survey.   Section \ref{sec:methods} presents the detailed descriptions of our analysis techniques and present our results. Finally, in Section \ref{sec:discussion} we discuss the main implications of our results and summarise them in Section \ref{sec:summary}.

For this work, we assume a flat $\Lambda$CDM cosmology with $\Omega_{M}$=0.3, $\Omega_{\Lambda}$=0.7, and $h$=0.7. 

 \begin{figure*}
\centering
\tiny
\includegraphics[scale=0.35, trim=0.0cm 0.0cm 0.0cm 0.0cm,clip=true]{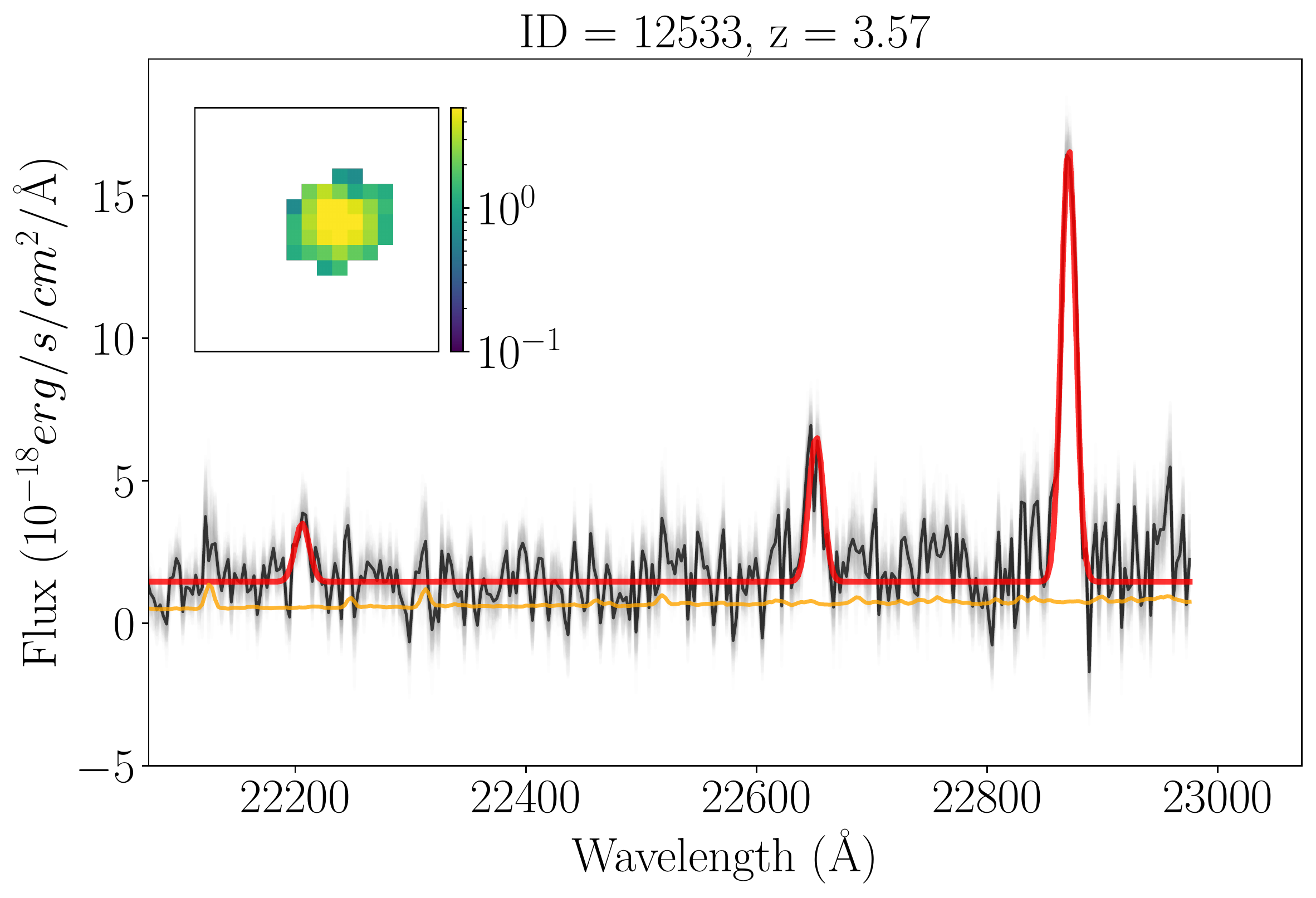}
\includegraphics[scale=0.35, trim=0.0cm 0.0cm 0.0cm 0.0cm,clip=true]{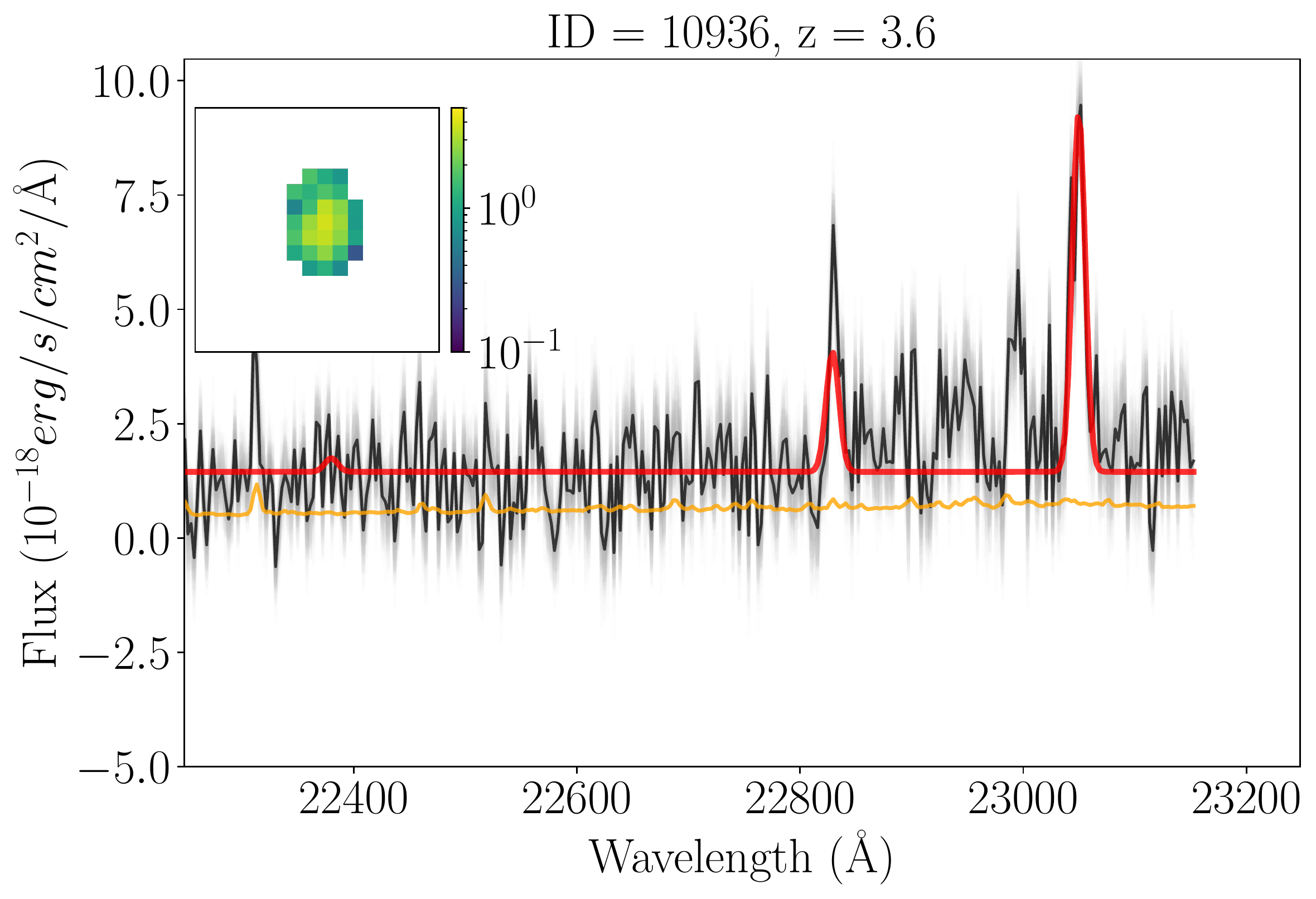}
\caption{Sample spectra for two EoR analogues from KMOS observations. The black curves are the 1D spectra in the observed frame, orange is the noise spectra, and the grey shaded region is the bootstrapped iterations. The red curves correspond to the best-fit curves to the \oiii${, 4959}$ and \hb\ emission lines. The inset image shows the wavelength collapsed \oiii\ emission maps.  }
\label{fig:spectra}
\end{figure*}

\section{Observations}\label{sec:observations}

\subsection{\mosel\ Survey}\label{sec:mosel_survey}

Our sample is part of the \mosel\ survey, which is a spectroscopic follow-up of the $z\sim 3$ galaxies selected from the FourStar Galaxy Evolution survey \citep[ZFOURGE;][]{Straatman2016}. We refer the readers to \cite{Tran2020} for a detailed description of the survey design.  In brief, the \mosel\ survey acquires near-infrared spectra of emission-line galaxies at $3.0<z<3.8$ to understand their contribution to the star formation history of the universe. Targets were selected to have composite \oiii+\hb\ EW$_0$ $> 230$\,\AA\ based on medium- and broad-band photometric data \citep{Forrest2018}. This paper builds on the sample presented in \cite{Tran2020} by specifically targeting 19 extreme emission-line galaxies (\oiii+\hb\ EW$_0$ $> 600$\,\AA) with KMOS/VLT.

 \subsection{Keck/MOSFIRE observations and data reduction}\label{sec:mosfire}
 
 Please refer to \cite{Gupta2020, Tran2020} for a detailed description of MOSFIRE observations and data reduction. In summary, a total of 5 masks were observed in the COSMOS field and one mask in the CDFS field using the $K$-band filter covering a wavelength of $1.93-2.38\,\micron$. The spectral dispersion is 2.17\,\AA/pixel. The seeing was $\sim 0.7''$. 
 
 A total of 95 galaxies were targeted between $0.9<z<4.8$, with the highest priority given to the emission line galaxies with  \oiii+\hb\ equivalent width  $>230$\,\AA\ (38 galaxies) between $2.5<z<4.0$. Possible active galactic nuclei (AGN) contaminants were removed using the \cite{Cowley2016} catalog. The data was reduced using the MOSFIRE data reduction pipeline\footnote{http://keck-datareductionpipelines.github.io/MosfireDRP} and flux calibration was performed using the \zfire\ data reduction pipeline \citep{Tran2015, Nanayakkara2016}.  We spectroscopically confirm 48 galaxies between $2.9<z<3.8$ of which 11 are extreme emission-line galaxies (\oiii+\hb\ EW$_0$ $>800$\AA), 13 are strong emission-line galaxies (\oiii+\hb\ EW$_0$ $230-800$\AA), and 24 are star-forming galaxies \citep[\oiii+\hb\ EW$_0$ $<230$\AA, ][]{Tran2020}.

\subsection{VLT/KMOS observations and data reduction}

We obtained additional observations for 19 galaxies with composite photometric \oiii\ +\hb\ EW$_0$ $>600$\AA\ \citep{Forrest2018} using KMOS/VLT  between November 15-18, 2019 (project code: 0104.B-0559, PI Gupta).  We observed a single field with 19 targets in the CDFS \citep{giacconi2002}.  Our primary targets had  photometric redshifts between $3.0<z<3.8$, but two galaxies at slightly lower redshifts ($z\sim 2.6$) were added to optimise the KMOS arm configuration. Targets were selected based on their composite photometric \oiii\ +\hb\ EW$_0$ irrespective of  their stellar mass or SFRs. AGNs were removed using the \cite{Cowley2016} catalogues that relies on near-infrared, X-ray and radio emission. The \cite{Cowley2016} catalogues are only 80\% mass complete for \logmstar$>9.5$ at $z=3$, therefore we cannot completely rule out AGNs from our sample. Although, we expect the contribution of AGNs to be minimal based on the stellar mass distribution of our sample.  The average stellar mass and star-formation rate of the full KMOS targets is \logmstar$=8.78\pm0.4$ and SFR$= 38 \pm 23$\msun/yr respectively.  

We use the $K$-band filter giving us a wavelength coverage of $1.93-2.49\mu$m at a spectral resolution of 2.8\,\AA/pixel. Each individual exposure is of 300 seconds. To maximize the on-source time, we dither to a sky position a third of the time resulting in the following strategy ``O-S-O-O-S" strategy (O: object, S: sky), reacquiring every 3 hours. To ensure that pointing remains well behaved during long exposures, we positioned three bright stars roughly equally spaced across the 24 KMOS arms. Throughout the exposure, the two-dimensional image of these stars remained well behaved. We reach a total on-source integration time of 12 hours. The seeing conditions ranged from $\sim 0.5 -1.0''$.

The data were reduced using the official KMOS/ESOReflex pipeline \citep{Freudling2013}. Each science frame is reduced individually using the nearest sky frame and the inbuilt KMOS sky rescaling routines. Each sky subtracted frame is flux calibrated using the observation of standard stars taken either during the beginning or at end of the night. The flux calibrated frames from individual nights are stacked using clipped average providing a flux and wavelength calibrated datacube for each target. We use a simple averaging routine to combine data from all three nights.

\subsection{Emission line flux measurement}\label{sec:em_line}

\subsubsection{KMOS/VLT}\label{sec:kmos_em_extraction}
Each datacube was visually inspected to identify the emission lines.  We detect bright emission lines in 16 out of 19 targeted galaxies, where 2 galaxies have H$\alpha$ instead of \oiii\ as the brightest line. In 8 galaxies, we detect the full triplet of \oiii$, 4959$ and \hb\ emission lines. For the rest, we detect the doublet \oiii$, 4959$, giving us reliable spectroscopic redshifts. 

To extract 1D-spectra, we first create a wavelength collapsed image within $\pm5$\AA\ of the expected \oiii\ emission line. We fit a 2D-Gaussian to the \oiii\ emission map and sum over spaxels within $\pm$ full-width half maximum (FWHM) from the centroid to generate the 1D spectrum for each galaxy (Figure \ref{fig:spectra}). The typical FWHM of \oiii\ emission profile is between $0.6-0.7''$, similar to the seeing at the time of the observations. Thus, our targets are not spatially resolved. Our results do not change significantly if we integrate flux within $0.5\,\times\,$FWHM. We also generate a 1D noise spectrum by summing the square of the noise spectrum of individual spaxels.  The extracted 1D spectra of the full KMOS sample are available in Appendix \ref{sec:all_spec_appendix}.   

We simultaneously fit the \oiii\,$, 4959$ and \hb\,$4861$ emission lines with three Gaussians and five free parameters: redshift, flux-\oiii, flux-\hb, width, and continuum level, for galaxies where we detect all three lines (Figure \ref{fig:spectra}: left). The $\rm{[OIII]}\,4959$ flux is fixed to be flux(\oiii)/3.  In case only \oiii\,$,4959$ are detected with signal-to-noise (S/N) greater than 3, we fit two Gaussian with four free parameters:  redshift, flux-\oiii, width, and continuum.  The average instrumental broadening is $\sim$2.6\AA\ at $z\sim 3.5$ (33 $\rm{km\, s}^{-1}$), measured from the width of skylines in the error spectrum. While fitting emission lines, we subtract the instrumental broadening in quadrature from the Gaussian line width. For each galaxy, we create 1000 realisations of the flux spectrum by perturbing the flux spectrum according to the noise spectrum. For each realisation, we perform the previously described fitting routine and re-estimate the best-fit parameters. We used the median and standard deviation from the bootstrapped realisation to represent the value and errors in the best-fit parameters. 

\subsubsection{MOSFIRE/Keck}

A detailed description of emission line extraction with the MOSFIRE spectrograph is given in \cite{Gupta2020}. In summary, we collapse the 2D slit spectra along the wavelength axis within $\pm5$\AA of the detected \oiii\ emission line to generate the spatial profile and fit a Gaussian. To generate the 1D spectra, we sum the 2D slit spectra within $\pm$FWHM from the centroid of the spatial profile. We use a similar procedure to Section \ref{sec:kmos_em_extraction} to extract emission line fluxes. The instrumental broadening for MOSFIRE is 2.5\AA\ (32 $\rm{km\, s}^{-1}$ at $z=3.5$). Thus, galaxies observed with both MOSFIRE and KMOS spectrograph have a similar spectral resolution.

\section{Analysis and results}\label{sec:methods}

\subsection{Spectral energy distribution fitting} \label{sec:sed_fit}
We fit the photometry of galaxies using the procedure described in \citet[][see Section 4.2]{Forrest2018}. Briefly,  we use the FAST \citep{Kriek2011} spectral energy distribution (SED) fitting code and the \citet[BC03]{Bruzual2003} stellar populations models with emission lines. The emission lines from \lya\ to 1-$\mu$m are calculated at varied ionisation, metallicity, and hydrogen density using the CLOUDY 08.00 photoionisation models \citep{Inoue2011, Salmon2015} and added to the BC03 stellar models. 

We use a \citet{Chabrier2003} IMF, \cite{Kriek2011} dust-law, solar metallicity ($Z=0.02$), and an exponentially declining star formation history. We also test our results assuming a sub-solar metallicity ($Z=0.004$), which changes the continuum level slightly without significantly affecting the main conclusion of this paper based on the emission-line properties. A detailed photometric analysis of the extreme emission line emitters with different stellar populations and emission-line libraries will be done in a future paper.  

\begin{figure}
	\centering
	\tiny
	\includegraphics[scale=0.35, trim=0.0cm 0.0cm 0.0cm 0.0cm,clip=true]{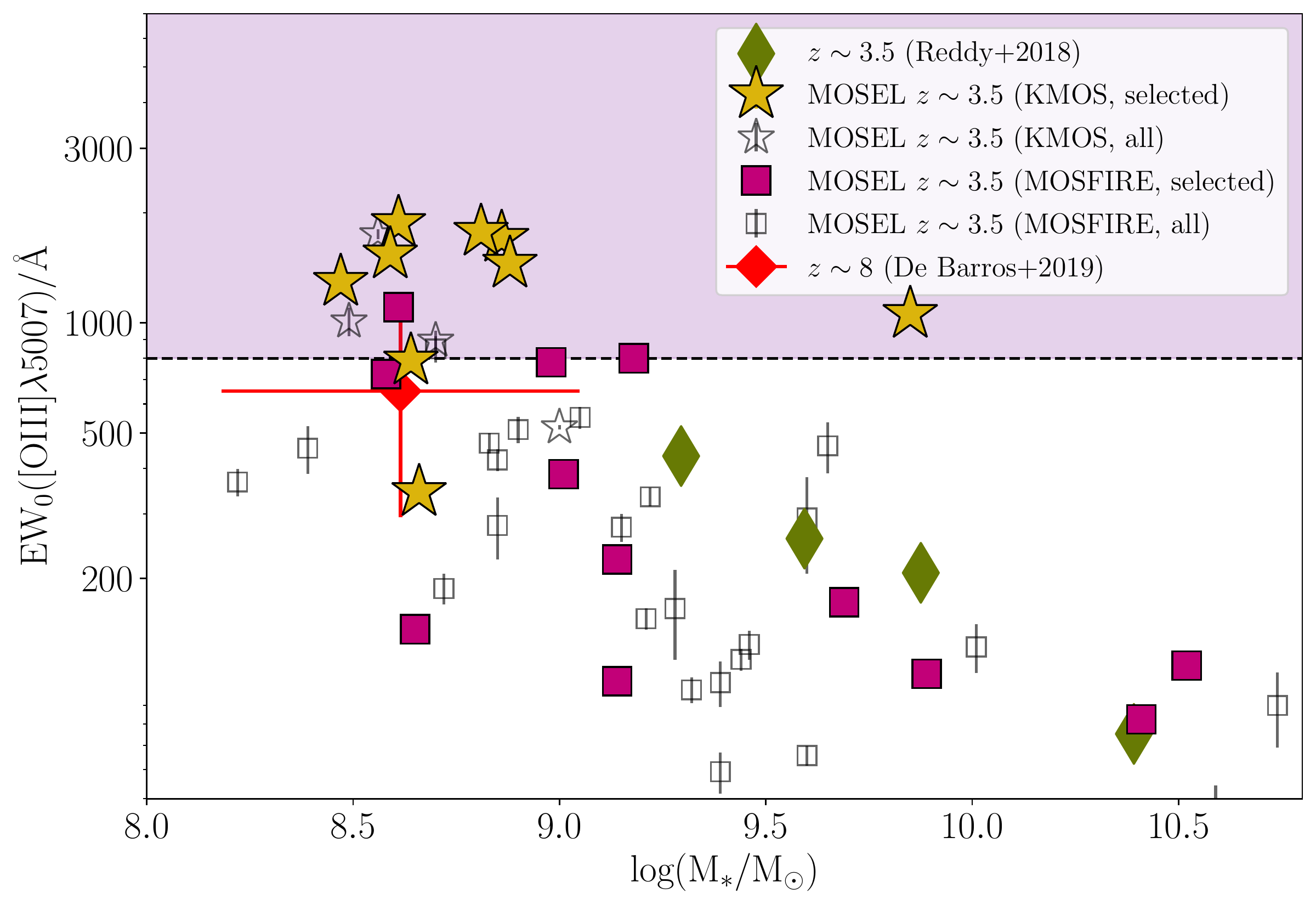}	
		
	\caption{Comparison between the rest-frame \oiii\ EW$_0$ and stellar mass distribution of the \mosel\ sample (KMOS: stars; MOSFIRE: squares) with the typical star-forming galaxies at $z\sim3$ \citep[][green diamonds]{Reddy2018}. Only galaxies with S/N$>10$ and minimal sky contamination {\bf within $\pm250$\,km/s} from the centroid will be used for the analysis in this paper (filled stars and filled squares, See section \ref{sec:sample_selection}). 
	The red diamond shows the median and 16-84$^{th}$ percentile for $z\sim 8$ galaxies from the \citet{DeBarros2019}. The black dashed line and purple shaded region highlight the \oiii\ EW$_0$ of $z>6$ galaxies from \citet{Roberts-Borsani2016}. The stellar mass and \oiii\ EW$_0$ of the \mosel\ sample is similar to $z>6$ galaxies. }
	\label{fig:o3_ew}
\end{figure}

The median stellar mass for the full sample is \logmstar$=9.14$ (see table \ref{tb:sample_info} for additional information about the sample).  Accounting for the emission lines strength in the SED fit reduces  the stellar mass by about $0.5\,$ dex for most of the \mosel\ sample \citep{Salmon2015, Forrest2018, Cohn2018}.  To estimate the \oiii\ EW$_0$, we use a top-hat filter of width 150\AA\ around 4675\AA\ and 5200\AA\ on the best-fit SEDs \citep{Tran2020}.   The median  \oiii\ EW$_0$ is 279 \AA\ (Figure \ref{fig:o3_ew}).  All except one of the KMOS targets and 50\% of the MOSFIRE targets have \oiii\ EW$_0$ and stellar mass similar to $z>6$ galaxies \citep{Roberts-Borsani2016, DeBarros2019}.

\begin{figure*}
	\centering
	\tiny
	\includegraphics[scale=0.45, trim=0.0cm 0.0cm 0.0cm 0.0cm,clip=true]{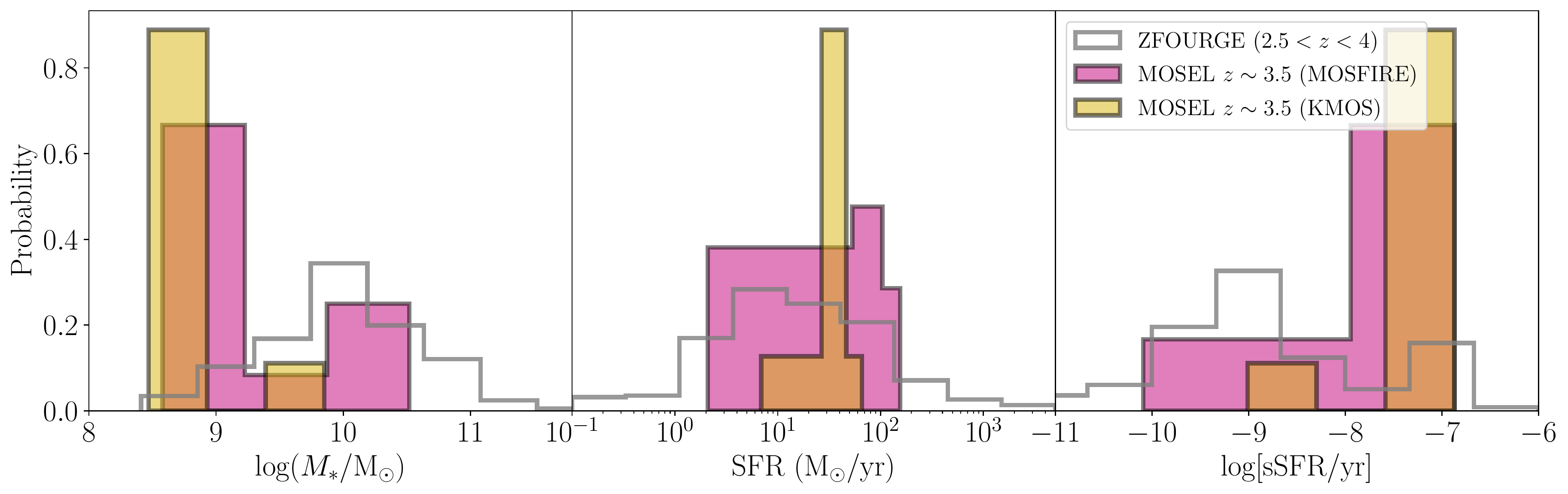}	
		
	\caption{The stellar mass (left), SFR (middle) and sSFR (right) distribution of the selected galaxies from the KMOS (gold) and MOSFIRE (magenta) observations. The grey histogram represents the control population selected from the \zfourge\ survey ($z_{\rm phot}=2.4-4$ and K-band magnitude $< 24$. The selected \mosel\ galaxies have 1.5\,dex lower stellar mass and almost 2\,dex higher sSFR than the typical $z\sim 3$ galaxies.  }
	\label{fig:eelgs_properties}
\end{figure*} 

\subsection{Final Sample Selection}\label{sec:sample_selection}
 
Traditionally multi-component Gaussian fits are used to characterise properties of outflows in galaxies. However, the multi-component fit
to individual galaxies does not significantly improve the residuals
either due to the minimal outflow component and/or the low spectral resolution of KMOS and MOSFIRE observations. Therefore, we characterize the outflow properties by stacking the 1D spectra. 

Stacking techniques by design are highly sensitive to residual sky contamination. Thus, we remove all galaxies with any significant sky contamination between $-250 - 250\,{\rm km/s}$ from the line centre. We further restrict our sample by selecting galaxies with \oiii\ S/N $>10$. One additional galaxy was removed from the KMOS sample because it had obvious signs of merger activity.   The compact and unresolved nature of our sample even in HST imaging makes it difficult to completely rule-out mergers from our sample (See Appendix \ref{sec:images_appendix}). We end up with a clean sample of 9 and 12 galaxies from KMOS and MOSFIRE observations respectively. We bootstrap the spectra using the noise to estimate median and 1-sigma errors. 

The restriction on S/N biases the final sample to galaxies with higher \oiii\ flux (Figure \ref{fig:o3_ew}). Galaxies observed with the MOSFIRE observations have a wider range of \oiii\ EW$_0$ because of the greater diversity of targets. The median stellar mass and \oiii\ EW$_0$ for the selected sample is \logmstar$\,=8.88$ and 780 \AA\  respectively (Table \ref{tb:sample_info}). 

Figure \ref{fig:eelgs_properties} compares the stellar mass, SFR and sSFR distribution of the selected sample with  galaxies in the redshift range $2.4<z<4$ from the \zfourge\ survey. We use the SED fitting template described in Section \ref{sec:sed_fit} to derive physical properties of all galaxies with photometric redshifts between $2.4<z<4$ and K-band magnitude brighter than $<24.0$. The selected \mosel\ sample has almost 1.5\,dex lower stellar mass and almost 2\,dex higher sSFR than the control sample.  Thus, the selected \mosel\ galaxies are sampling the low mass extremely star-forming galaxy population at $z=3-4$ and are similar to the $z>6$ galaxies (Figure \ref{fig:o3_ew} \& \ref{fig:eelgs_properties}).

\begin{table}
	\small
	\begin{center}
		\caption{Summary of target properties}
		\label{tb:sample_info}
		\begin{tabular}{ | *{5}{c|}  }
			\hline
			\hline
			Galaxy properties & \multicolumn{2}{c|}{KMOS }  & \multicolumn{2}{c|}{MOSFIRE }  \\
			& All [13]$^a$ & Selected [9]$^a$   & All [45]$^a$ & Selected [12]$^a$ \\
			\hline
			\vspace{0.2cm}

			\logmstar$^b$ & $8.66\substack{+0.2 \\ -0.1}$ & $8.66\substack{+0.2 \\ -0.1}$ &  $9.32\substack{+1.2 \\ -0.6}$ & $9.14\substack{+0.9 \\ -0.5}$\\
			
			\vspace{0.2cm}
			
			\oiii\ EW$_0^c$ (\AA) & $1288\substack{+461 \\ -521}$ & $1446\substack{+305 \\ -582}$ & $145\substack{+323 \\ -108}$ & $198\substack{+587 \\ -90}$\\ 
			
            SFR$^b$ (\msun/yr) & $40\substack{+13 \\ -12}$ & $40\substack{+4 \\ -4}$ & $30\substack{+79 \\ -27}$
            & $57\substack{+52 \\ -53}$\\
			\hline 
			\hline
		\end{tabular}
	\end{center}
	
	\begin{flushleft}
	\textbf{Notes:} $^a$ The number quoted within square brackets corresponds to the total number of galaxies in each sample. \\
	$^b$ Median, and $16^{\rm th}$ and $84^{\rm th}$ percentiles from the best-fit SEDs. \\
	$^c$ Rest-frame spectroscopic EW.

	\end{flushleft}
	
\end{table}%

\subsection{Spectral Stacking}\label{sec:stacking}

The spectral stacks are created by stacking the spectrum in the rest-frame using the best-fit redshifts from Section \ref{sec:em_line}.  We first fit a constant to remove the background between $\pm 1000\,$km/s after masking the wavelength region within $\pm\ 3\sigma$ from the centroid of the \oiii\ emission. We then normalise each spectrum based on the peak \oiii\ flux. The normalisation is necessary to avoid biasing the stacked spectrum towards bright \oiii\ emitters. After background removal and normalisation, each spectrum is interpolated on a uniform wavelength grid with 0.6\AA\ wavelength resolution, which is similar to the rest-frame wavelength dispersion of the KMOS observations (0.48\AA\ for MOSFIRE observations). 

We use jackknife resampling to estimate the variance in the stacked spectrum. Each realisation is created by averaging all but one galaxy spectrum from the sample. We use mean and variance from jackknife replicates to represent the final mean and error in the stacked spectrum. We also use individual jackknife replicates to estimate the reliability of our best-fit parameters in section \ref{sec:stacking_results}.  

\subsection{Stacking Results} \label{sec:stacking_results}

Stacking is done by subdividing the sample based on stellar mass and \oiii\ EW$_0$ and using the procedure described in Section \ref{sec:stacking}. We use the stellar mass \logmstar$\,=\,9.0$ and \oiii\ EW$_0$ $= 650$\AA\ to subdivide our sample into two bins. This results in a roughly equal number of galaxies in each bin. 
Figure \ref{fig:stack_spectra_mass} and Figure \ref{fig:stack_spectra_oiii} show the final stacked spectrum based on stellar mass and \oiii\ EW$_0$ respectively.  We use the python package LMFIT \citep{Newville2014} that uses a non-linear optimisation to fit a single and two-component Gaussian profile to each stacked spectrum. The broad component parameters are allowed to vary by the following amounts: velocity = $\pm100$, F$_{\rm broad}$ = $\rm 0 - F_{narrow}$, and $\sigma_{\rm broad}\ =\ 1 - 10 \times \sigma_{\rm narrow}$. 

\subsubsection{Stacking based on stellar mass}

Choosing a stellar mass cut of \logmstar$=\,9.0$, we end up with 12 galaxies in the low mass bin (8: KMOS and 4: MOSFIRE) and 9 galaxies in the high mass bin (1: KMOS and 8: MOSFIRE). The average S/N of the stacked spectrum is $\sim 50$. 

\begin{figure*}
	\centering
	\tiny
	\includegraphics[scale=0.45, trim=0.0cm 0.0cm 0.0cm 0.0cm,clip=true]{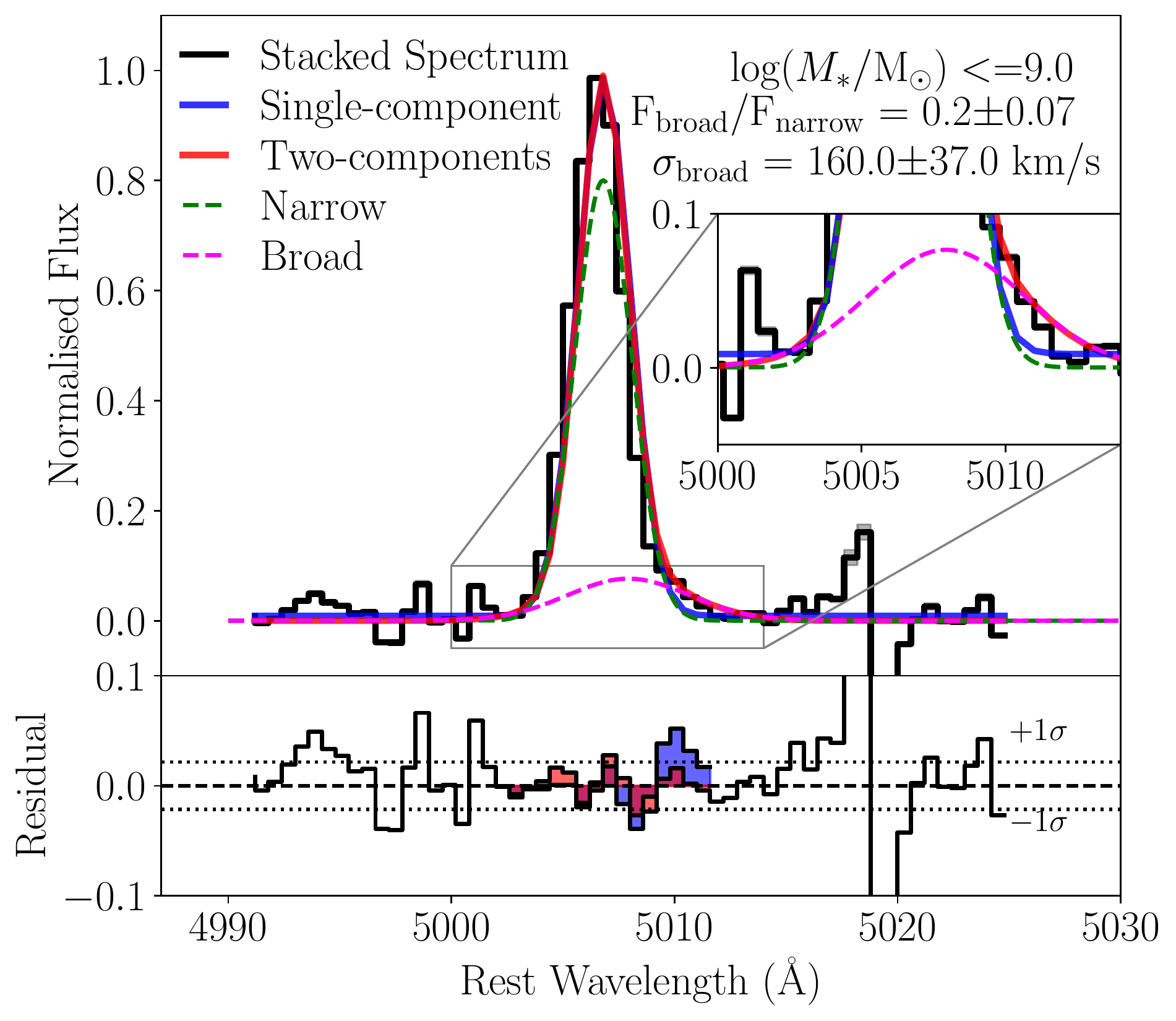}	
    \includegraphics[scale=0.45, trim=0.0cm 0.0cm 0.0cm 0.0cm,clip=true]{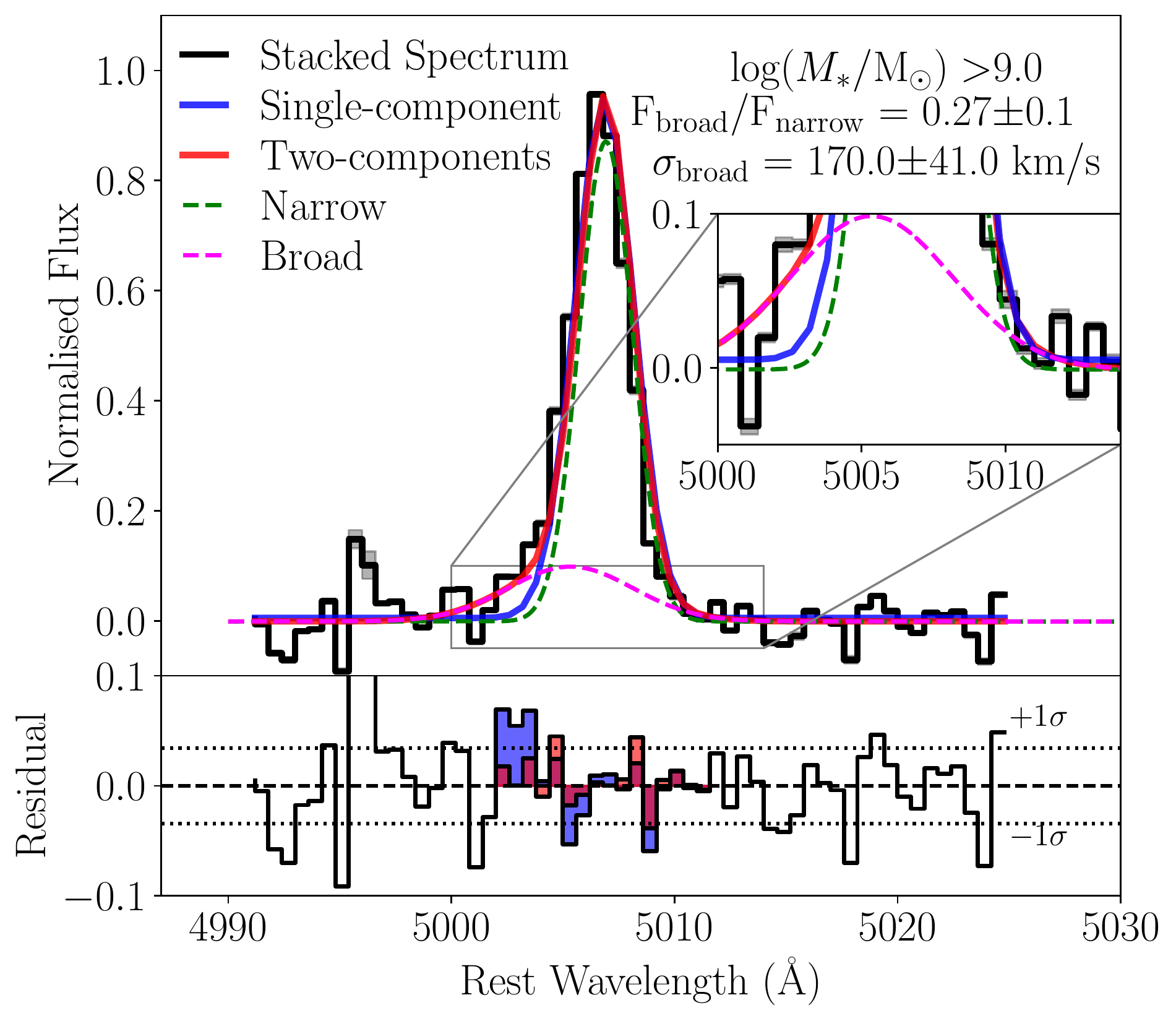}	

	\caption{The black line and shaded region indicate the mean and standard deviation in the bootstrapped spectra for the galaxies with  \logmstar$<=9.0$ (left, N$=12$) and \logmstar$>9.0$ (right, N$=9$).  The solid blue and red curves indicate the best-fit result from the one and two-component emission line fitting respectively. The green dashed and pink dotted lines shows the narrow and broad sub-components in the best-fit two-component emission profile (red line). The bottom plot in each panel indicates the residuals after subtracting the best-fit one- (blue) and two-components curves (red) from the stacked spectrum.   The inset plot in each panel zooms in on the base spectra around $\pm 7$\AA\ from the central peak. The broad component detected in low mass galaxies is redshifted with respect to the central peak, whereas it is blueshifted in the high mass sample.}
	\label{fig:stack_spectra_mass}
\end{figure*}

\begin{figure*}
	\centering
	\tiny

	\includegraphics[scale=0.45, trim=0.0cm 0.0cm 0.0cm 0.0cm,clip=true]{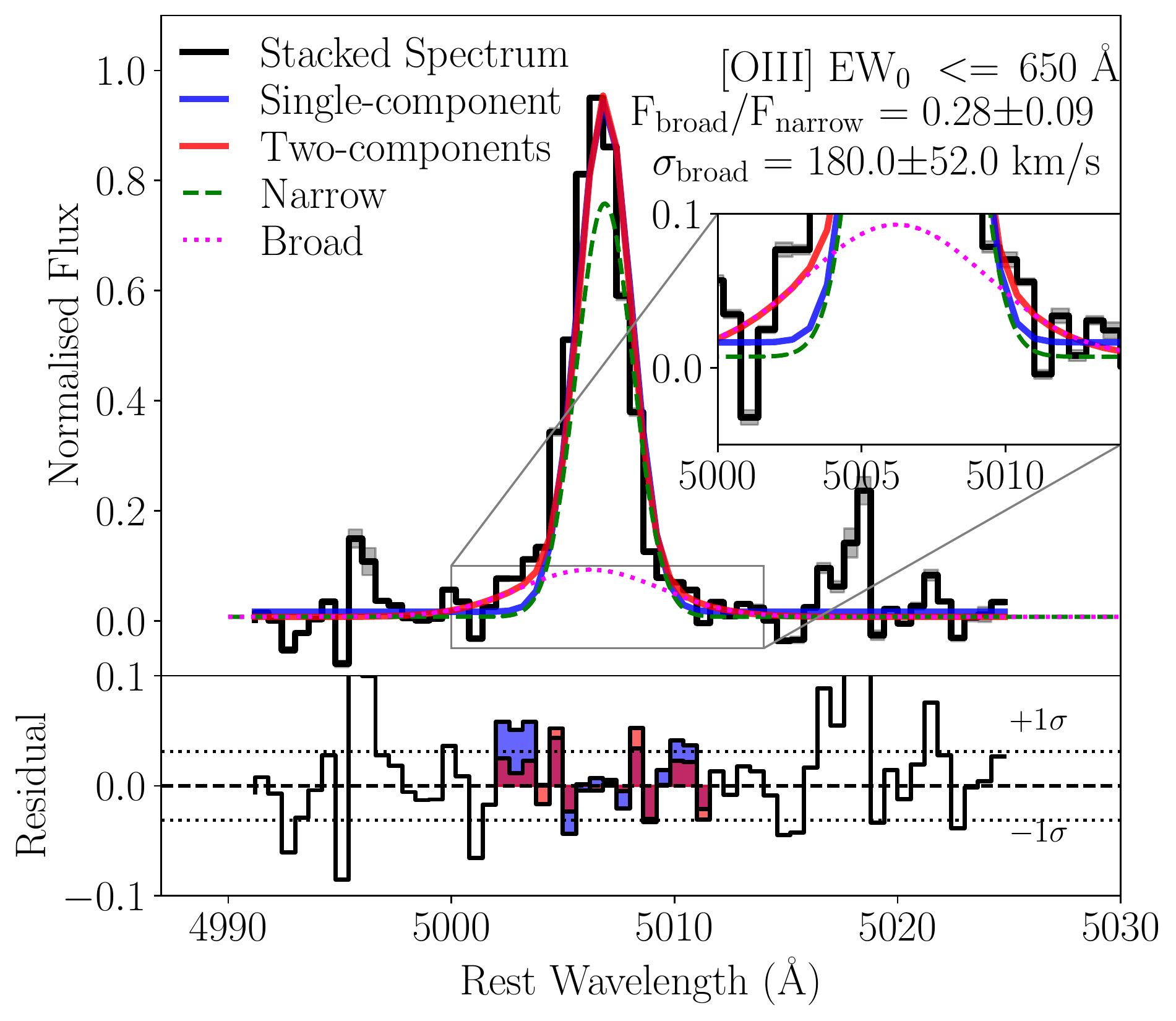}	
	\includegraphics[scale=0.45, trim=0.0cm 0.0cm 0.0cm 0.0cm,clip=true]{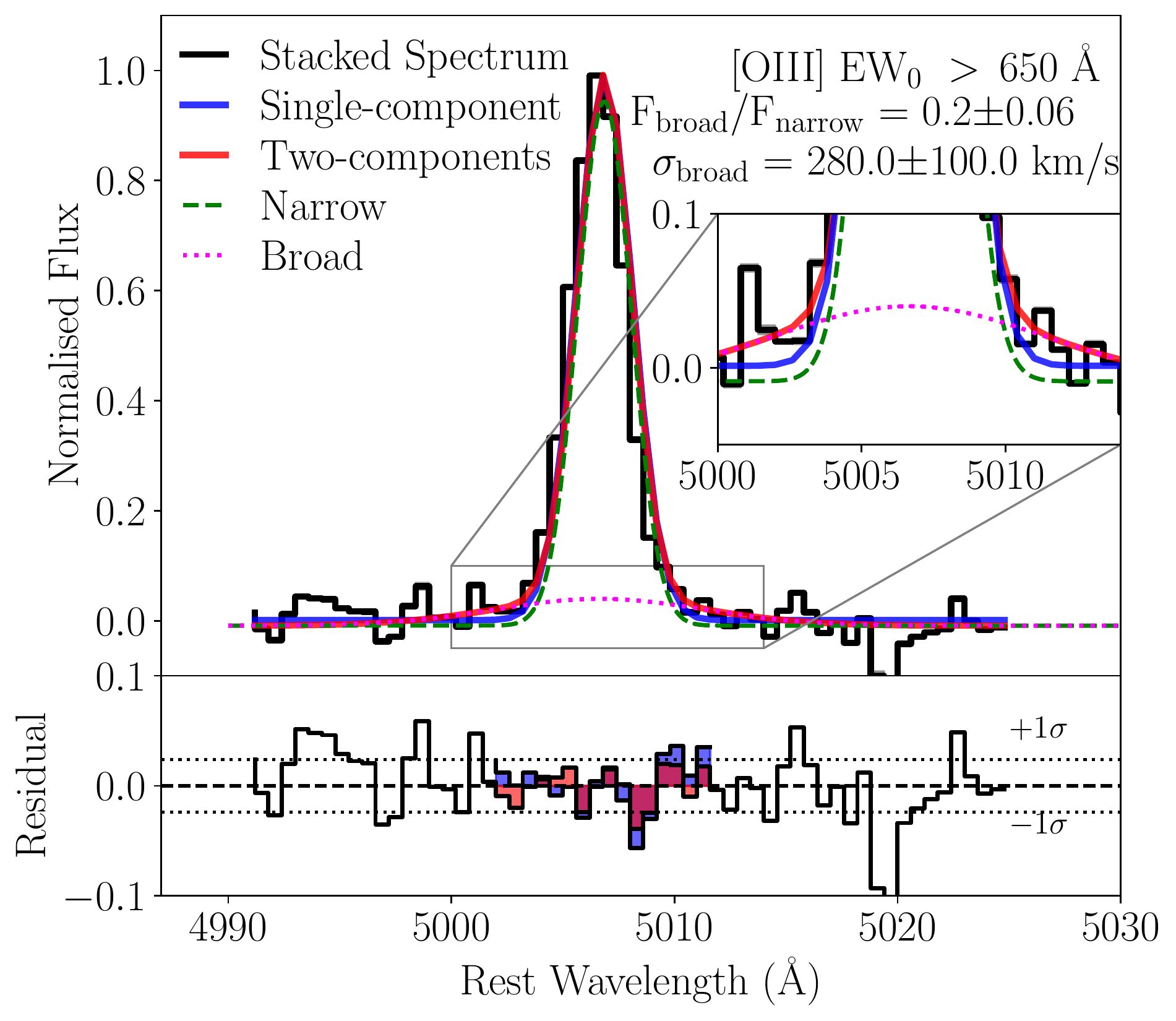}	
	\caption{Spectral stacks based on the \oiii\ EW$_0$ $<=650$\,\AA\ (left, N$=9$) and   \oiii\ EW$_0$ $>650$\,\AA\ (right, N$=12$). The colour scheme same as Figure \ref{fig:stack_spectra_mass}.  We detect a weak secondary broad component in both samples. }
	\label{fig:stack_spectra_oiii}
\end{figure*}

We detect a weak signature of a secondary component in low mass galaxies  (Figure \ref{fig:stack_spectra_mass}: left panel). The residual spectrum after fitting a single-component Gaussian exhibits multiple peaks above $1\sigma$ level redshifted from central the \oiii\ peak. The residual flux reduces by 33\% after adding a secondary broad component.  We use the Baysian Information Criteria (BIC) to test if we are overfitting the data by adding secondary Gaussian component. The lower BIC-score for the double component fit (289) than the single component fit (296) suggests that the double component Gaussian profile is a better model of the stacked spectrum. For low mass galaxies,  the ratio of flux in the broad component to narrow component (${\rm F_{broad}/F_{narrow}}$) is $ 0.2\pm0.07$ and the broad component has a velocity offset ($v_{cen}$) of  $67 \pm 39$ km/s compared to the narrow component. The FWHM of the narrow and broad components for this sample is $140\pm 7$ and $367\pm 88 $\,km/s respectively.

For galaxies in the high-mass bin, the residual spectrum after fitting a single-component Gaussian exhibit a clear peak towards the blue-side of the central \oiii\ peak (Figure \ref{fig:stack_spectra_mass}: right panel). The residual flux within $\pm 5$\,\AA\ from the central peak reduce by about 38\%  after adding adding a secondary component and almost all residual peaks fall below the 1-$\sigma$ level.  The BIC-score also improves slightly for the double component fit (261 from 264). The ratio of flux in the broad component ${\rm F_{broad}/F_{narrow}} = 0.27\pm0.1$ and $v_{\rm cen} = -95\pm68$\,km/s compared to the narrow component. The FWHM of the narrow and broad components is $150\,\pm\,12$ and $390\pm96$\,km/s respectively.       

 Thus, we detect a broad component at about $20\%$ flux level and velocity width almost 2.5 times the narrow component irrespective of the stellar mass. To check if the stacked spectra are biased towards a few galaxies with significant secondary components, we fit a two-component emission profile to each jackknife repetitions. We use the input parameters from the final stacked spectra in Figure \ref{fig:stack_spectra_mass} to fit individual repetitions. For low mass galaxies, we estimate a median ${\rm F_{broad}/F_{narrow}} = 0.2\pm 0.02$, $\sigma_{\rm broad} = 161\pm 18$\,km/s and $v_{\rm cen} = 71\pm 13$\,km/s across various jackknife repetitions.  Similarly, ${\rm F_{broad}/F_{narrow}} = 0.29\pm 0.14$, $\sigma_{\rm broad} = 155\pm 63$\,km/s and the velocity offset ($v_{\rm cen}$) is $ = -87\pm 17$\,km/s across various jackknife repetitions for high mass galaxies.  The slightly higher standard errors in the high mass repetitions is because one of the repeats have zero flux in the broad component. Therefore, our broad-component measurements in Figure \ref{fig:stack_spectra_mass} are not biased towards few galaxies and represents a mean of the final stacked populations.

\subsubsection{Stacking based on \oiii\ EW$_0$}

We use \oiii\ EW$_0$ of 650\,\AA\ to subdivide the selected sample, similar to mean \oiii\ EW$_0$ of $z>6$ galaxies in \cite{DeBarros2019}. We end up with a sample of 12 galaxies with high \oiii\ EW$_0$  (8: KMOS, 4: MOSFIRE) and 9 galaxies with low \oiii\ EW$_0$ (1: KMOS, 8: MOSFIRE). We achieve an average S/N of $\sim 40$ in each stacked spectra. 

We detect a clear secondary component in the low \oiii\ EW$_0$ galaxies (Figure \ref{fig:stack_spectra_oiii}: left panel). The residual spectrum has clear blue-shifted and redshifted peaks after fitting a single component Gaussian. After adding a secondary broad component, most peaks in the residual spectrum fall below the $1\sigma$ level with a total reduction in residual flux by 28\%.  The BIC-score improves from 248 to 245 after adding the secondary Gaussian component. The  FWHM of broad component and narrow component is $ 412\pm112$\,km/s and  $137\pm 10$\,km/s respectively. The broad component has negligible velocity offset from the line centre ($-37 \pm 31$\,km/s).   

In contrast, the secondary component in high \oiii\ EW$_0$ galaxies is extremely weak (Figure \ref{fig:stack_spectra_oiii}: right panel). The residual spectrum after fitting a single component Gaussian exhibit some weak peaks towards the red-side, although their flux is close to $1\sigma$ noise level.  The flux of residual peaks reduces slightly after fitting a secondary broad component with a 27\% reduction in total flux in the residual spectrum within $\pm 5$\AA.  Moreover the BIC-score remains the same after adding the secondary Gaussian component, suggesting that single component Gaussian might be a better model of the stacked \oiii\ spectrum.  The ratio of flux in the broad component to narrow component is $0.2\pm 0.06$ with almost negligible velocity offset ($v_{\rm cen} = -12\pm 53$\,km/s). The FWHM of the broad and narrow components is $131\pm8$ and $518\pm 161$\,km/s.

We again fit a secondary broad component to each jackknife repetitions to estimate bias in the stacked measurements towards some galaxies. For low \oiii\ EW$_0$ galaxies, ${\rm F_{broad}/F_{narrow}} = 0.29\pm 0.02$, $\sigma_{\rm broad} = 182\pm 22$\,km/s and $v_{\rm cen} = -41\pm 8$\,km/  after performing a two-component fit to each jackknife repetitions. Similarly,  we estimate ${\rm F_{broad}/F_{narrow}} = 0.23\pm 0.03$, $\sigma_{\rm broad} = 325\pm 101$\,km/s and $v_{\rm cen} = -42\pm 55$\,km/s for each jackknifed repetitions in the high \oiii\ EW$_0$ galaxies. Relatively small errors in various measurements suggests that our stacks in Figure \ref{fig:stack_spectra_oiii} are not biased towards few galaxies and indeed represents average properties of the stacked populations. 

\subsection{Detection threshold for the broad component}

The lack of strong broad component detected in the stacked spectra could also be due to our limited spectral resolution and/or S/N. We simulate a double Gaussian profile where two components have zero velocity offset, and the flux ratio of the broad component to narrow component is 0.3 and  velocity dispersion for the narrow component of $70\,$km/s, similar to the typical narrow component width in stacked spectra after accounting for the velocity dispersion.   We also add random noise to reach S/N of $=40-50$, similar to the S/N in the stacked spectra. We vary the velocity width of the broad component between $0.5-5$ times the narrow component and fit a double Gaussian profile for each iteration. 

\begin{figure}
	\centering
	\tiny
	\includegraphics[scale=0.42, trim=0.0cm 0.0cm 0.0cm 0.0cm,clip=true]{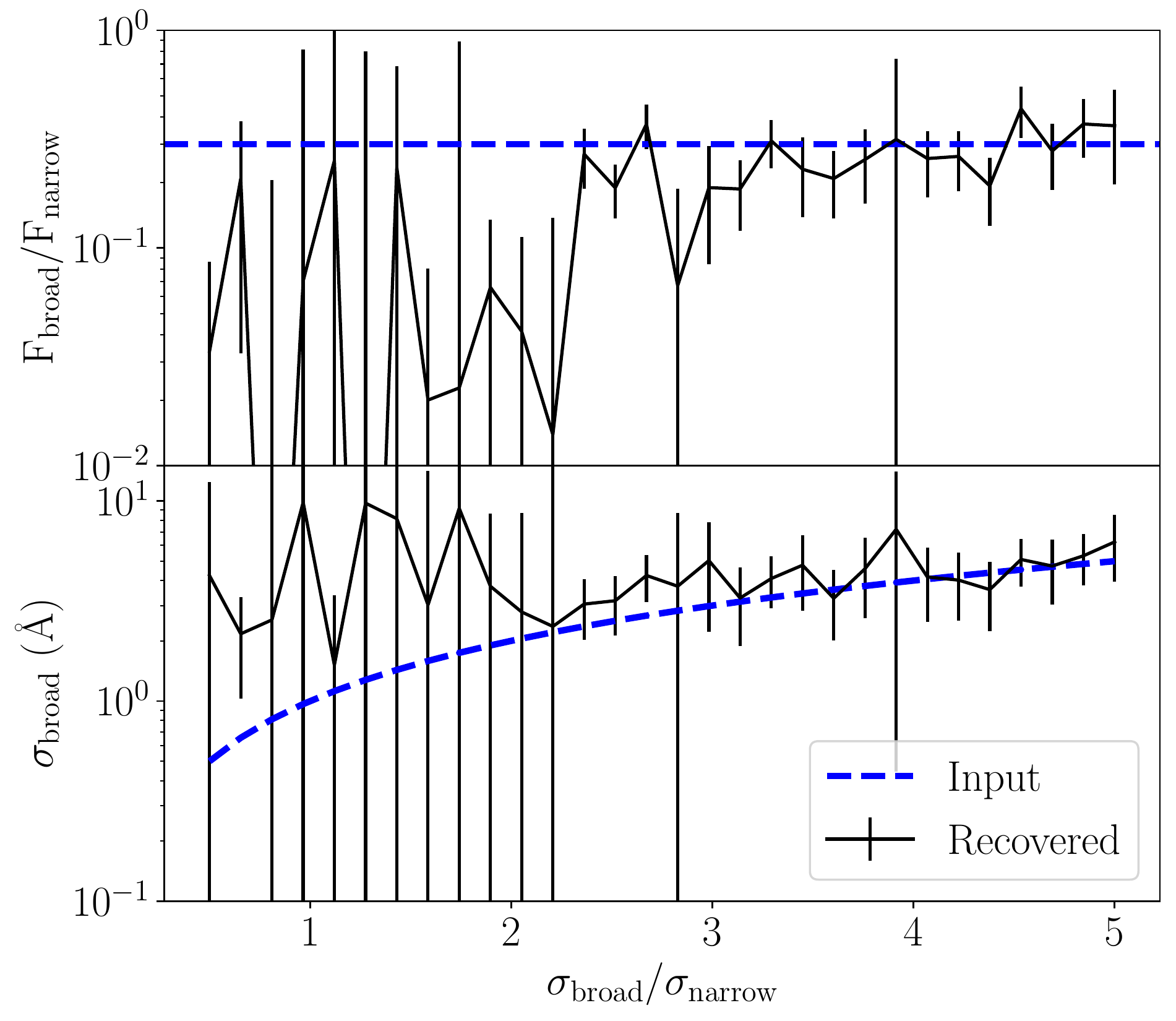}	
	\caption{The ratio of flux (top) and velocity (bottom) recovered after fitting a two-component Gaussian profile (black) as a function of the ratio of the velocity width of the broad to narrow component for the simulated \oiii\ profile. The blue dashed line indicates the input parameters. We can recover the input parameters if the velocity width of the broad component is twice the velocity width of the narrow component.}
	
	\label{fig:recovered_properties}
\end{figure}

Figure \ref{fig:recovered_properties} shows the flux ratio and velocity width of the broad component recovered after fitting a double component Gaussian profile.  These simple simulations illustrate that if the velocity width of the broad component is greater than twice the narrow component, we can safely recover the underlying broad component at the S/N and spectral resolution of our observations. Our observations will not detect low energy outflows with velocity width less than twice the velocity width of the narrow component ($\sigma_{\rm broad} \lesssim 120$km/s).  The typical velocity width of the broad component detected in the stacked spectra (Figure \ref{fig:stack_spectra_mass} \& \ref{fig:stack_spectra_oiii}) is around 2.5-4 times the velocity width of the narrow component. Therefore, we can assume that we are not missing significant flux from the broad component. 

\subsection{Testing the broad component in simulated data}\label{sec:mock_simulations}

We use the technique by \citet{Concas2022} to test if a single exponential rotating disk can produce a broad component in the integrated spectra due to observational effects such as the intrinsic rotation of galaxies, instrument broadening and/or beam-smearing. Briefly, we create mock datacubes of \oiii\ emission from $z\sim 3.5$ galaxies using the KINematic Molecular Simulator \citep[KinMS,][]{Davis2013}. The mock datacube is created using the PSF $=\,0.6''$, pixel size $=\,0.2''$ and spectral resolution of $33 $\,km/s to match with the KMOS observations. 

The galaxy is assumed to have \logmstar$=9.5$ ($> $ stellar mass for 80\% of our selected sample), $V_{\rm max} = 101\,$km/s \citep{DiTeodoro2016} and gas velocity dispersion $\sigma_{\rm gas} = 57$\,km/s \citep{Ubler2019}. We use \citet{VanderWel2014} catalogues to estimate the average scale radius of our galaxies. We limit the sample to galaxies whose sizes have been determined with good fidelity \citep[$f=0$, ][]{VanDerWel2012}. Only 11 out of 21 galaxies have accurate size estimates, resulting in an average scale radius of $\sim 0.15''$ for our sample.

\begin{figure*}
	\centering
	\tiny
	\includegraphics[scale=0.42, trim=0.0cm 0.0cm 0.0cm 0.0cm,clip=true]{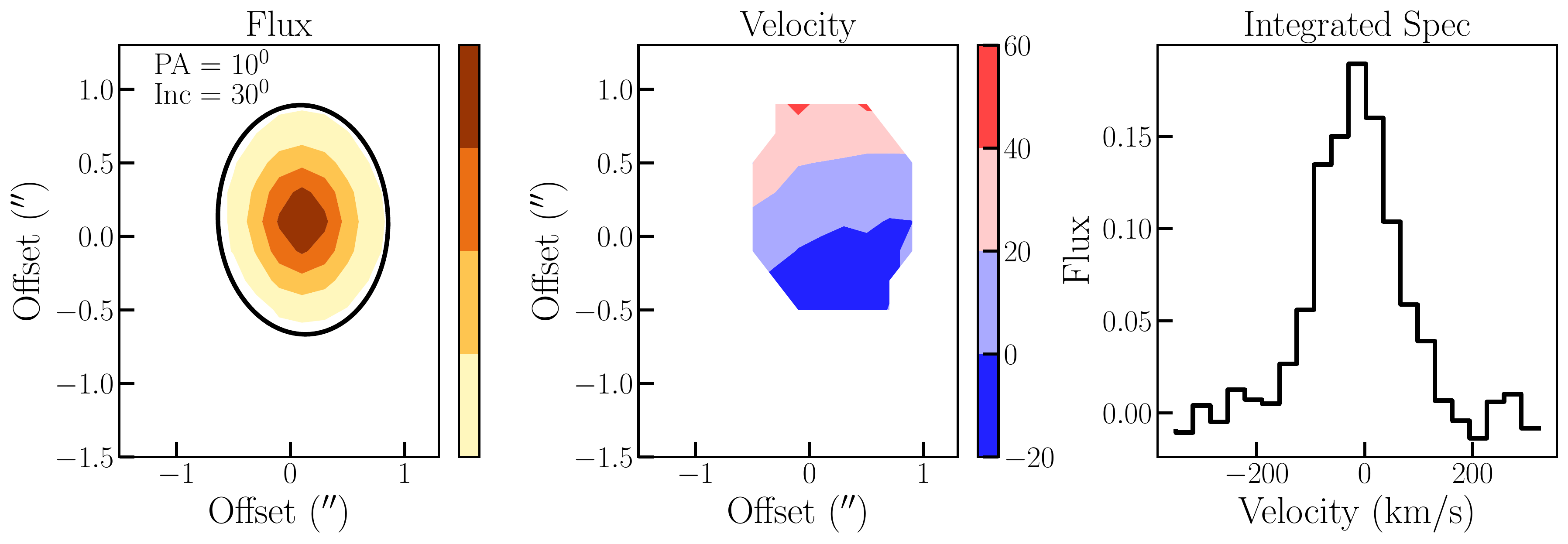}	
	\caption{The \oiii\ surface brightness (left) and the rotation velocity (middle) map from the mock datacube of a galaxy at PA=$10\deg$ and inc$=30\deg$.  The right panel shows the integrated 1D-spectrum extracted after summing over flux within $\pm$FWHM from the peak surface brightness (black ellipse in left panel).  }
	
	\label{fig:mock_cube}
\end{figure*}

Figure \ref{fig:mock_cube} shows the flux and velocity map for an example mock datacube. Note that we barely resolve the kinematics in the mock datacubes because of the compact nature of our sample. We create mock datacubes for 10 galaxies (average number of galaxies in the stacked spectra in Section \ref{sec:stacking_results}) assuming a uniform random distribution for their position angle and inclination on the sky.  We extract the mock 1D-spectra from mock datacubes using the procedure described in Section \ref{sec:kmos_em_extraction} and add random noise to reach a S/N of $\sim 10$, in line with the S/N cut-off of the selected spectra (Figure \ref{fig:mock_cube}). From the mock 1D-spectra of 10 galaxies, we create a mock stacked spectrum using the jackknife technique (\ref{sec:stacking}), and fit double and single-component Gaussian profiles to the stacked spectra (Section \ref{sec:stacking_results}).

\begin{figure}
	\centering
	\tiny
	\includegraphics[scale=0.33, trim=0.0cm 0.0cm 0.0cm 0.0cm,clip=true]{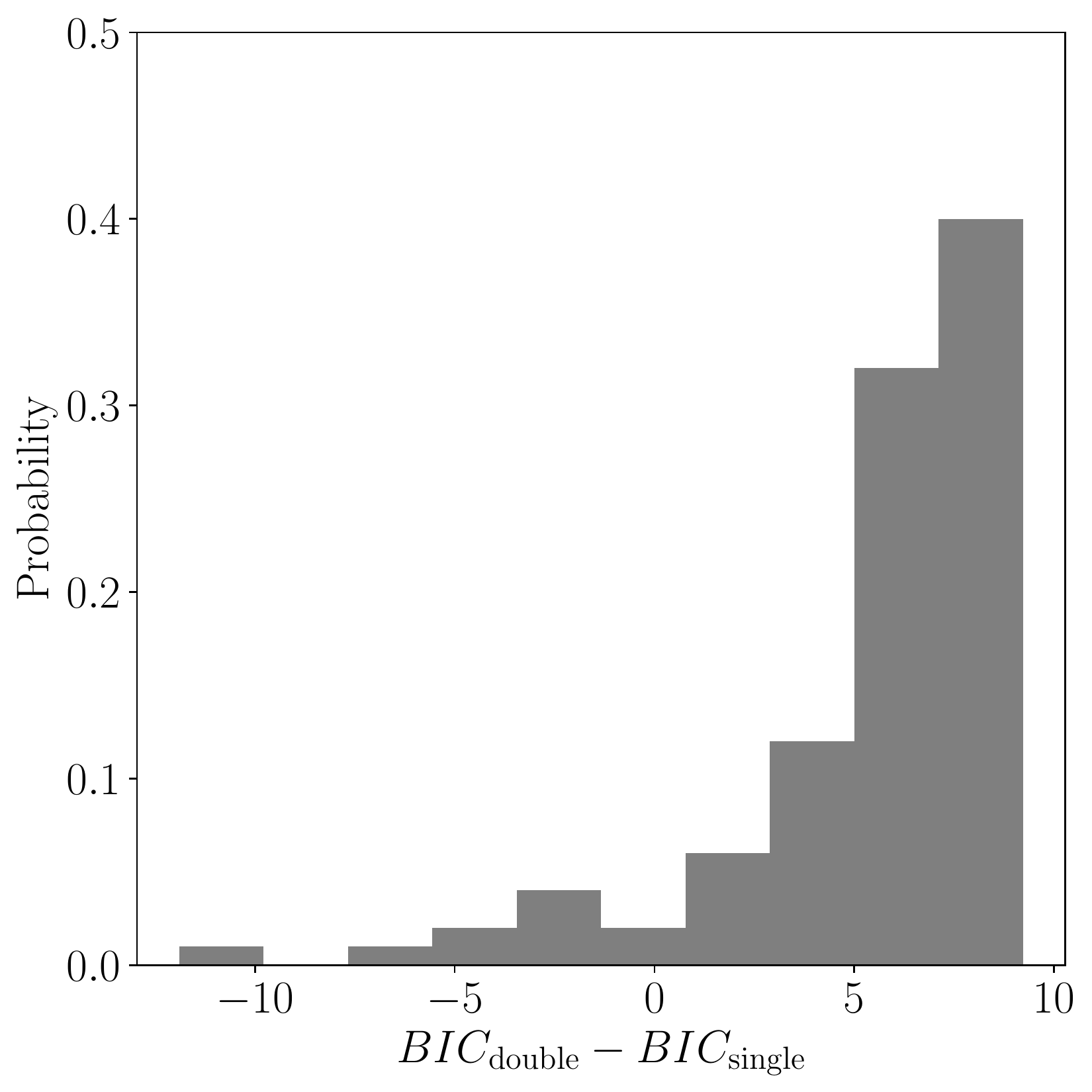}	
	\caption{ The difference between the BIC score from a double component and single component fit to the stacked spectra from mock datacubes. Only $<10\%$ of iterations have a lower BIC score  for the double component fit than the single component. 
   }
	
	\label{fig:mock_fit_comp}
\end{figure}

We create 100 mock stacked spectra using the procedure described above, and fitting single and double Gaussian emission profiles to each. Figure \ref{fig:mock_fit_comp} shows the distribution of the difference between the BIC score between double and single component fit. The BIC score of the double component fit is lower in only $<10\%$ of iterations, suggesting that the single Gaussian better fits the profile for  $>90\%$ of the mock stacked spectra. 

We suspect that our mock integrated spectra are dominated by broadening introduced by the gas velocity dispersion that washes out any non-Gaussianity introduced by the rotation and/or beam smearing in the integrated spectra. Therefore, the broad components detected in the observed spectra are most likely to trace intrinsic non-circular motions associated with outflows. Note that the resolved kinematics measurements at $z>2$ in \cite{Ubler2019} are for galaxies with \logmstar$>10$, therefore the gas velocity dispersion in our mock stacks might be overestimated. Based on our model, we think the intrinsic rotation of galaxies, instrument broadening and/or beam-smearing cannot explain the broad component detected in Section \ref{sec:stacking_results}.   

\section{Discussion}\label{sec:discussion}

This work analyses the prevalence of galactic-scale outflows in the extreme emission-line galaxies at $z=3-4$ by stacking the \oiii\ emission profiles of individual galaxies.  The extreme emission-line galaxies are selected from the \mosel\ survey and have an average spectroscopic \oiii\ EW$_0$ ($780$\AA) and stellar mass (\logmstar$=8.88$) equivalent to the galaxies detected at $z>6$ \citep{DeBarros2019}.  The results presented in this paper are only based on the stacked \oiii\ spectra because  \hb\ detected in only 65\% of the sample with  an average S/N of $\sim 5$, insufficient to detect the weak broad component.
 We stack our sample into two bins of stellar mass and \oiii\ EW$_0$. We detect a weak broad component that accounts for approximately 20\% of the flux in the \oiii\ line in all cases except for the high \oiii\ EW stack. 

\subsection{Outflow properties}\label{sec:outflow_properties}

We calculate  the mass loading factor and outflow velocity to determine the efficiency of star-formation driven outflows in samples with significant  broad component detections (Figure \ref{fig:stack_spectra_mass} \& \ref{fig:stack_spectra_oiii}).  Calculating the mass loading factor from the \oiii\ emission profile is not straightforward. \cite{Freeman2019, Concas2022} shows that \halpha\ and \oiii\ emission lines have a similar ratio of flux in the broad and narrow component, especially for galaxies below $M_* <10^{10}\,$\msun\ at $z\sim 2.0$. Therefore, we can use the parametrisation developed for the \halpha\ emission line to estimate the mass loading factors for our galaxies. 

We use equation 6 in \cite{Davies2019b},
\begin{equation}\label{eq:mass_loading_factor}
\eta = \left(\frac{3.15\times2.1\times10^{48}}{1.99\times10^{33}}\right) \frac{1.36m_{\rm H}}{\gamma_{\rm H\alpha} n_e} \left(\frac{f_{\rm broad}}{f_{\rm narrow}}\right) \frac{v_{\rm out}}{R_{\rm out}}, 
\end{equation}
here the outflow dependent parameters are: $f_{\rm broad}/f_{\rm narrow} $ = flux ratio of the broad to narrow component, $v_{\rm out}$ = outflow velocity, and $R_{\rm out}$ = the extent of the outflow. $1.36m_{\rm H}$ is the effective nucleon mass of gas with 10\% Helium, $\gamma_{\rm H\alpha} = 3.56 \times 10^{-25} {\rm erg\ s^{-1}\ cm^3}$ is the \halpha\ emissivity at $10^4$\,K and $n_e$ is the local electron density in the outflow.  To keep our results comparable with \cite{Davies2019b}, we use the same electron density ($n_e = 380 {\rm cm^{-3}}$).  The outflow velocity is measured as $v_{\rm out} = v_{\rm center, broad} - 2\sigma_{\rm broad}$ \citep[][]{Genzel2011, Davies2019b}.

Estimating the $R_{\rm out}$ is not obvious for our galaxies because they are not resolved in the KMOS observations.  We assume that the extent of the outflow is the same as the effective radius of galaxies. Using the galaxies whose sizes have been measured with good fidelity in \citet[$f=0$, ][]{VanDerWel2012} catalogues, the $R_{\rm out} \sim 0.15''\,=1.2$\,kpc. This is slightly smaller than the $R_{\rm out}$ assumed by \cite{Davies2019b} (1.7\,Kpc), but our conclusions do not change significantly if we use 1.7\,Kpc instead.  

\begin{figure}
	\centering
	\tiny
	\includegraphics[scale=0.45, trim=0.0cm 0.0cm 0.0cm 0.0cm,clip=true]{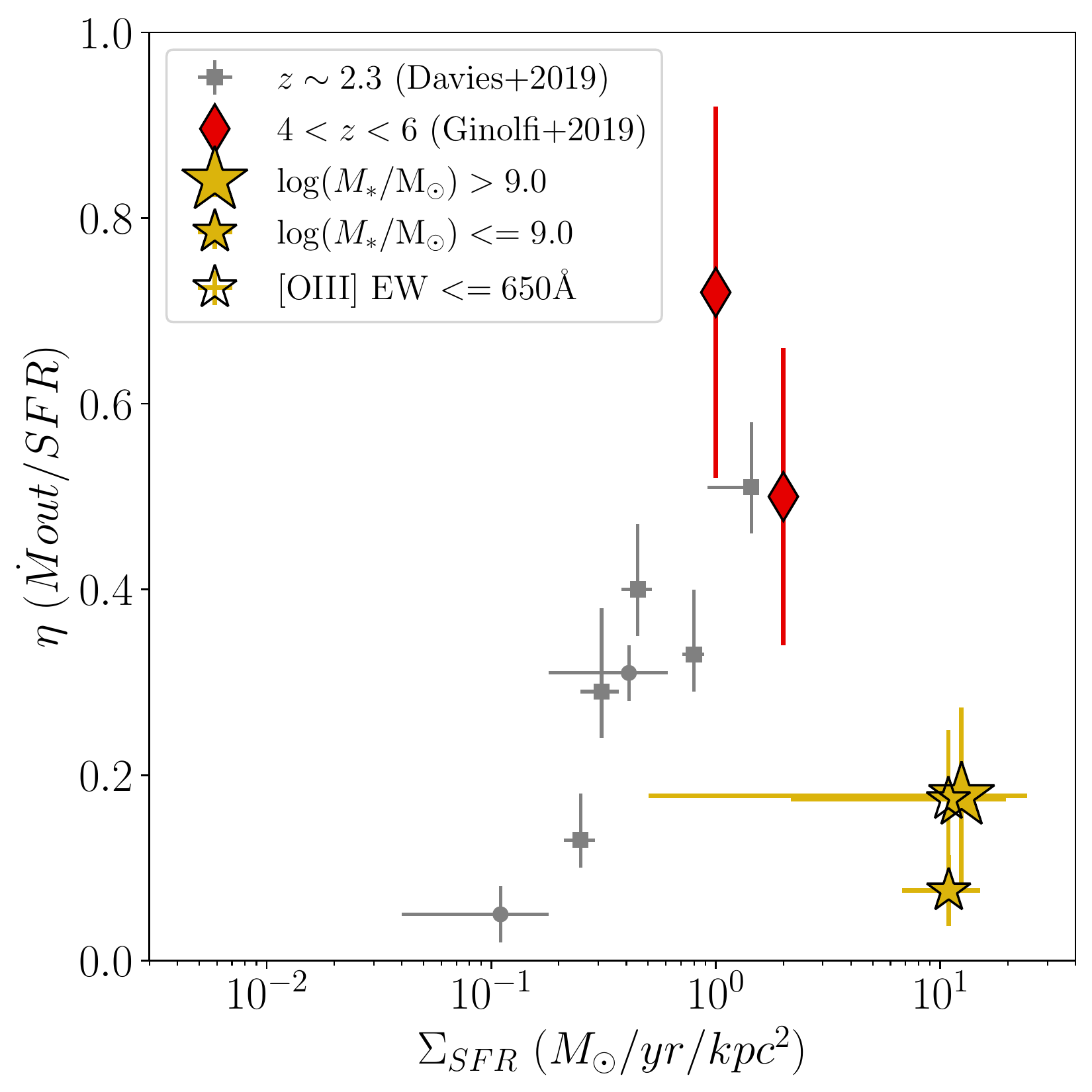}
	\caption{Mass loading factor ($\eta$) as a function of star formation surface density from stacked spectra of galaxies with \logmstar$>9.0$ (large solid star), \logmstar$<=9.0$ (small solid star) and EW$_0<=650$\,\AA\ (small open star). The errors bars represents the 1-sigma error from the best-fit parameters. The grey squares and circles represent the $\eta-\Sigma_{\rm SFR}$ relation for $z\sim 2.3$ galaxies \citep{Davies2019b} and red diamonds represent the median mass loading factor of \ciiradio\ emitters at $4<z<4$ \citep{Ginolfi2019}.}
	\label{fig:mass_loading_factor}
\end{figure}

Figure \ref{fig:mass_loading_factor} shows the mass loading factor as a function of star-formation surface density ($\Sigma_{\rm SFR}$). Errors in the mass loading factor are translated from the 1-sigma uncertainty in the best-fit parameters. They do not account for the systematic uncertainties due to $n_e$ and $R_{\rm out}$.  Our sample has extremely low mass loading factor than expected based on the $\eta-\Sigma_{\rm SFR}$ relation using star-forming galaxies at $z\sim2.3$ and  $z>5$ \citep{Davies2019b, Ginolfi2019}. Note that the average stellar mass of our sample is \logmstar$\sim8.88$, whereas both  \citet[\logmstar$=10.1$]{Davies2019b} and \citet[ \logmstar$=9.95$]{Ginolfi2019} are targeting galaxies that are nearly 1.0\,dex more massive than our sample. We need an average electron density of $30\,{\rm cm}^{-3}$ at $R_{\rm out} = 1.2$\,kpc to push the mass loading factor of our galaxies to close to unity. However, such low electron densities are atypical for high redshift galaxies with high SFRs even at low stellar masses \citep[$\sim 100-350 {\rm cm}^{-3} $, ][]{Onodera2016, Kaasinen2017, Harshan2020, Davies2021a}. Reducing the $R_{\rm out}$ by 25\% increases the mass loading factor by 30\%, therefore it is possible that outflows exist at significantly smaller scale than the full extent of the galactic-disk in our sample.

We are using the SFR calculated from the SED fits in Section \ref{sec:sed_fit} and the typical effective radius of our galaxies ($\sim\, 0.15''$) to estimate the $\Sigma_{\rm SFR}$. Whereas \cite{Davies2019b} uses spatially resolved \halpha, and \cite{Ginolfi2019} relies on empirical conversion between infrared luminosity at $158\mu m$ to infer luminosity between 8 to 1000\,$\mu m$ \cite{Bethermin2020}.  The SFRs measured from infrared luminosity are consistent with those derived using \halpha, albeit with significant scatter \citep[0.2\,dex,][]{DominguezSanchez2012, Shivaei2016, Onodera2016, Wisnioski2019}. The SFRs measured from our best-fit SEDs after including emission lines are similar to the SFRs measured by infrared luminosity. 

Direct detection of \halpha\ emission were impossible beyond $z>2.8$, until \textit{JWST} became available.  The SFRs measured from SED fitting after including emission lines are higher by 0.3\,dex than SFRs measured from UV continuum for a sample of extreme emission-line galaxies at $z\sim 3$ \citep{Onodera2020}. A recent analysis by \cite{Fetherolf2021} showed that UV continuum   underestimates the SFR by about 0.2-0.3\,dex compared to the \halpha\ emission line for galaxies with SFRs $> 2$\,\msun/yr at $z\sim2$. Moreover, we would need to reduce the \halpha\ SFR by at least a factor of 10 for \mosel\ galaxies to follow the $\eta - \Sigma_{\rm SFR}$ relation by \citep{Davies2019b}. \cite{Cohn2018} showed that extreme emission line galaxies are undergoing their first burst of star formation. Therefore, it is unlikely that \halpha\ SFRs for \mosel\ galaxies ($\sim 10$ Myr timescale) are 10 times smaller than SFRs measured from the full SED fitting ($\sim 100$ Myr timescale). 

Our observations suggest that galactic-scale outflows are extremely weak in our extreme emission-line galaxies.  Previous studies at $z>1$ detect prominent galactic-scale outflows at high star formation surface densities, with $\eta \sim 1-2$ at $\Sigma_{\rm SFR}$ similar to our sample \citep{Newman2012, Davies2019b}. Whereas, the mass loading factor of our sample is similar to the low stellar mass galaxies (\logmstar$\sim 9$) at $z>1$ \citep{Freeman2019, Swinbank2019, Concas2022}, even if they have almost 10 times higher $\Sigma_{\rm SFR}$.

Galactic-scale outflows are typically invoked to explain the chemical enrichment of the intergalactic medium (IGM) and shut down the SFR in galaxies undergoing a burst of star formation. Extremely low mass loading factors in the low mass galaxies with high SFRs implies that outflows alone are inefficient at pushing gas out to large galacto-centric distances in such systems. We speculate that rapid consumption of gas supply might be primarily responsible for shutting down SFR in extreme star-forming galaxies. 

\subsection{Outflows in extreme emission-line galaxies}\label{sec:outflows_eelgs}

Lack of or reduced efficiency of galactic-scale outflows detected in our observations could be  due to the  multi-phase nature of the outflows. The \oiii\ emission line only probes the outflowing component in the $10^4-10^5$\,K gas.  Observations using other tracers such as CO, \ciiradio find significant outflowing gas component in colder phases of the ISM \citep{Bischetti2019, Herrera-Camus2021, Veilleux2020}.
Observations of high ionisation species such as \civ\ along the quasar sightlines up to 200\,kpc from any nearby \lya\ emitters suggest hot phases ($10^6-10^7$\,K) of enriched gas being pushed out from the interstellar medium \citep{Bielby2020, Diaz2020, Bischetti2022}. Thus, observations relying just on the optical emission lines might be missing large components of the outflowing gas. 

Hydrodynamic cosmological simulations suggests a time delay of $\sim60$\, Myrs between the star formation peak and galactic-scale gas outflows \citep{Muratov2015}. The extreme emission line galaxies are at the peak in their SFHs \citep{Cohn2018} and might be at the pre-outflow stage. Most previous observations find a strong correlation between the instantaneous star formation rate ($<10$\,Myrs) and the mass loading factor  \citep{Newman2012, Freeman2019, Davies2019b}, which cannot be explained by a time-delay  of $60-100$\,Myrs between starburst and outflow event. 

Another possibility is galactic-winds might be suppressed in extreme star-forming conditions. The young and compact starbursts in the local universe (Green pea galaxies) show extremely narrow \lya\ emission profiles and the low ionisation species have almost negligible velocity offsets \citep{Jaskot2017}. The minimal systematic velocity offset between the HI gas and metal ions such as \SiII, \OI\ suggests that catastrophic cooling and high pressure in extreme starbursts is suppressing the galactic winds from stellar feedback and supernovae explosions \citep{Silich2007, Silich2017}.

Semi-analytic models of winds indicate that bubbles in highly star-forming galaxies transition rapidly (<1 Gyrs) from the energy-bounded to the momentum-bounded state,  leading to efficient radiative cooling \citep{Lochhaas2018}. The galactic winds in a catastrophically cooling scenario can only  reach a few pcs from the launch site before radiatively cooling and falling back onto the galaxy. Thus, the low global mass loading factors for the highly star-forming galaxies like green peas could be due to the catastrophic cooling of the galactic winds \citep{Jaskot2013, Jaskot2017, Berg2019a}. 

Most extreme star-forming galaxies produce an excess flux in the emission lines corresponding to the highly ionised species such as \civ, \heii\ etc \citep{Berg2019, Tang2020, Berg2021} that are typically associated with harder photoionising sources \citep{Jaskot2013, Berg2019a, Berg2021}.  Recent models show that bubbles in a catastrophic cooling scenario around super star-clusters might be partially or completely density-bounded instead of radiation-bounded as is typical in \hii\ regions \citep{Gray2019}. The density-bounded nature of the \hii\ nebula can produce higher ionisation species without the requirement for the exotic photoionising sources \citep{Gray2019, danehkar2021}. Thus, the prevalence of the high ionisation species in the spectra of extreme star-forming galaxies could also be indicative of the suppression or catastrophic cooling of galactic winds in such systems. Observations at rest-UV wavelength are required to quantify the flux of high ionisation emission lines in the \mosel\ sample.

\subsection{Possibility of gas inflows in \mosel\ sample}\label{sec:inflows}

The secondary broad component detected in low mass galaxies (Figure \ref{fig:stack_spectra_mass}) is redshifted compared to the central \oiii\ peak. The outflowing gas is typically detected blue-shifted compared to the central ISM because photons from outflowing gas moving away from the observer are severely attenuated by the intervening ISM. The geometry of the outflow and projection effects can lead to the detection of a redshifted broad component for individual galaxies \citep{Herrera-Camus2021}, but unlikely in a stacked profile of multiple galaxies.   On the other hand, the gas inflows on the observer side would be redshifted relative to the central ISM and are more likely to be observed. We speculate that the photoionisation of the inflowing gas could be producing the secondary broad component in low mass galaxies.     

Accretion of gas near the cosmic noon is extremely likely, especially in high star-forming galaxies \citep{Bouche2013}.   Using simple analytic assumptions \citet{Kacprzak2016} showed that cold-mode gas accretion is responsible for the relatively lower metallicity of the high star-forming galaxies ($>10\,$\msun/yr).  Observations of extended \lya\ halos  also indicate large reservoirs of the neutral hydrogen in the circum-galactic medium of high-redshift galaxies \citep{Wisotzki2018, Leclercq2017, Leclercq2020}.

Simulations of \lya\ emission around high redshift galaxies ($z=3.5$) show that at a small galactocentric distance ($r<0.5\,R_{\rm vir}$) the inflowing gas accounts for 70\% of the neutral gas component \citep{Mitchell2021}.  However, the total fraction of gas influenced by stellar feedback (both inflowing and outflowing) remains relatively low ($\sim 7\%$) across all radii in these simulations.  The inflowing gas needs to be photoionised by the stellar feedback for it to explain the redshifted broad component in the \oiii\ emission profile (Figure \ref{fig:stack_spectra_mass}). Ionised gas inflows have been observed for some galaxies in the local universe  \citep{Cameron2021}. Detailed photoionisation models of the circum-galactic medium around high-redshift galaxies are required to understand the fraction and surface brightness of the gas inflows in different phases of the ISM (neutral/ionised).

We use archival MUSE spectra to gain insights into the kinematic of gas around \mosel\ sample using \lya\ emission.  Four out of 12 galaxies with \logmstar$<=9.0$ have archival MUSE spectra from the MUSE-Wide survey \citep{Herenz2017} of which only two have been modelled by \cite{Gronke2017} using an expanding shell model. The S/N of the \lya\ emission is insufficient for the other two galaxies. The shell velocities for these two galaxies are $-350\ {\rm and}\ 35$\,km/s. In contrast, the typical expansion velocity for $z=3-4$ \lya\ emitters in the MUSE-Wide survey is 150\,km/s. 

The negative or close to zero expansion velocity suggests that the \lya\ emission in the two \mosel\ galaxies is either coming from an inflowing or a minimally expanding shell. Thus, these two galaxies could be experiencing strong gas inflows. Observations of \lya\ emission profiles for the full \mosel\ sample are necessary to confirm the prevalence of inflowing shells in the extreme emission line emitters.  

\section{Summary and Conclusions}\label{sec:summary}

We use deep K-band spectroscopic observations from the \mosel\ survey and stack the \oiii\ emission profile to quantify the prevalence of galactic-scale outflows in extreme emission-line galaxies at $z=3-4$. The targets selected in this work have \oiii\ EW$_0$ ($\sim 600$\AA) and stellar mass (\logmstar$=8.98$) comparable to galaxies at $z>6$, and thus are analogous to the $z>6$ galaxies (EoR analogues, Figure \ref{fig:o3_ew}). 

 We find a weak signature of a broad component in the stacked spectra of both low (\logmstar$<=9.0$) and high mass galaxies (\logmstar$>9.0$, Figure \ref{fig:stack_spectra_mass}). The broad component is not significantly detected in the stacked spectra of galaxies with \oiii\ EW$_0$ $>650$\,\AA\ but clearly detected for galaxies with \oiii\ EW$_0$ $<650$\,\AA\  (Figure \ref{fig:stack_spectra_oiii}). Each stacked spectra has an average S/N of nearly 40 and velocity resolution of 36\,km/s. Only 20-25\% of the total \oiii\ flux in the stacked spectra is in the broad component, and the FWHM of the broad component is nearly 400 km/s (3 times the FWHM of narrow component). We mock-simulate the \oiii\ emission profile from \logmstar$=9.5$ galaxy with scale radius $=0.15''$ (similar to our sample) to rule out that observational effects (rotation, instrument resolution and/or beam smearing) can explain the broad  component in our stacked spectra (Section \ref{sec:mock_simulations}). 
 
  Using parametrisation by \citet{Davies2019b} and assuming the broad component of \oiii\ and \halpha\ have similar properties, we estimate that the mass loading factor for the EoR analogues is $\sim 0.2$ (Figure \ref{fig:mass_loading_factor}). The mass loading factor that we derive for EoR analogues is less than or equal to comparable mass galaxies \citep{Freeman2019, Swinbank2019}, despite galaxies in our sample having 10 times higher star-formation rates.

Our observations indicate that the galactic-scale outflows are extremely weak in the galaxies analogous to the EoR galaxies. We hypothesise that either the multi-phase nature of the supernovae driven outflows limits us from detecting significant outflowing gas in the warm ionised medium ($10^4-10^5$\,K, Section \ref{sec:outflows_eelgs}), or the catastrophic collapse of galactic-winds in extreme star-forming regimes is responsible for the low globally averaged mass loading factors. A similar suggestion of suppressed superwinds have been found in green-pea galaxies in the local universe \citep{Jaskot2013, Jaskot2017, Berg2019a}. The weak outflow signature in the low mass galaxies suggests that massive galaxies might be driving the early chemical enrichment ($z>6$) of the intergalactic medium.  

Moreover, we detect a redshifted broad component for galaxies with stellar mass \logmstar$<=9.0$ (Figure \ref{fig:stack_spectra_mass}), tentatively suggesting that inflowing gas rather than outflows  might be producing the  broad component (Section \ref{sec:inflows}). Gas inflows are typically detected through detailed kinematic analysis of \lya\ emission around high redshift galaxies \citep{Martin}.  If true, then our observations will be the first detection of gas inflows using the optical emission lines in the high redshift universe.  The EoR analogues are a unique sample of highly star-forming, low stellar mass galaxies (Figure \ref{fig:eelgs_properties}) that are undergoing their first burst of star formations \citep{Cohn2018}, therefore likely to be experiencing strong gas inflows. Deeper \lya\ emission maps with MUSE and/or higher spectral resolution with future instruments such as MAVIS \citep{McDermid2020} will confirm the prevalence and relative strength of gas inflows around EoR analogues.  

\section*{Acknowledgements}
The authors thank the referee for their extremely valuable feedback and suggestions.
This research were supported by the Australian Research Council Centre of Excellence for All Sky Astrophysics in 3 Dimensions (ASTRO 3D), through project number CE170100013. 

Based on observations collected at the European Organisation for Astronomical Research in the Southern Hemisphere under ESO programme 0104.B-0559. Some of the data presented herein were obtained at the W. M. Keck Observatory, which is operated as a scientific partnership among the California Institute of Technology, the University of California and the National Aeronautics and Space Administration. The Observatory was made possible by the generous financial support of the W. M. Keck Foundation. The authors wish to recognize and acknowledge the very significant cultural role and reverence that the summit of Maunakea has always had within the indigenous Hawaiian community.  We are most fortunate to have the opportunity to conduct observations from this mountain.

\section*{Data Availability}

The data underlying this article will be shared on reasonable request to the corresponding author.



\bibliographystyle{mnras}

\begin{thebibliography}{}
	\makeatletter
	\relax
	\def\mn@urlcharsother{\let\do\@makeother \do\$\do\&\do\#\do\^\do\_\do\%\do\~}
	\def\mn@doi{\begingroup\mn@urlcharsother \@ifnextchar [ {\mn@doi@}
		{\mn@doi@[]}}
	\def\mn@doi@[#1]#2{\def\@tempa{#1}\ifx\@tempa\@empty \href
		{http://dx.doi.org/#2} {doi:#2}\else \href {http://dx.doi.org/#2} {#1}\fi
		\endgroup}
	\def\mn@eprint#1#2{\mn@eprint@#1:#2::\@nil}
	\def\mn@eprint@arXiv#1{\href {http://arxiv.org/abs/#1} {{\tt arXiv:#1}}}
	\def\mn@eprint@dblp#1{\href {http://dblp.uni-trier.de/rec/bibtex/#1.xml}
		{dblp:#1}}
	\def\mn@eprint@#1:#2:#3:#4\@nil{\def\@tempa {#1}\def\@tempb {#2}\def\@tempc
		{#3}\ifx \@tempc \@empty \let \@tempc \@tempb \let \@tempb \@tempa \fi \ifx
		\@tempb \@empty \def\@tempb {arXiv}\fi \@ifundefined
		{mn@eprint@\@tempb}{\@tempb:\@tempc}{\expandafter \expandafter \csname
			mn@eprint@\@tempb\endcsname \expandafter{\@tempc}}}
	
	\bibitem[\protect\citeauthoryear{Berg, Erb, Henry, Skillman  \& McQuinn}{Berg
		et~al.}{2019a}]{Berg2019a}
	Berg D.~A.,  Erb D.~K.,  Henry R. B.~C.,  Skillman E.~D.,   McQuinn K. B.~W.,
	2019a, \mn@doi [The Astrophysical Journal] {10.3847/1538-4357/ab020a}, 874,
	93
	
	\bibitem[\protect\citeauthoryear{Berg, Chisholm, Erb, Pogge, Henry  \&
		Olivier}{Berg et~al.}{2019b}]{Berg2019}
	Berg D.~A.,  Chisholm J.,  Erb D.~K.,  Pogge R.,  Henry A.,   Olivier G.~M.,
	2019b, \mn@doi [The Astrophysical Journal] {10.3847/2041-8213/ab21dc}, 878,
	L3
	
	\bibitem[\protect\citeauthoryear{Berg, Chisholm, Erb, Skillman, Pogge  \&
		Olivier}{Berg et~al.}{2021}]{Berg2021}
	Berg D.~A.,  Chisholm J.,  Erb D.~K.,  Skillman E.~D.,  Pogge R.~W.,   Olivier
	G.~M.,  2021, \mn@doi [The Astrophysical Journal] {10.3847/1538-4357/ac141b},
	922, 170
	
	\bibitem[\protect\citeauthoryear{B{\'{e}}thermin et~al.,}{B{\'{e}}thermin
		et~al.}{2020}]{Bethermin2020}
	B{\'{e}}thermin M.,  et~al., 2020, \mn@doi [Astronomy & Astrophysics]
	{10.1051/0004-6361/202037649}, 643, A2
	
	\bibitem[\protect\citeauthoryear{Bielby et~al.,}{Bielby
		et~al.}{2020}]{Bielby2020}
	Bielby R.~M.,  et~al., 2020, \mn@doi [Monthly Notices of the Royal Astronomical
	Society] {10.1093/mnras/staa546}, 493, 5336
	
	\bibitem[\protect\citeauthoryear{Bischetti et~al.,}{Bischetti
		et~al.}{2019}]{Bischetti2019}
	Bischetti M.,  et~al., 2019, \mn@doi [Astronomy and Astrophysics]
	{10.1051/0004-6361/201935524}, 628, 1
	
	\bibitem[\protect\citeauthoryear{Bischetti et~al.,}{Bischetti
		et~al.}{2022}]{Bischetti2022}
	Bischetti M.,  et~al., 2022, ] {10.1038/s41586-022-04608-1}, 6
	
	\bibitem[\protect\citeauthoryear{Bouch{\'{e}}, Murphy, Kacprzak, P{\'{e}}roux,
		Contini, Martin  \& Dessauges-Zavadsky}{Bouch{\'{e}}
		et~al.}{2013}]{Bouche2013}
	Bouch{\'{e}} N.,  Murphy M.~T.,  Kacprzak G.~G.,  P{\'{e}}roux C.,  Contini T.,
	Martin C.~L.,   Dessauges-Zavadsky M.,  2013, \mn@doi [Science]
	{10.1126/science.1234209}, 341, 50
	
	\bibitem[\protect\citeauthoryear{Bruzual \& Charlot}{Bruzual \&
		Charlot}{2003}]{Bruzual2003}
	Bruzual G.,  Charlot S.,  2003, \mn@doi [Monthly Notices of the Royal
	Astronomical Society] {10.1046/j.1365-8711.2003.06897.x}, 344, 1000
	
	\bibitem[\protect\citeauthoryear{Cameron et~al.,}{Cameron
		et~al.}{2021}]{Cameron2021}
	Cameron A.~J.,  et~al., 2021, \mn@doi [The Astrophysical Journal Letters]
	{10.3847/2041-8213/ac18ca}, 918, L16
	
	\bibitem[\protect\citeauthoryear{Chabrier}{Chabrier}{2003}]{Chabrier2003}
	Chabrier G.,  2003, \mn@doi [Publications of the Astronomical Society of the
	Pacific] {10.1086/376392}, 115, 763
	
	\bibitem[\protect\citeauthoryear{Chevalier \& Clegg}{Chevalier \&
		Clegg}{1985}]{Chevalier1985}
	Chevalier R.~A.,  Clegg A.~W.,  1985, \mn@doi [Nature] {10.1038/317044a0}, 317,
	44
	
	\bibitem[\protect\citeauthoryear{Cohn et~al.,}{Cohn et~al.}{2018}]{Cohn2018}
	Cohn J.~H.,  et~al., 2018, \mn@doi [The Astrophysical Journal]
	{10.3847/1538-4357/aaed3d}, 869, 141
	
	\bibitem[\protect\citeauthoryear{Concas et~al.,}{Concas
		et~al.}{2022}]{Concas2022}
	Concas A.,  et~al., 2022, \mn@doi [Monthly Notices of the Royal Astronomical
	Society] {10.1093/mnras/stac1026}, 513, 2535
	
	\bibitem[\protect\citeauthoryear{Cowley et~al.,}{Cowley
		et~al.}{2016}]{Cowley2016}
	Cowley M.~J.,  et~al., 2016, \mn@doi [Monthly Notices of the Royal Astronomical
	Society] {10.1093/mnras/stv2992}, 457, 629
	
	\bibitem[\protect\citeauthoryear{Danehkar, Oey  \& Gray}{Danehkar
		et~al.}{2021}]{danehkar2021}
	Danehkar A.,  Oey M.~S.,   Gray W.~J.,  2021, \mn@doi [The Astrophysical
	Journal] {10.3847/1538-4357/ac1a76}, 921, 91
	
	\bibitem[\protect\citeauthoryear{Davies et~al.,}{Davies
		et~al.}{2019}]{Davies2019b}
	Davies R.~L.,  et~al., 2019, \mn@doi [The Astrophysical Journal]
	{10.3847/1538-4357/ab06f1}, 873, 122
	
	\bibitem[\protect\citeauthoryear{Davies et~al.,}{Davies
		et~al.}{2021}]{Davies2021a}
	Davies R.~L.,  et~al., 2021, \mn@doi [The Astrophysical Journal]
	{10.3847/1538-4357/abd551}, 909, 78
	
	\bibitem[\protect\citeauthoryear{Davis et~al.,}{Davis et~al.}{2013}]{Davis2013}
	Davis T.~A.,  et~al., 2013, \mn@doi [Monthly Notices of the Royal Astronomical
	Society] {10.1093/mnras/sts353}, 429, 534
	
	\bibitem[\protect\citeauthoryear{{De Barros}, Oesch, Labb{\'{e}}, Stefanon,
		Gonz{\'{a}}lez, Smit, Bouwens  \& Illingworth}{{De Barros}
		et~al.}{2019}]{DeBarros2019}
	{De Barros} S.,  Oesch P.~A.,  Labb{\'{e}} I.,  Stefanon M.,  Gonz{\'{a}}lez
	V.,  Smit R.,  Bouwens R.~J.,   Illingworth G.~D.,  2019, \mn@doi [Monthly
	Notices of the Royal Astronomical Society] {10.1093/mnras/stz940}, 489, 2355
	
	\bibitem[\protect\citeauthoryear{Dekel \& Silk}{Dekel \&
		Silk}{1986}]{Dekel1986}
	Dekel A.,  Silk J.,  1986, \mn@doi [The Astrophysical Journal]
	{10.1086/164050}, 303, 39
	
	\bibitem[\protect\citeauthoryear{{Di Teodoro}, Fraternali  \& Miller}{{Di
			Teodoro} et~al.}{2016}]{DiTeodoro2016}
	{Di Teodoro} E.~M.,  Fraternali F.,   Miller S.~H.,  2016, \mn@doi [Astronomy &
	Astrophysics] {10.1051/0004-6361/201628315}, 594, A77
	
	\bibitem[\protect\citeauthoryear{D{\'{i}}az, Ryan-Weber, Karman, Caputi,
		Salvadori, Crighton, Ouchi  \& Vanzella}{D{\'{i}}az et~al.}{2021}]{Diaz2020}
	D{\'{i}}az C.~G.,  Ryan-Weber E.~V.,  Karman W.,  Caputi K.~I.,  Salvadori S.,
	Crighton N.~H.,  Ouchi M.,   Vanzella E.,  2021, \mn@doi [Monthly Notices of
	the Royal Astronomical Society] {10.1093/mnras/staa3129}, 502, 2645
	
	\bibitem[\protect\citeauthoryear{{Dom{\'{i}}nguez S{\'{a}}nchez}
		et~al.,}{{Dom{\'{i}}nguez S{\'{a}}nchez} et~al.}{2012}]{DominguezSanchez2012}
	{Dom{\'{i}}nguez S{\'{a}}nchez} H.,  et~al., 2012, \mn@doi [Monthly Notices of
	the Royal Astronomical Society] {10.1111/j.1365-2966.2012.21710.x}, 426, 330
	
	\bibitem[\protect\citeauthoryear{Du, Shapley, Tang, Stark, Martin, Mobasher,
		Topping  \& Chevallard}{Du et~al.}{2020}]{Du2019}
	Du X.,  Shapley A.~E.,  Tang M.,  Stark D.~P.,  Martin C.~L.,  Mobasher B.,
	Topping M.~W.,   Chevallard J.,  2020, \mn@doi [The Astrophysical Journal]
	{10.3847/1538-4357/ab67b8}, 890, 65
	
	\bibitem[\protect\citeauthoryear{Endsley, Stark, Chevallard  \&
		Charlot}{Endsley et~al.}{2020}]{Endsley2020}
	Endsley R.,  Stark D.~P.,  Chevallard J.,   Charlot S.,  2020, \mn@doi [Monthly
	Notices of the Royal Astronomical Society] {10.1093/mnras/staa3370}, 500,
	5229
	
	\bibitem[\protect\citeauthoryear{Faisst et~al.,}{Faisst
		et~al.}{2020}]{Faisst2020}
	Faisst A.~L.,  et~al., 2020, \mn@doi [The Astrophysical Journal Supplement
	Series] {10.3847/1538-4365/ab7ccd}, 247, 61
	
	\bibitem[\protect\citeauthoryear{Fetherolf et~al.,}{Fetherolf
		et~al.}{2021}]{Fetherolf2021}
	Fetherolf T.,  et~al., 2021, \mn@doi [Monthly Notices of the Royal Astronomical
	Society] {10.1093/mnras/stab2570}, 508, 1431
	
	\bibitem[\protect\citeauthoryear{Forrest et~al.,}{Forrest
		et~al.}{2017}]{Forrest2017}
	Forrest B.,  et~al., 2017, \mn@doi [The Astrophysical Journal]
	{10.3847/2041-8213/aa653b}, 838, L12
	
	\bibitem[\protect\citeauthoryear{Forrest et~al.,}{Forrest
		et~al.}{2018}]{Forrest2018}
	Forrest B.,  et~al., 2018, \mn@doi [The Astrophysical Journal]
	{10.3847/1538-4357/aad232}, 863, 131
	
	\bibitem[\protect\citeauthoryear{Freeman et~al.,}{Freeman
		et~al.}{2019}]{Freeman2019}
	Freeman W.~R.,  et~al., 2019, \mn@doi [The Astrophysical Journal]
	{10.3847/1538-4357/ab0655}, 873, 102
	
	\bibitem[\protect\citeauthoryear{Freudling, Romaniello, Bramich, Ballester,
		Forchi, Garc{\'{i}}a-Dabl{\'{o}}, Moehler  \& Neeser}{Freudling
		et~al.}{2013}]{Freudling2013}
	Freudling W.,  Romaniello M.,  Bramich D.~M.,  Ballester P.,  Forchi V.,
	Garc{\'{i}}a-Dabl{\'{o}} C.~E.,  Moehler S.,   Neeser M.~J.,  2013, \mn@doi
	[Astronomy & Astrophysics] {10.1051/0004-6361/201322494}, 559, A96
	
	\bibitem[\protect\citeauthoryear{Fujimoto et~al.,}{Fujimoto
		et~al.}{2019}]{Fujimoto2019b}
	Fujimoto S.,  et~al., 2019, \mn@doi [The Astrophysical Journal]
	{10.3847/1538-4357/ab480f}, 887, 107
	
	\bibitem[\protect\citeauthoryear{Genzel et~al.,}{Genzel
		et~al.}{2011}]{Genzel2011}
	Genzel R.,  et~al., 2011, \mn@doi [The Astrophysical Journal]
	{10.1088/0004-637X/733/2/101}, 733, 101
	
	\bibitem[\protect\citeauthoryear{Giacconi et~al.,}{Giacconi
		et~al.}{2002}]{giacconi2002}
	Giacconi R.,  et~al., 2002, \mn@doi [The Astrophysical Journal Supplement
	Series] {10.1086/338927}, 139, 369
	
	\bibitem[\protect\citeauthoryear{Ginolfi et~al.,}{Ginolfi
		et~al.}{2019}]{Ginolfi2019a}
	Ginolfi M.,  et~al., 2019, \mn@doi [Astronomy & Astrophysics]
	{10.1051/0004-6361/201936872}, 633, A90
	
	\bibitem[\protect\citeauthoryear{Ginolfi, Hunt, Tortora, Schneider  \&
		Cresci}{Ginolfi et~al.}{2020}]{Ginolfi2019}
	Ginolfi M.,  Hunt L.~K.,  Tortora C.,  Schneider R.,   Cresci G.,  2020,
	\mn@doi [Astronomy & Astrophysics] {10.1051/0004-6361/201936304}, 638, A4
	
	\bibitem[\protect\citeauthoryear{Gray, Oey, Silich  \& Scannapieco}{Gray
		et~al.}{2019}]{Gray2019}
	Gray W.~J.,  Oey M.~S.,  Silich S.,   Scannapieco E.,  2019, \mn@doi [The
	Astrophysical Journal] {10.3847/1538-4357/ab510d}, 887, 161
	
	\bibitem[\protect\citeauthoryear{Gronke}{Gronke}{2017}]{Gronke2017}
	Gronke M.,  2017, \mn@doi [Astronomy and Astrophysics]
	{10.1051/0004-6361/201731791}, 608, 1
	
	\bibitem[\protect\citeauthoryear{Gupta et~al.,}{Gupta et~al.}{2020}]{Gupta2020}
	Gupta A.,  et~al., 2020, \mn@doi [The Astrophysical Journal]
	{10.3847/1538-4357/ab7b6d}, 893, 23
	
	\bibitem[\protect\citeauthoryear{Harshan et~al.,}{Harshan
		et~al.}{2020}]{Harshan2020}
	Harshan A.,  et~al., 2020, \mn@doi [The Astrophysical Journal]
	{10.3847/1538-4357/ab76cf}, 892, 77
	
	\bibitem[\protect\citeauthoryear{Herenz et~al.,}{Herenz
		et~al.}{2017}]{Herenz2017}
	Herenz E.~C.,  et~al., 2017, \mn@doi [Astronomy & Astrophysics]
	{10.1051/0004-6361/201731055}, 606, A12
	
	\bibitem[\protect\citeauthoryear{Herrera-Camus et~al.,}{Herrera-Camus
		et~al.}{2021}]{Herrera-Camus2021}
	Herrera-Camus R.,  et~al., 2021, \mn@doi [Astronomy and Astrophysics]
	{10.1051/0004-6361/202039704}, 649, 1
	
	\bibitem[\protect\citeauthoryear{Hutchison et~al.,}{Hutchison
		et~al.}{2019}]{Hutchison2019}
	Hutchison T.~A.,  et~al., 2019, \mn@doi [The Astrophysical Journal]
	{10.3847/1538-4357/ab22a2}, 879, 70
	
	\bibitem[\protect\citeauthoryear{Inoue}{Inoue}{2011}]{Inoue2011}
	Inoue A.~K.,  2011, \mn@doi [Monthly Notices of the Royal Astronomical Society]
	{10.1111/j.1365-2966.2011.18906.x}, 415, 2920
	
	\bibitem[\protect\citeauthoryear{Jaskot \& Oey}{Jaskot \&
		Oey}{2013}]{Jaskot2013}
	Jaskot A.~E.,  Oey M.~S.,  2013, \mn@doi [The Astrophysical Journal]
	{10.1088/0004-637X/766/2/91}, 766, 91
	
	\bibitem[\protect\citeauthoryear{Jaskot, Oey, Scarlata  \& Dowd}{Jaskot
		et~al.}{2017}]{Jaskot2017}
	Jaskot A.~E.,  Oey M.~S.,  Scarlata C.,   Dowd T.,  2017, \mn@doi [The
	Astrophysical Journal] {10.3847/2041-8213/aa9d83}, 851, L9
	
	\bibitem[\protect\citeauthoryear{Kaasinen, Bian, Groves, Kewley  \&
		Gupta}{Kaasinen et~al.}{2016}]{Kaasinen2017}
	Kaasinen M.,  Bian F.,  Groves B.,  Kewley L.,   Gupta A.,  2016, \mn@doi
	[Monthly Notices of the Royal Astronomical Society] {10.1093/mnras/stw2827},
	465, 3220
	
	\bibitem[\protect\citeauthoryear{Kacprzak et~al.,}{Kacprzak
		et~al.}{2016}]{Kacprzak2016}
	Kacprzak G.~G.,  et~al., 2016, \mn@doi [The Astrophysical Journal]
	{10.3847/2041-8205/826/1/L11}, 826, L11
	
	\bibitem[\protect\citeauthoryear{Kriek, van Dokkum, Whitaker, Labb{\'{e}},
		Franx  \& Brammer}{Kriek et~al.}{2011}]{Kriek2011}
	Kriek M.,  van Dokkum P.~G.,  Whitaker K.~E.,  Labb{\'{e}} I.,  Franx M.,
	Brammer G.~B.,  2011, \mn@doi [The Astrophysical Journal]
	{10.1088/0004-637X/743/2/168}, 743, 168
	
	\bibitem[\protect\citeauthoryear{Krumholz, Burkhart, Forbes  \&
		Crocker}{Krumholz et~al.}{2018}]{Krumholz2018}
	Krumholz M.~R.,  Burkhart B.,  Forbes J.~C.,   Crocker R.~M.,  2018, \mn@doi
	[Monthly Notices of the Royal Astronomical Society] {10.1093/mnras/sty852},
	477, 2716
	
	\bibitem[\protect\citeauthoryear{Labb{\'{e}} et~al.,}{Labb{\'{e}}
		et~al.}{2013}]{Labbe2013}
	Labb{\'{e}} I.,  et~al., 2013, \mn@doi [The Astrophysical Journal]
	{10.1088/2041-8205/777/2/L19}, 777, L19
	
	\bibitem[\protect\citeauthoryear{Larson}{Larson}{1974}]{Larson1974}
	Larson R.~B.,  1974, \mn@doi [Monthly Notices of the Royal Astronomical
	Society] {10.1093/mnras/169.2.229}, 169, 229
	
	\bibitem[\protect\citeauthoryear{{Le F{\`{e}}vre} et~al.,}{{Le F{\`{e}}vre}
		et~al.}{2020}]{Fevre2019}
	{Le F{\`{e}}vre} O.,  et~al., 2020, \mn@doi [Astronomy & Astrophysics]
	{10.1051/0004-6361/201936965}, 643, A1
	
	\bibitem[\protect\citeauthoryear{Leclercq et~al.,}{Leclercq
		et~al.}{2017}]{Leclercq2017}
	Leclercq F.,  et~al., 2017, \mn@doi [Astronomy & Astrophysics]
	{10.1051/0004-6361/201731480}, 608, A8
	
	\bibitem[\protect\citeauthoryear{Leclercq et~al.,}{Leclercq
		et~al.}{2020}]{Leclercq2020}
	Leclercq F.,  et~al., 2020, \mn@doi [Astronomy & Astrophysics]
	{10.1051/0004-6361/201937339}, 635, A82
	
	\bibitem[\protect\citeauthoryear{Livermore, Finkelstein  \& Lotz}{Livermore
		et~al.}{2017}]{Livermore2017}
	Livermore R.~C.,  Finkelstein S.~L.,   Lotz J.~M.,  2017, \mn@doi [The
	Astrophysical Journal] {10.3847/1538-4357/835/2/113}, 835, 113
	
	\bibitem[\protect\citeauthoryear{Lochhaas, Thompson, Quataert  \&
		Weinberg}{Lochhaas et~al.}{2018}]{Lochhaas2018}
	Lochhaas C.,  Thompson T.~A.,  Quataert E.,   Weinberg D.~H.,  2018, \mn@doi
	[Monthly Notices of the Royal Astronomical Society] {10.1093/mnras/sty2421},
	481, 1873
	
	\bibitem[\protect\citeauthoryear{{Mac Low} \& Ferrara}{{Mac Low} \&
		Ferrara}{1999}]{MacLow1999}
	{Mac Low} M.,  Ferrara A.,  1999, \mn@doi [The Astrophysical Journal]
	{10.1086/306832}, 513, 142
	
	\bibitem[\protect\citeauthoryear{Mainali et~al.,}{Mainali
		et~al.}{2019}]{Mainali2020}
	Mainali R.,  et~al., 2019, \mn@doi [Monthly Notices of the Royal Astronomical
	Society] {10.1093/mnras/staa751}, 494, 719
	
	\bibitem[\protect\citeauthoryear{Marcolini, Strickland, D'Ercole, Heckman  \&
		Hoopes}{Marcolini et~al.}{2005}]{Marcolini2005}
	Marcolini A.,  Strickland D.~K.,  D'Ercole A.,  Heckman T.~M.,   Hoopes C.~G.,
	2005, \mn@doi [Monthly Notices of the Royal Astronomical Society]
	{10.1111/j.1365-2966.2005.09343.x}, 362, 626
	
	\bibitem[\protect\citeauthoryear{Martin et~al.,}{Martin et~al.}{2019}]{Martin}
	Martin D.~C.,  et~al., 2019, \mn@doi [Nature Astronomy]
	{10.1038/s41550-019-0791-2}, 3, 822
	
	\bibitem[\protect\citeauthoryear{Maseda et~al.,}{Maseda
		et~al.}{2014}]{Maseda2014}
	Maseda M.~V.,  et~al., 2014, \mn@doi [Astrophysical Journal]
	{10.1088/0004-637X/791/1/17}, 791, 17
	
	\bibitem[\protect\citeauthoryear{McDermid et~al.,}{McDermid
		et~al.}{2020}]{McDermid2020}
	McDermid R.~M.,  et~al., 2020, ] {10.25949/zdaw-rx65}, pp 1--141
	
	\bibitem[\protect\citeauthoryear{Mitchell, Schaye, Bower  \& Crain}{Mitchell
		et~al.}{2020}]{Mitchell2019}
	Mitchell P.~D.,  Schaye J.,  Bower R.~G.,   Crain R.~A.,  2020, \mn@doi
	[Monthly Notices of the Royal Astronomical Society] {10.1093/mnras/staa938},
	23, 1
	
	\bibitem[\protect\citeauthoryear{Mitchell, Blaizot, Cadiou, Dubois, Garel  \&
		Rosdahl}{Mitchell et~al.}{2021}]{Mitchell2021}
	Mitchell P.~D.,  Blaizot J.,  Cadiou C.,  Dubois Y.,  Garel T.,   Rosdahl J.,
	2021, \mn@doi [Monthly Notices of the Royal Astronomical Society]
	{10.1093/mnras/stab035}, 501, 5757
	
	\bibitem[\protect\citeauthoryear{Muratov, Keres, Faucher-Giguere, Hopkins,
		Quataert  \& Murray}{Muratov et~al.}{2015}]{Muratov2015}
	Muratov A.~L.,  Keres D.,  Faucher-Giguere C.-A.,  Hopkins P.~F.,  Quataert E.,
	Murray N.,  2015, \mn@doi [Monthly Notices of the Royal Astronomical
	Society] {10.1093/mnras/stv2126}, 454, 2691
	
	\bibitem[\protect\citeauthoryear{Murray, Quataert  \& Thompson}{Murray
		et~al.}{2005}]{Murray2005}
	Murray N.,  Quataert E.,   Thompson T.~A.,  2005, \mn@doi [The Astrophysical
	Journal] {10.1086/426067}, 618, 569
	
	\bibitem[\protect\citeauthoryear{Nanayakkara et~al.,}{Nanayakkara
		et~al.}{2016}]{Nanayakkara2016}
	Nanayakkara T.,  et~al., 2016, \mn@doi [The Astrophysical Journal]
	{10.3847/0004-637X/828/1/21}, 828, 1
	
	\bibitem[\protect\citeauthoryear{Nelson et~al.,}{Nelson
		et~al.}{2019}]{Nelson2019}
	Nelson D.,  et~al., 2019, \mn@doi [Monthly Notices of the Royal Astronomical
	Society] {10.1093/mnras/stz2306}, 490, 3234
	
	\bibitem[\protect\citeauthoryear{Newman et~al.,}{Newman
		et~al.}{2012}]{Newman2012}
	Newman S.~F.,  et~al., 2012, \mn@doi [Astrophysical Journal]
	{10.1088/0004-637X/761/1/43}, 761
	
	\bibitem[\protect\citeauthoryear{Newville, Stensitzki, Allen  \&
		Ingargiola}{Newville et~al.}{2014}]{Newville2014}
	Newville M.,  Stensitzki T.,  Allen D.~B.,   Ingargiola A.,  2014, {LMFIT:
		Non-Linear Least-Square Minimization and Curve-Fitting for Python},
	\mn@doi{10.5281/zenodo.11813}, \url {https://doi.org/10.5281/zenodo.11813}
	
	\bibitem[\protect\citeauthoryear{Onodera et~al.,}{Onodera
		et~al.}{2016}]{Onodera2016}
	Onodera M.,  et~al., 2016, \mn@doi [The Astrophysical Journal]
	{10.3847/0004-637X/822/1/42}, 822, 42
	
	\bibitem[\protect\citeauthoryear{Onodera et~al.,}{Onodera
		et~al.}{2020}]{Onodera2020}
	Onodera M.,  et~al., 2020, \mn@doi [The Astrophysical Journal]
	{10.3847/1538-4357/abc174}, 904, 180
	
	\bibitem[\protect\citeauthoryear{Pillepich et~al.,}{Pillepich
		et~al.}{2019}]{Pillepich2019}
	Pillepich A.,  et~al., 2019, \mn@doi [Monthly Notices of the Royal Astronomical
	Society] {10.1093/mnras/stz2338}, 490, 3196
	
	\bibitem[\protect\citeauthoryear{Reddy et~al.,}{Reddy et~al.}{2018}]{Reddy2018}
	Reddy N.~A.,  et~al., 2018, \mn@doi [The Astrophysical Journal]
	{10.3847/1538-4357/aaed1e}, 869, 92
	
	\bibitem[\protect\citeauthoryear{Rizzo, Vegetti, Fraternali, Stacey  \&
		Powell}{Rizzo et~al.}{2021}]{Rizzo2021}
	Rizzo F.,  Vegetti S.,  Fraternali F.,  Stacey H.~R.,   Powell D.,  2021,
	\mn@doi [Monthly Notices of the Royal Astronomical Society]
	{10.1093/mnras/stab2295}, 507, 3952
	
	\bibitem[\protect\citeauthoryear{Roberts-Borsani et~al.,}{Roberts-Borsani
		et~al.}{2016}]{Roberts-Borsani2016}
	Roberts-Borsani G.~W.,  et~al., 2016, \mn@doi [The Astrophysical Journal]
	{10.3847/0004-637x/823/2/143}, 823, 143
	
	\bibitem[\protect\citeauthoryear{Salmon et~al.,}{Salmon
		et~al.}{2015}]{Salmon2015}
	Salmon B.,  et~al., 2015, \mn@doi [Astrophysical Journal]
	{10.1088/0004-637X/799/2/183}, 799
	
	\bibitem[\protect\citeauthoryear{Sharples et~al.,}{Sharples
		et~al.}{2012}]{Sharples2012}
	Sharples R.,  et~al., 2012, in McLean I.~S.,  Ramsay S.~K.,   Takami H.,  eds,
	Vol. 8446, Ground-based and Airborne Instrumentation for Astronomy IV. p.
	84460K, \mn@doi{10.1117/12.926021}, \url
	{http://proceedings.spiedigitallibrary.org/proceeding.aspx?doi=10.1117/12.926021}
	
	\bibitem[\protect\citeauthoryear{Sharples et~al.,}{Sharples
		et~al.}{2013}]{Sharples2013}
	Sharples R.,  et~al., 2013, The Messenger, 151, 21
	
	\bibitem[\protect\citeauthoryear{Shivaei et~al.,}{Shivaei
		et~al.}{2016}]{Shivaei2016}
	Shivaei I.,  et~al., 2016, \mn@doi [The Astrophysical Journal]
	{10.3847/2041-8205/820/2/L23}, 820, L23
	
	\bibitem[\protect\citeauthoryear{Silich \& Tenorio-Tagle}{Silich \&
		Tenorio-Tagle}{2017}]{Silich2017}
	Silich S.,  Tenorio-Tagle G.,  2017, \mn@doi [Monthly Notices of the Royal
	Astronomical Society] {10.1093/mnras/stw2879}, 465, 1375
	
	\bibitem[\protect\citeauthoryear{Silich, Tenorio‐Tagle  \&
		Munoz‐Tunon}{Silich et~al.}{2007}]{Silich2007}
	Silich S.,  Tenorio‐Tagle G.,   Munoz‐Tunon C.,  2007, \mn@doi [The
	Astrophysical Journal] {10.1086/521706}, 669, 952
	
	\bibitem[\protect\citeauthoryear{Smit et~al.,}{Smit et~al.}{2018}]{Smit2018}
	Smit R.,  et~al., 2018, \mn@doi [Nature] {10.1038/nature24631}, 553, 179
	
	\bibitem[\protect\citeauthoryear{Stark}{Stark}{2016}]{Stark2016a}
	Stark D.~P.,  2016, \mn@doi [Annual Review of Astronomy and Astrophysics]
	{10.1146/annurev-astro-081915-023417}, 54, 761
	
	\bibitem[\protect\citeauthoryear{Stark et~al.,}{Stark et~al.}{2016}]{Stark2017}
	Stark D.~P.,  et~al., 2016, \mn@doi [Monthly Notices of the Royal Astronomical
	Society] {10.1093/mnras/stw2233}, 464, 469
	
	\bibitem[\protect\citeauthoryear{Straatman et~al.,}{Straatman
		et~al.}{2016}]{Straatman2016}
	Straatman C. M.~S.,  et~al., 2016, \mn@doi [The Astrophysical Journal]
	{10.3847/0004-637X/830/1/51}, 830, 51
	
	\bibitem[\protect\citeauthoryear{Strickland \& Stevens}{Strickland \&
		Stevens}{2000}]{Strickland2000}
	Strickland D.~K.,  Stevens I.~R.,  2000, \mn@doi [Monthly Notices of the Royal
	Astronomical Society] {10.1046/j.1365-8711.2000.03391.x}, 314, 511
	
	\bibitem[\protect\citeauthoryear{Swinbank et~al.,}{Swinbank
		et~al.}{2019}]{Swinbank2019}
	Swinbank A.~M.,  et~al., 2019, \mn@doi [Monthly Notices of the Royal
	Astronomical Society] {10.1093/mnras/stz1275}, 487, 381
	
	\bibitem[\protect\citeauthoryear{Tang, Stark, Chevallard  \& Charlot}{Tang
		et~al.}{2019}]{Tang2018}
	Tang M.,  Stark D.~P.,  Chevallard J.,   Charlot S.,  2019, \mn@doi [Monthly
	Notices of the Royal Astronomical Society] {10.1093/mnras/stz2236}, 489, 2572
	
	\bibitem[\protect\citeauthoryear{Tang, Stark, Chevallard, Charlot, Endsley  \&
		Congiu}{Tang et~al.}{2020}]{Tang2020}
	Tang M.,  Stark D.~P.,  Chevallard J.,  Charlot S.,  Endsley R.,   Congiu E.,
	2020, \mn@doi [Monthly Notices of the Royal Astronomical Society]
	{10.1093/mnras/staa3454}, 22, 1
	
	\bibitem[\protect\citeauthoryear{Tran et~al.,}{Tran et~al.}{2015}]{Tran2015}
	Tran K.-V.~H.,  et~al., 2015, \mn@doi [The Astrophysical Journal]
	{10.1088/0004-637X/811/1/28}, 811, 28
	
	\bibitem[\protect\citeauthoryear{Tran et~al.,}{Tran et~al.}{2020}]{Tran2020}
	Tran K.-V.~H.,  et~al., 2020, \mn@doi [The Astrophysical Journal]
	{10.3847/1538-4357/ab8cba}, 898, 45
	
	\bibitem[\protect\citeauthoryear{Tremonti et~al.,}{Tremonti
		et~al.}{2004}]{Tremonti2004}
	Tremonti C.~A.,  et~al., 2004, \mn@doi [The Astrophysical Journal]
	{10.1086/423264}, 613, 898
	
	\bibitem[\protect\citeauthoryear{{\"{U}}bler et~al.,}{{\"{U}}bler
		et~al.}{2019}]{Ubler2019}
	{\"{U}}bler H.,  et~al., 2019, \mn@doi [The Astrophysical Journal]
	{10.3847/1538-4357/ab27cc}, 880, 48
	
	\bibitem[\protect\citeauthoryear{{Van Der Wel} et~al.,}{{Van Der Wel}
		et~al.}{2012}]{VanDerWel2012}
	{Van Der Wel} A.,  et~al., 2012, \mn@doi [Astrophysical Journal, Supplement
	Series] {10.1088/0067-0049/203/2/24}, 203
	
	\bibitem[\protect\citeauthoryear{Veilleux, Maiolino, Bolatto  \&
		Aalto}{Veilleux et~al.}{2020}]{Veilleux2020}
	Veilleux S.,  Maiolino R.,  Bolatto A.~D.,   Aalto S.,  2020, \mn@doi
	[Astronomy and Astrophysics Review] {10.1007/s00159-019-0121-9}, 28
	
	\bibitem[\protect\citeauthoryear{Wisnioski et~al.,}{Wisnioski
		et~al.}{2019}]{Wisnioski2019}
	Wisnioski E.,  et~al., 2019, \mn@doi [The Astrophysical Journal]
	{10.3847/1538-4357/ab4db8}, 886, 124
	
	\bibitem[\protect\citeauthoryear{Wisotzki et~al.,}{Wisotzki
		et~al.}{2018}]{Wisotzki2018}
	Wisotzki L.,  et~al., 2018, \mn@doi [Nature] {10.1038/s41586-018-0564-6}, 562,
	229
	
	\bibitem[\protect\citeauthoryear{Witstok, Smit, Maiolino, Curti, Laporte,
		Massey, Richard  \& Swinbank}{Witstok et~al.}{2021}]{Witstok2021}
	Witstok J.,  Smit R.,  Maiolino R.,  Curti M.,  Laporte N.,  Massey R.,
	Richard J.,   Swinbank M.,  2021, \mn@doi [Monthly Notices of the Royal
	Astronomical Society] {10.1093/mnras/stab2591}, 508, 1686
	
	\bibitem[\protect\citeauthoryear{Yan et~al.,}{Yan et~al.}{2020}]{Yan2020}
	Yan L.,  et~al., 2020
	
	\bibitem[\protect\citeauthoryear{van~der Wel et~al.,}{van~der Wel
		et~al.}{2014}]{VanderWel2014}
	van~der Wel a.,  et~al., 2014, \mn@doi [The Astrophysical Journal]
	{10.1088/0004-637X/788/1/28}, 788, 28
	
	\makeatother
\end{thebibliography}



\appendix

\section{Stacked sample - 1D spectra}\label{sec:all_spec_appendix}

\begin{figure*}
	\centering
	\tiny
	\includegraphics[scale=0.3, trim=0.2cm 0.0cm 0.0cm 0.0cm,clip=true]{10936_o3_fit_2d_fits.pdf}
	\includegraphics[scale=0.3, trim=0.2cm 0.0cm 0.0cm 0.0cm,clip=true]{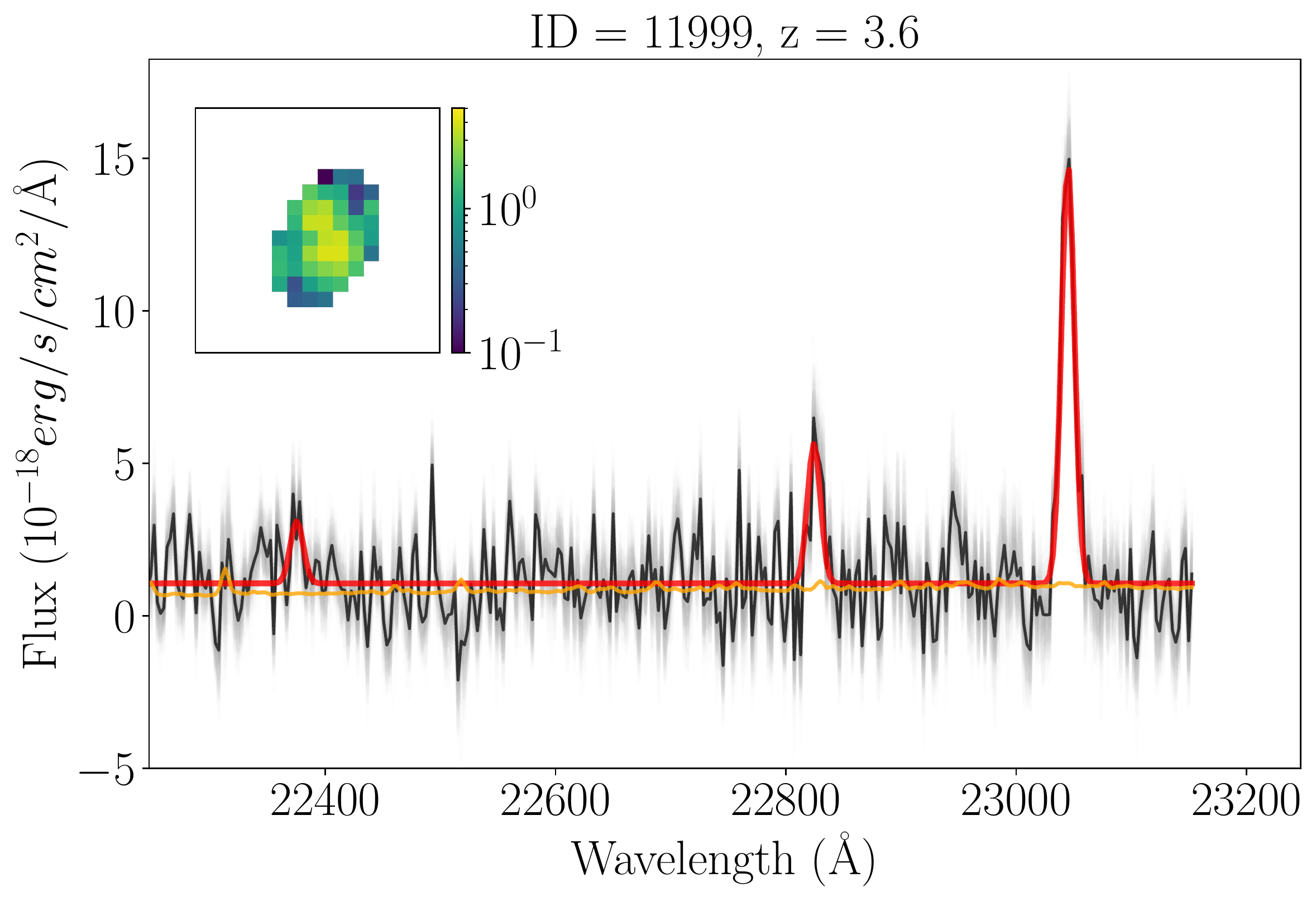}	
	\includegraphics[scale=0.3, trim=0.2cm 0.0cm 0.0cm 0.0cm,clip=true]{12533_o3_fit_2d_fits.pdf}	
	\includegraphics[scale=0.3, trim=0.2cm 0.0cm 0.0cm 0.0cm,clip=true]{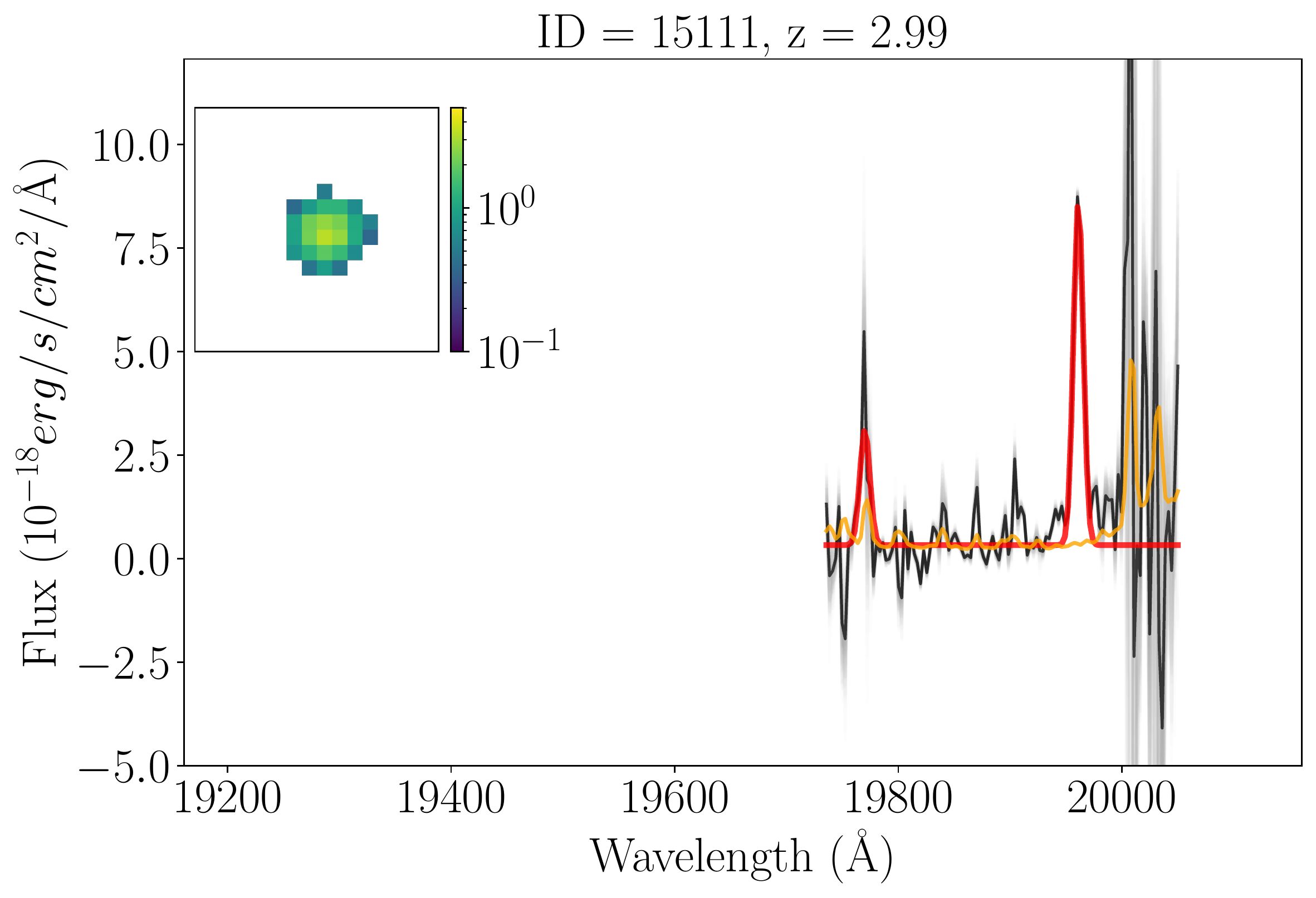}	
	\includegraphics[scale=0.3, trim=0.2cm 0.0cm 0.0cm 0.0cm,clip=true]{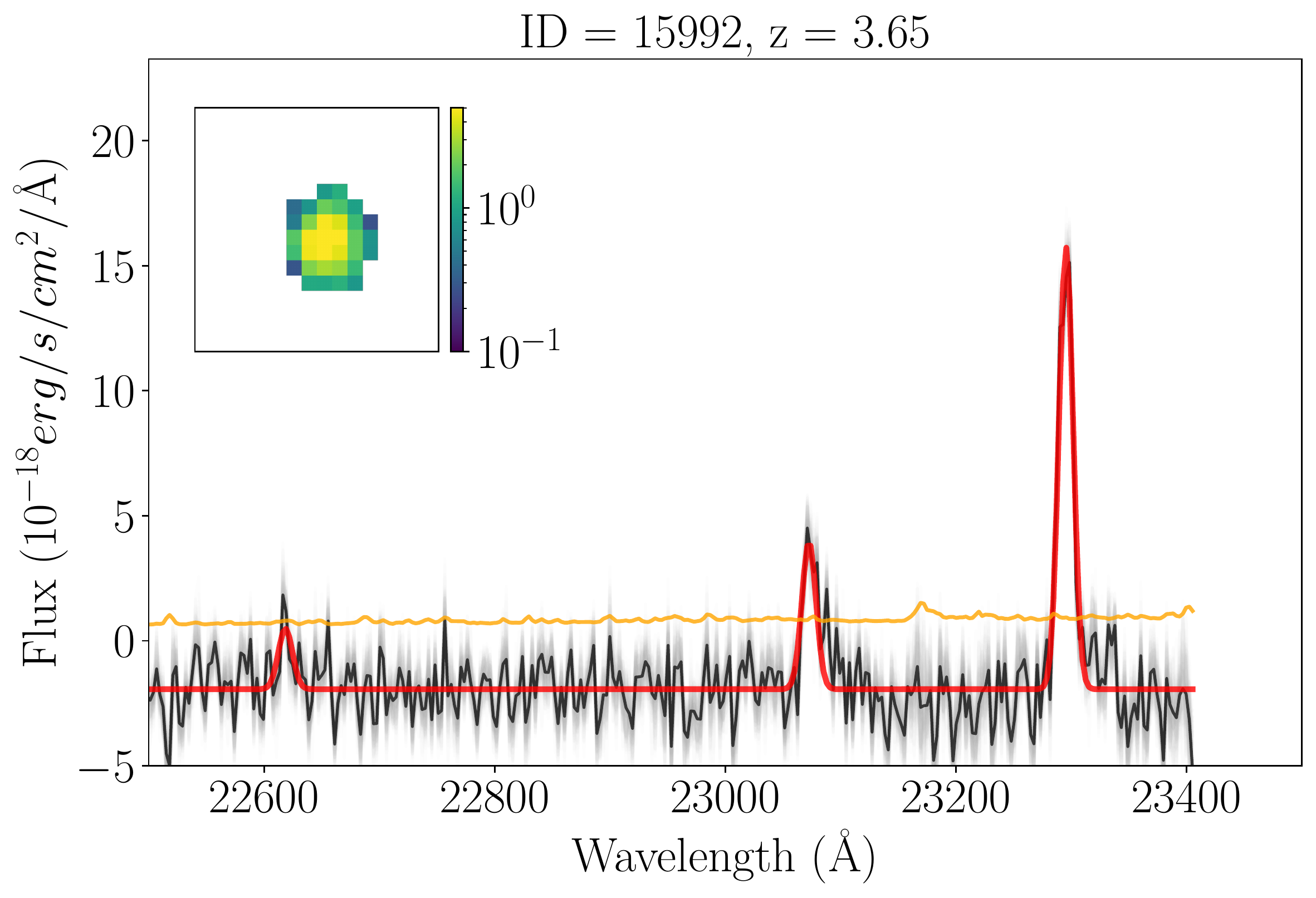}
	\includegraphics[scale=0.3, trim=0.2cm 0.0cm 0.0cm 0.0cm,clip=true]{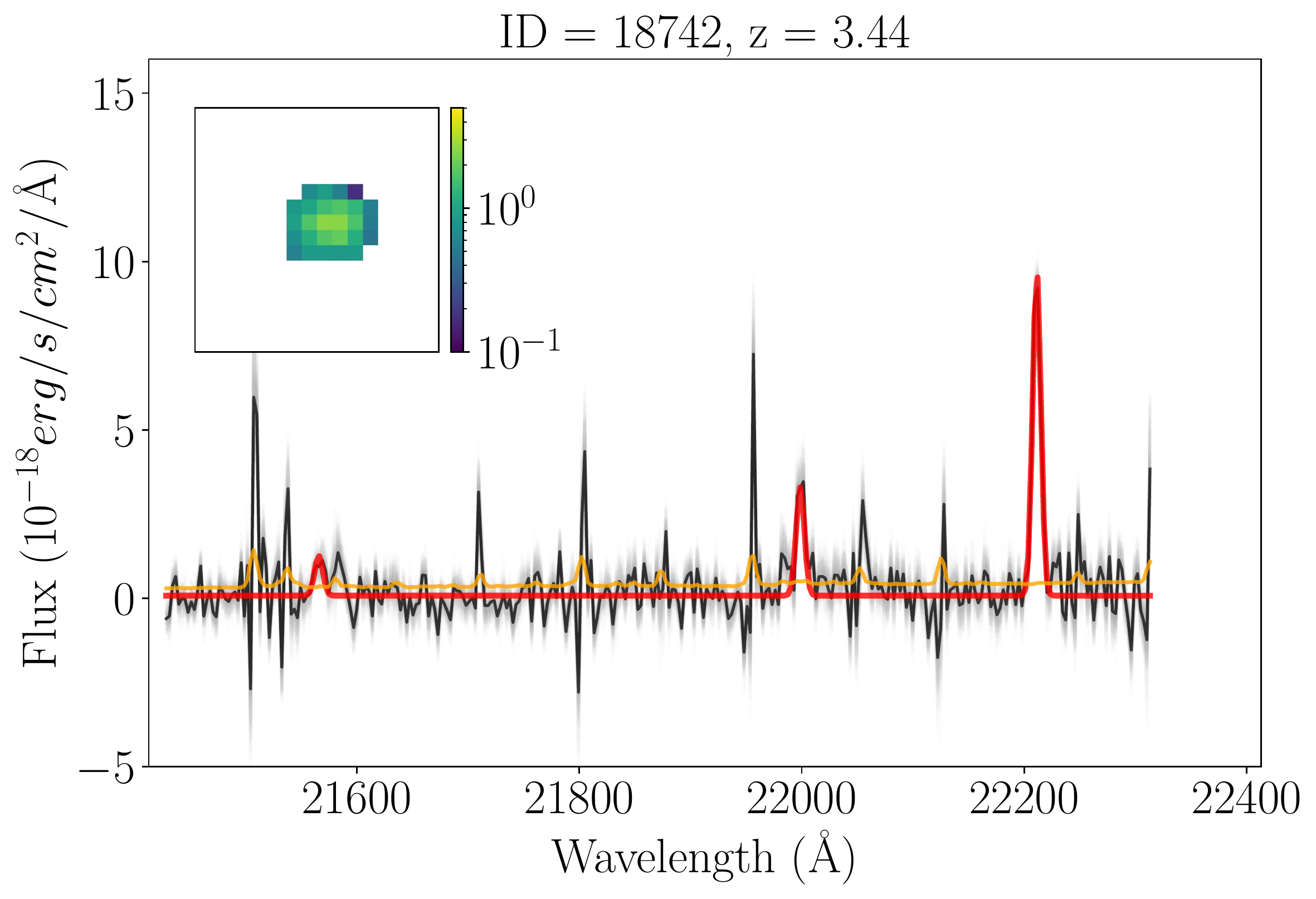}
	\includegraphics[scale=0.3, trim=0.2cm 0.0cm 0.0cm 0.0cm,clip=true]{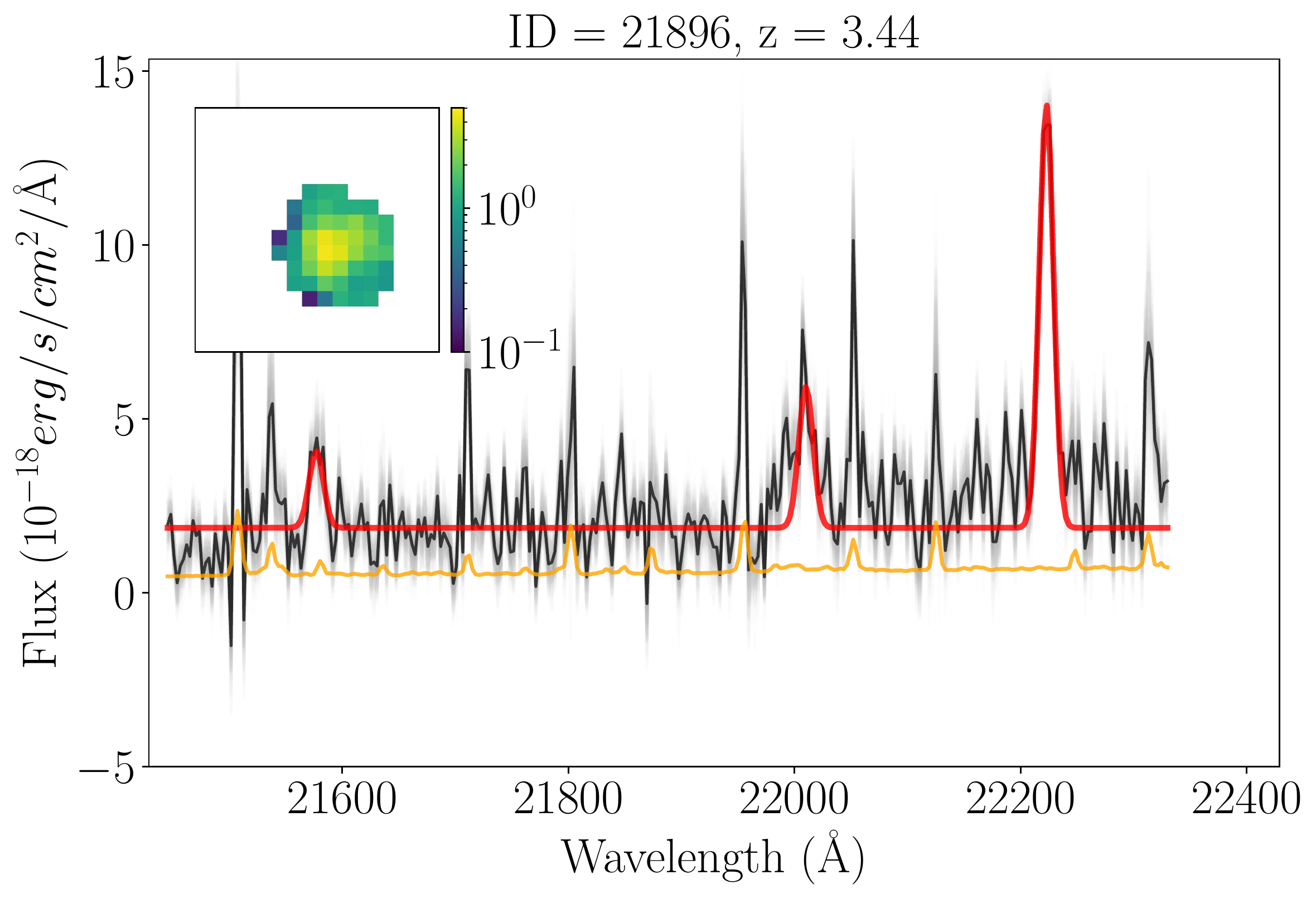}
	\includegraphics[scale=0.3, trim=0.2cm 0.0cm 0.0cm 0.0cm,clip=true]{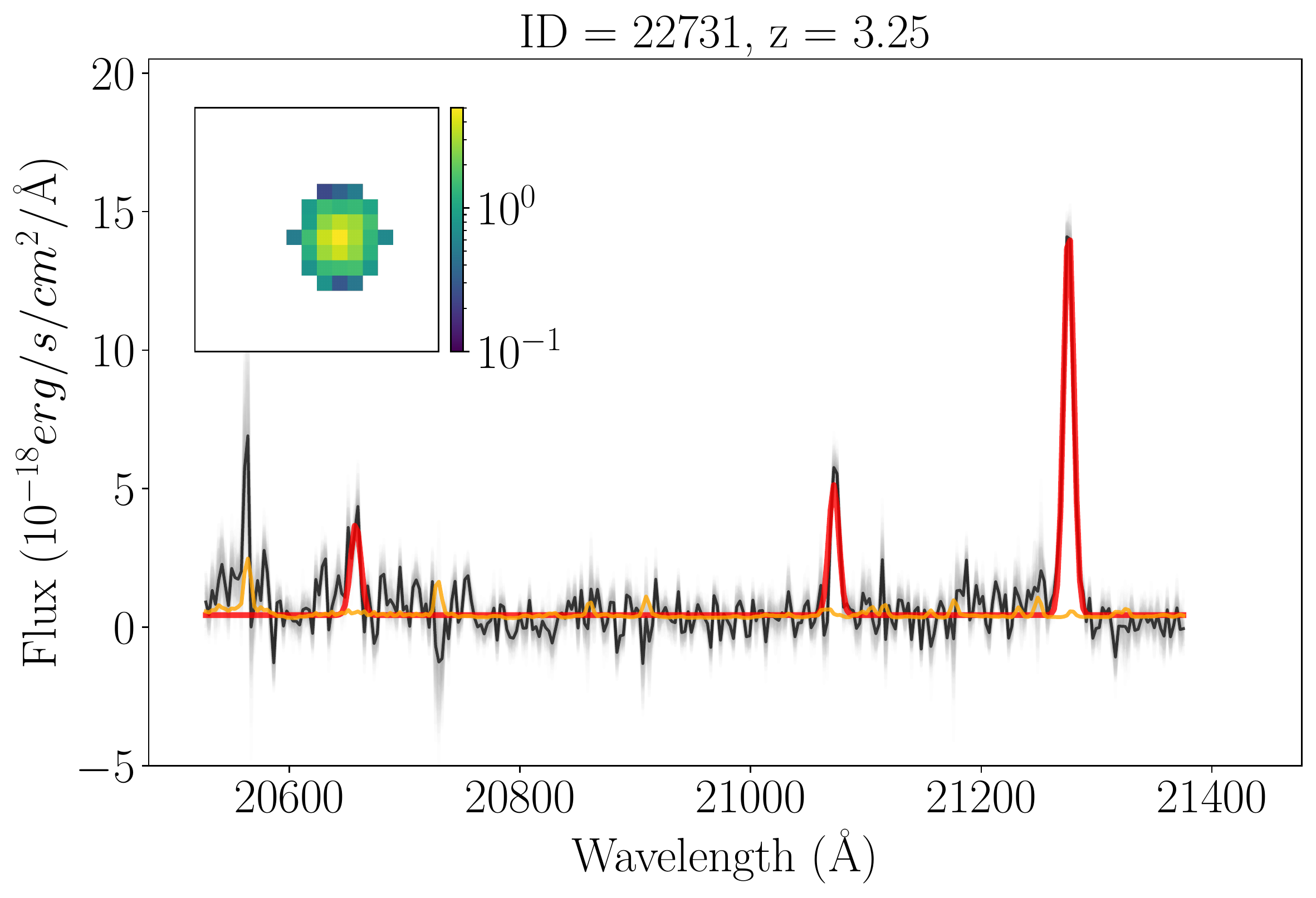}	
	\includegraphics[scale=0.3, trim=0.2cm 0.0cm 0.0cm 0.0cm,clip=true]{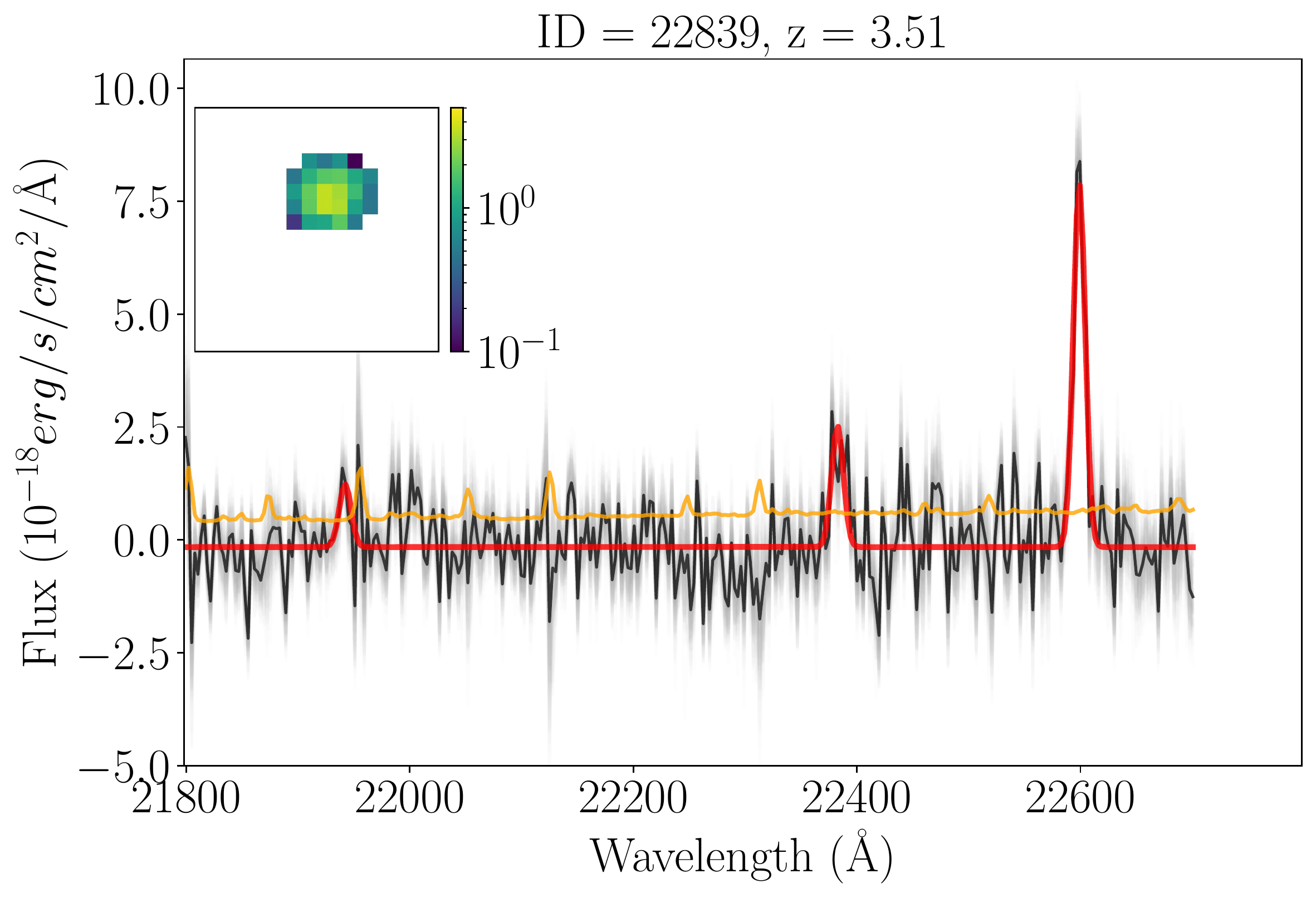}
		
	\caption{Spectra of galaxies selected for the stacking analysis from the KMOS observations. The colour scheme is same as Figure \ref{fig:spectra}.  }
	\label{fig:o3_fits_kmos}
\end{figure*}

\begin{figure*}
	\centering
	\tiny
	\includegraphics[scale=0.27, trim=0.2cm 0.0cm 0.0cm 0.0cm,clip=true]{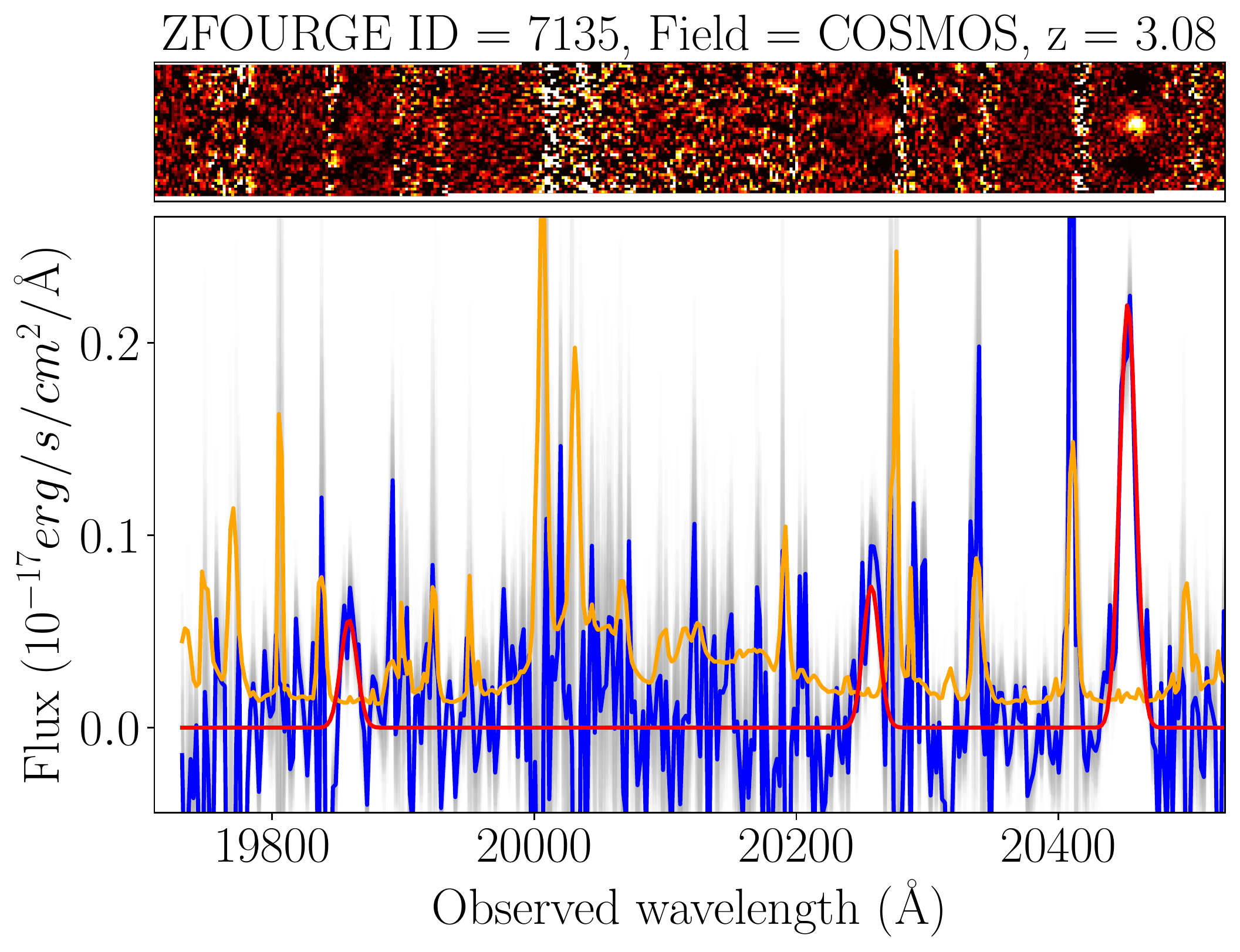}
	\includegraphics[scale=0.27, trim=0.2cm 0.0cm 0.0cm 0.0cm,clip=true]{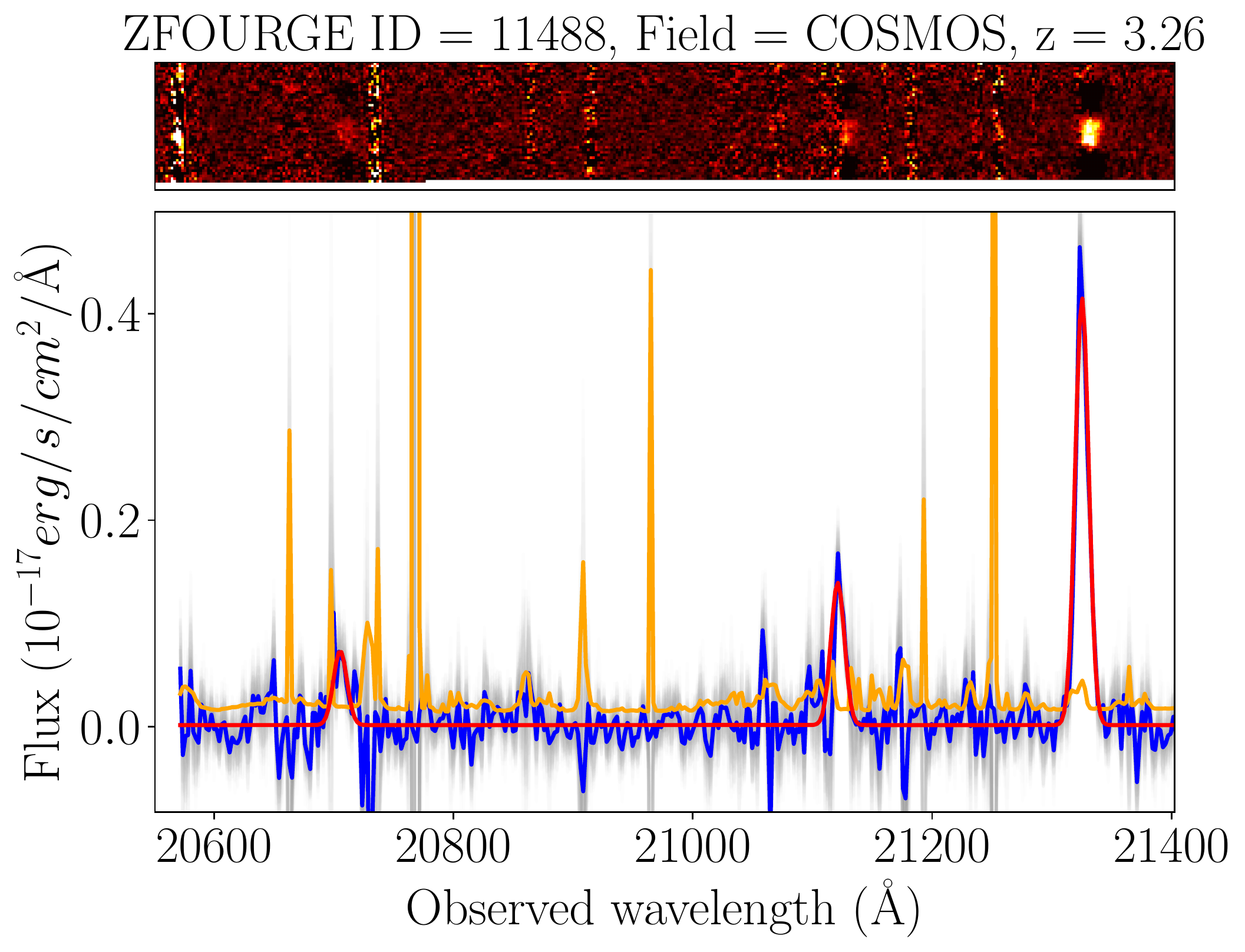}	
	\includegraphics[scale=0.27, trim=0.2cm 0.0cm 0.0cm 0.0cm,clip=true]{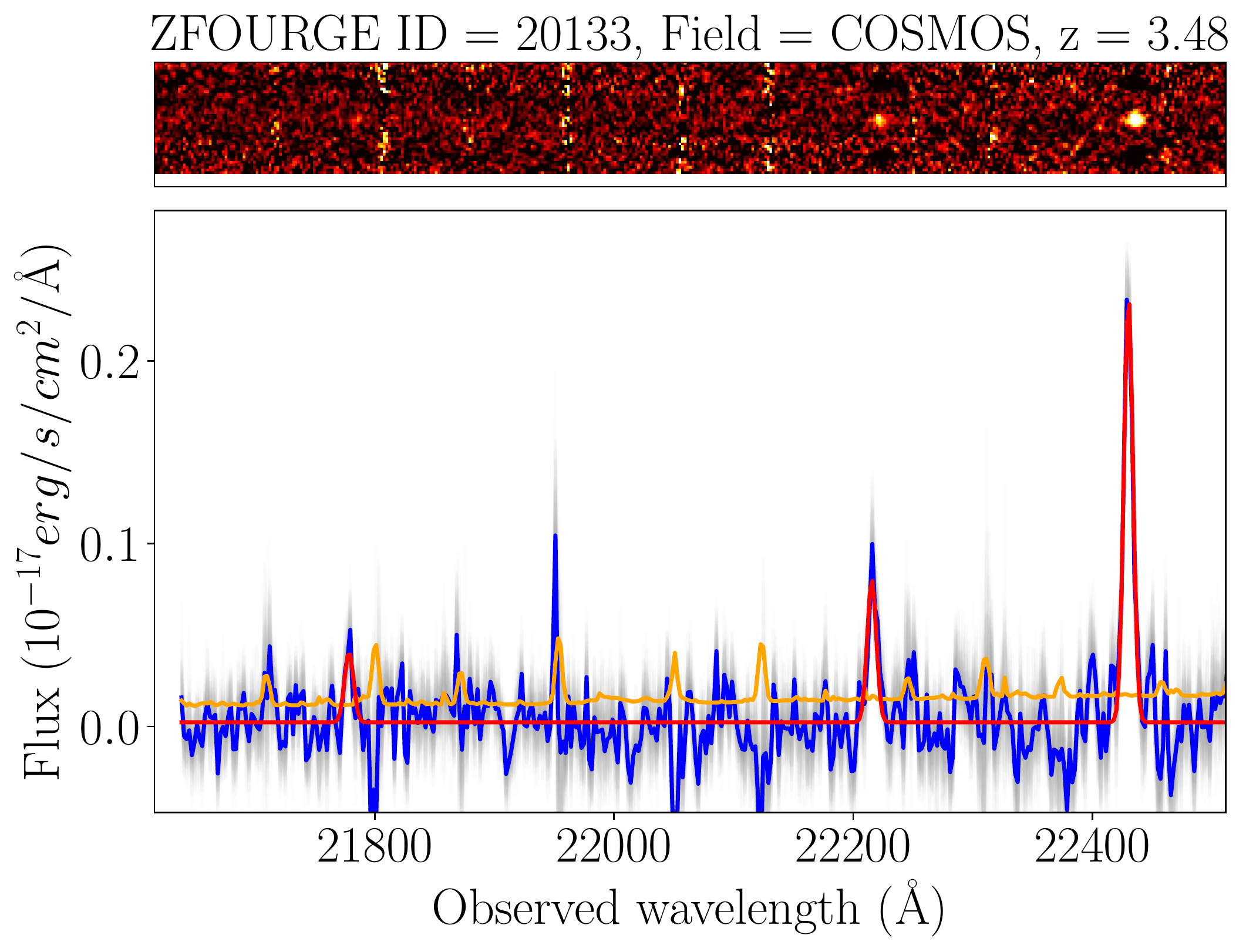}	
	\includegraphics[scale=0.27, trim=0.2cm 0.0cm 0.0cm 0.0cm,clip=true]{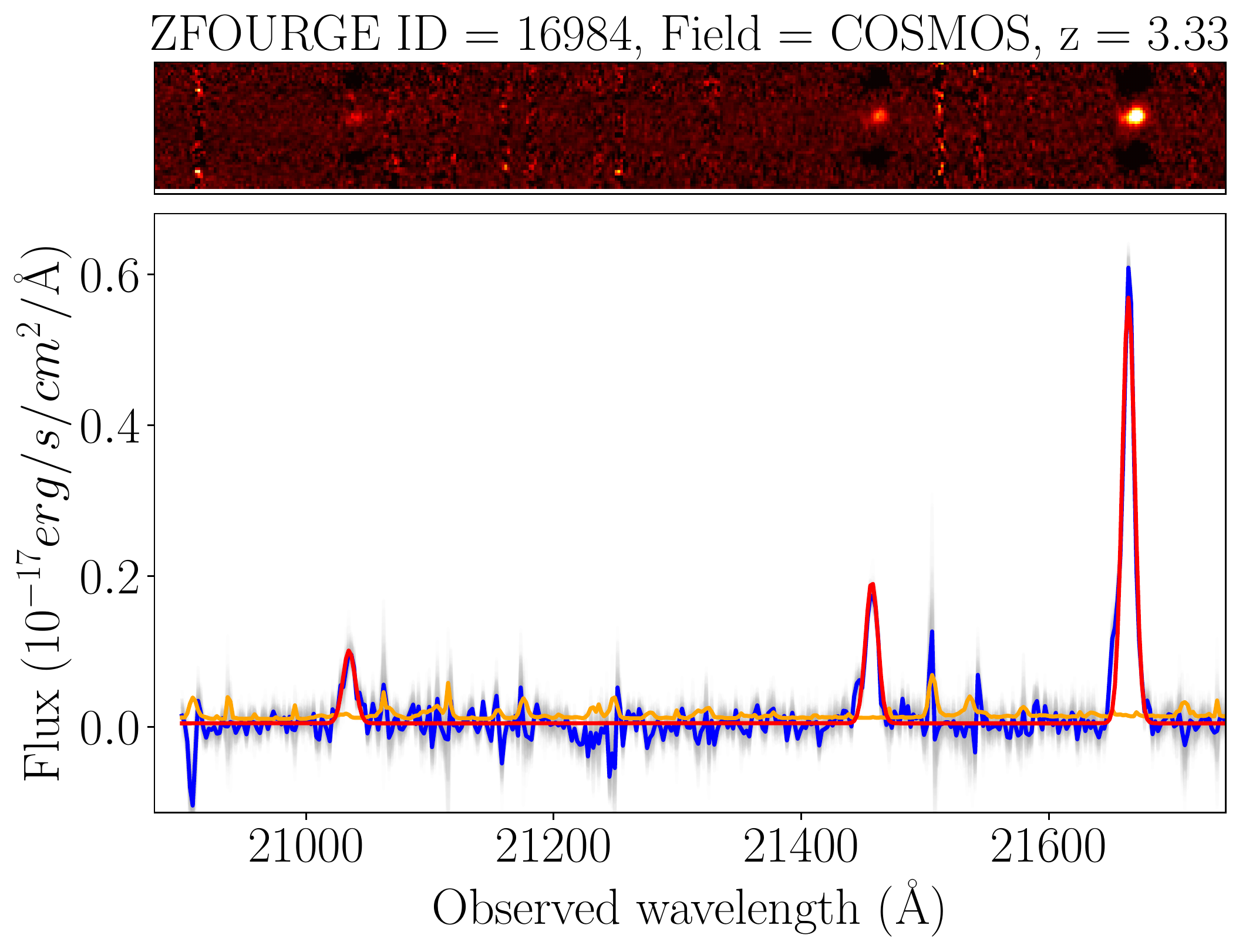}	
	\includegraphics[scale=0.27, trim=0.2cm 0.0cm 0.0cm 0.0cm,clip=true]{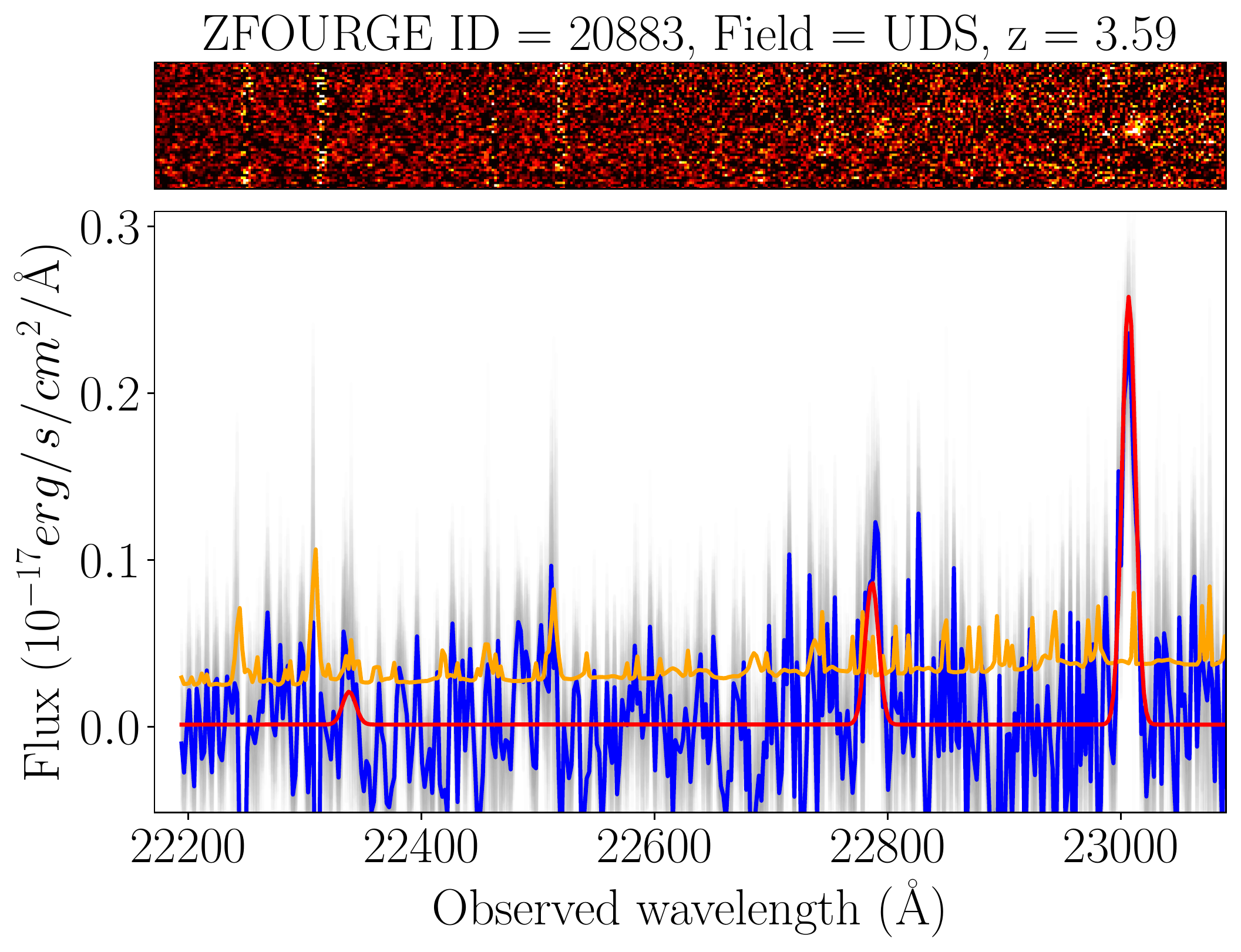}
	\includegraphics[scale=0.27, trim=0.2cm 0.0cm 0.0cm 0.0cm,clip=true]{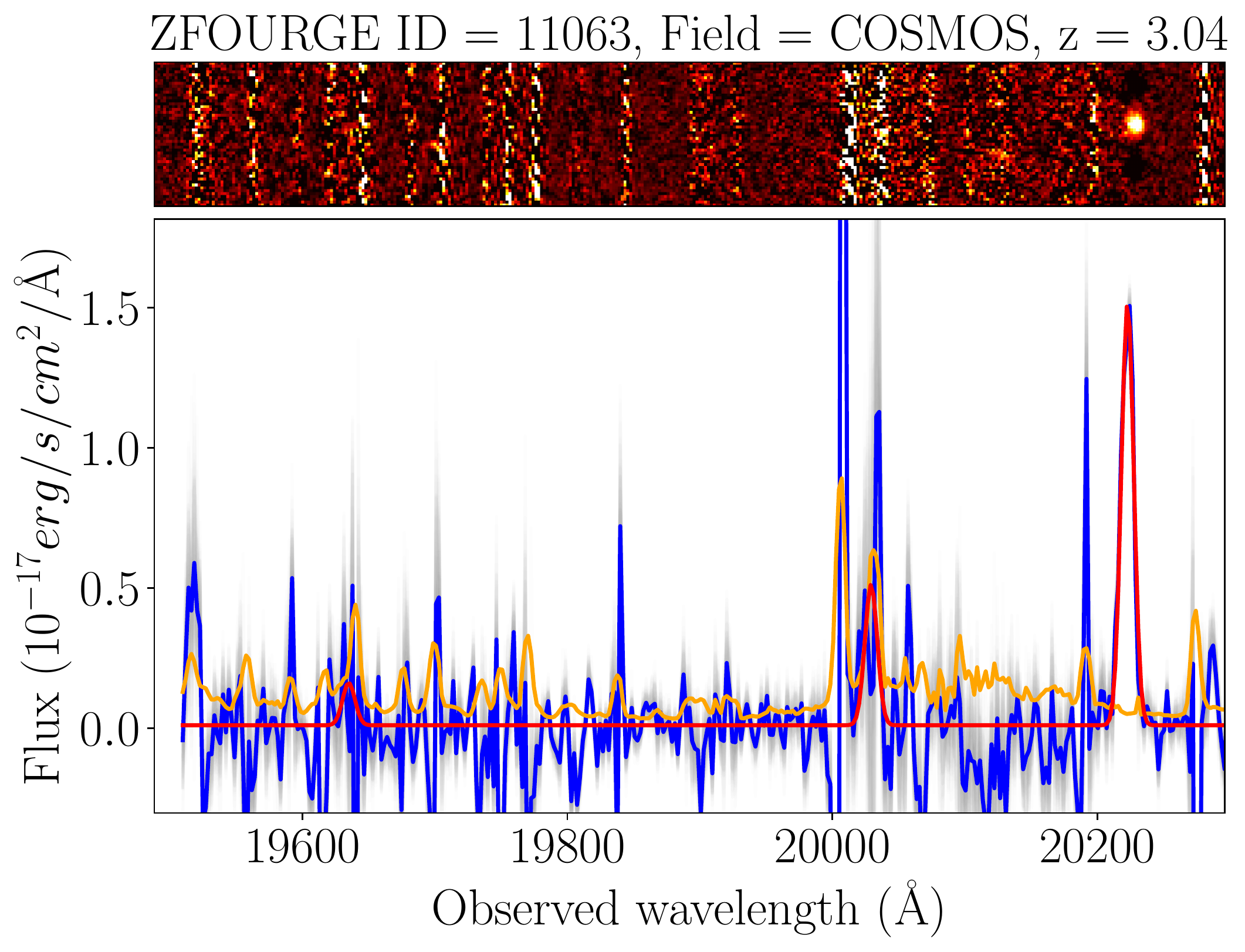}
	\includegraphics[scale=0.27, trim=0.2cm 0.0cm 0.0cm 0.0cm,clip=true]{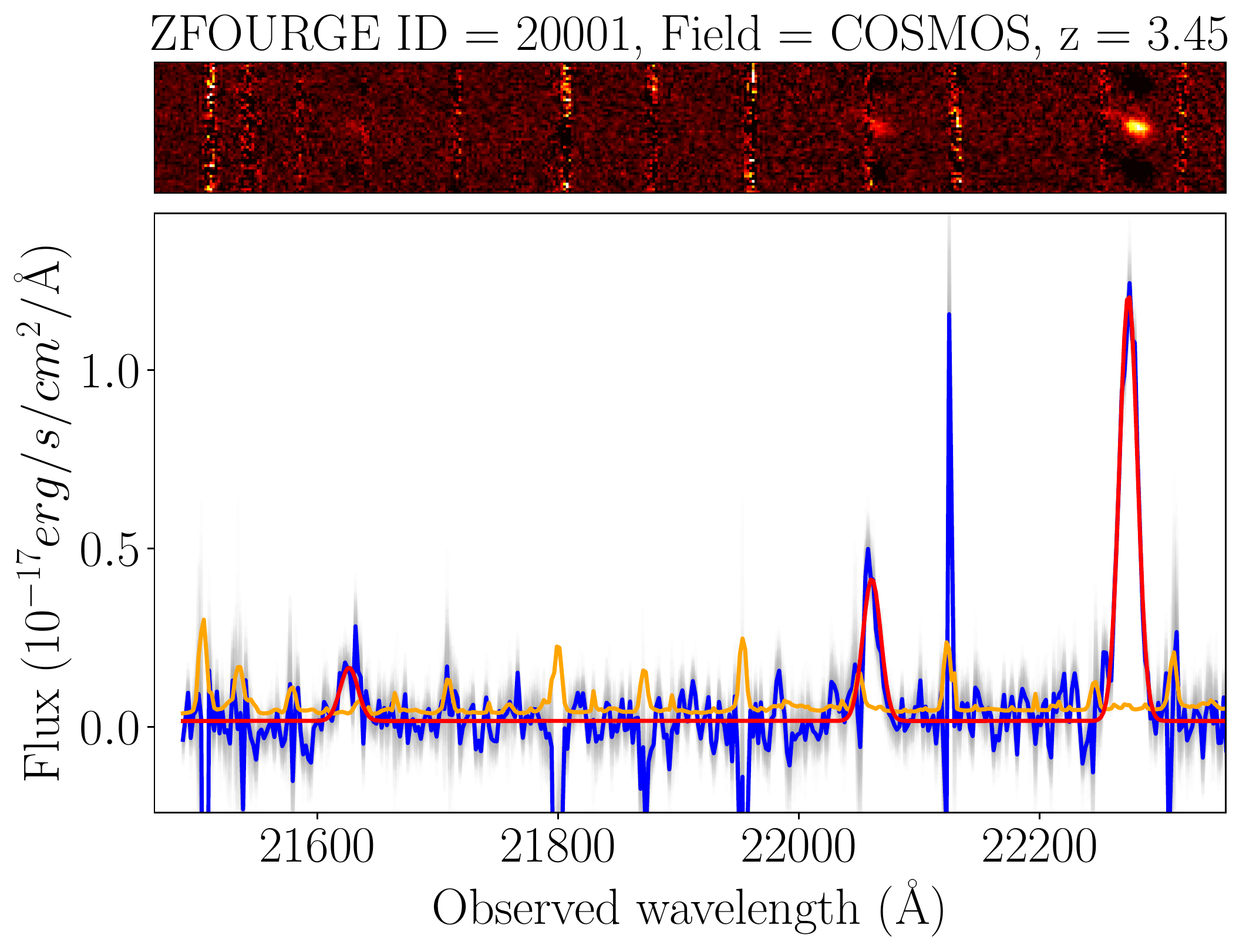}
	\includegraphics[scale=0.27, trim=0.2cm 0.0cm 0.0cm 0.0cm,clip=true]{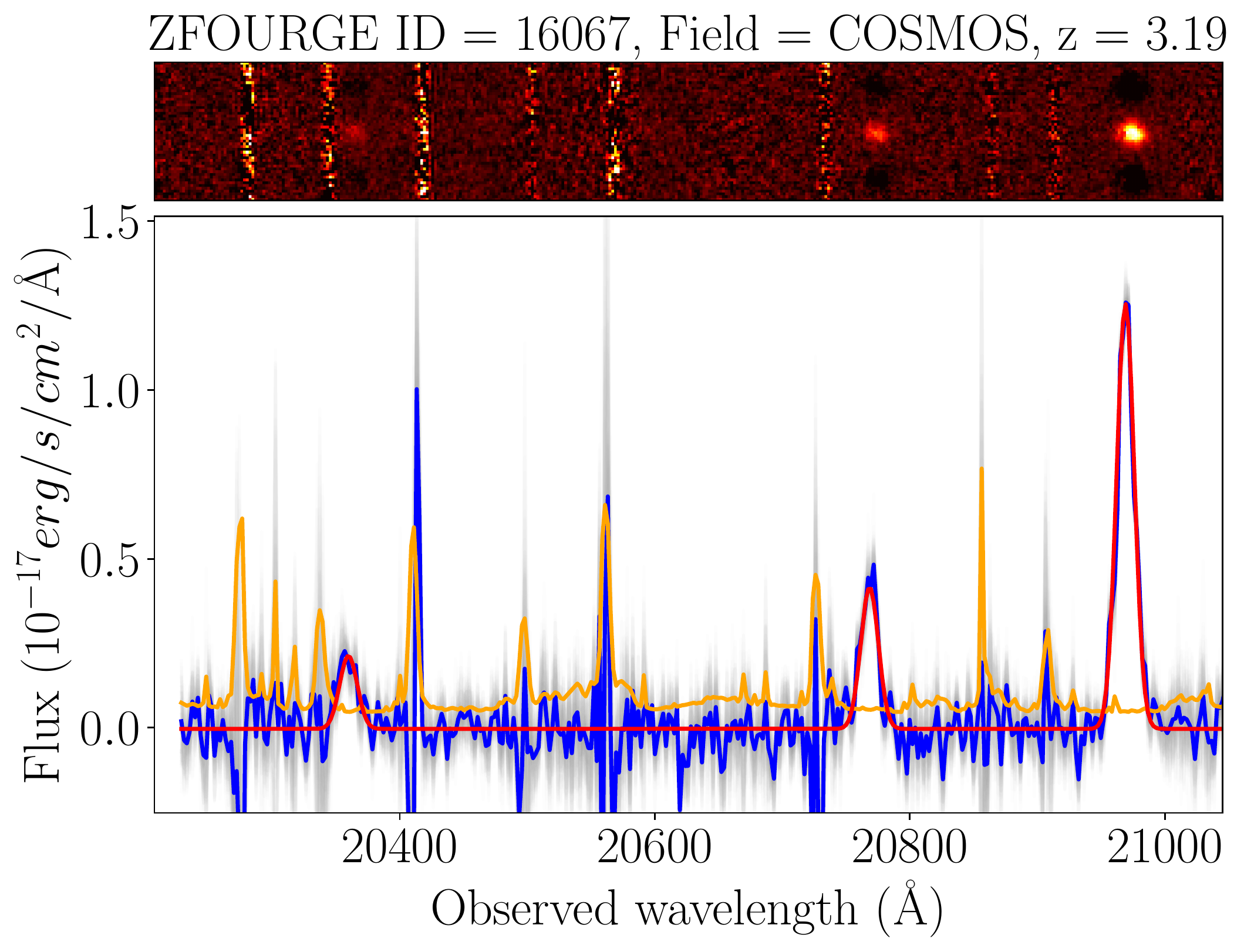}	
	\includegraphics[scale=0.27, trim=0.2cm 0.0cm 0.0cm 0.0cm,clip=true]{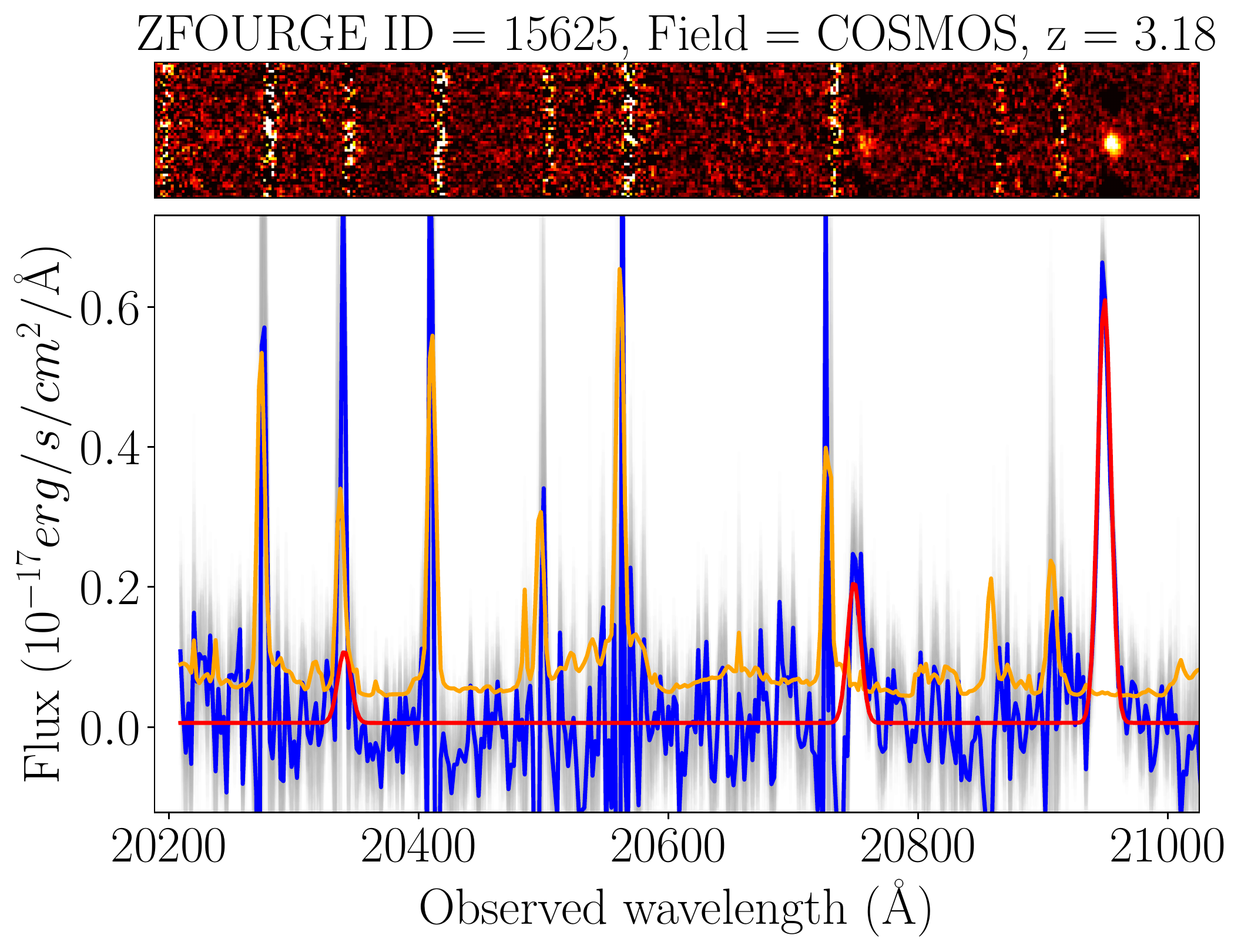}
	\includegraphics[scale=0.27, trim=0.2cm 0.0cm 0.0cm 0.0cm,clip=true]{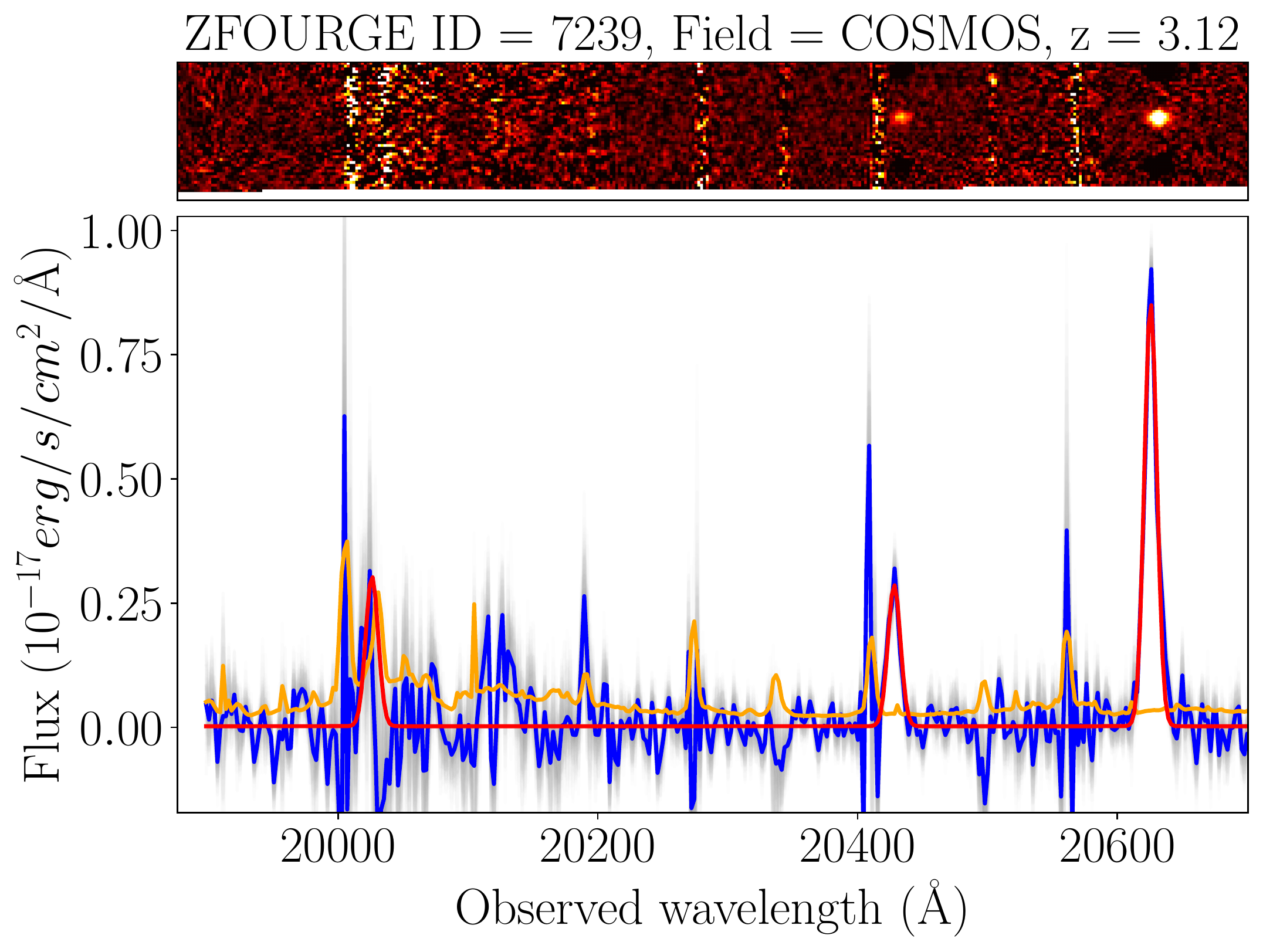}
	\includegraphics[scale=0.27, trim=0.2cm 0.0cm 0.0cm 0.0cm,clip=true]{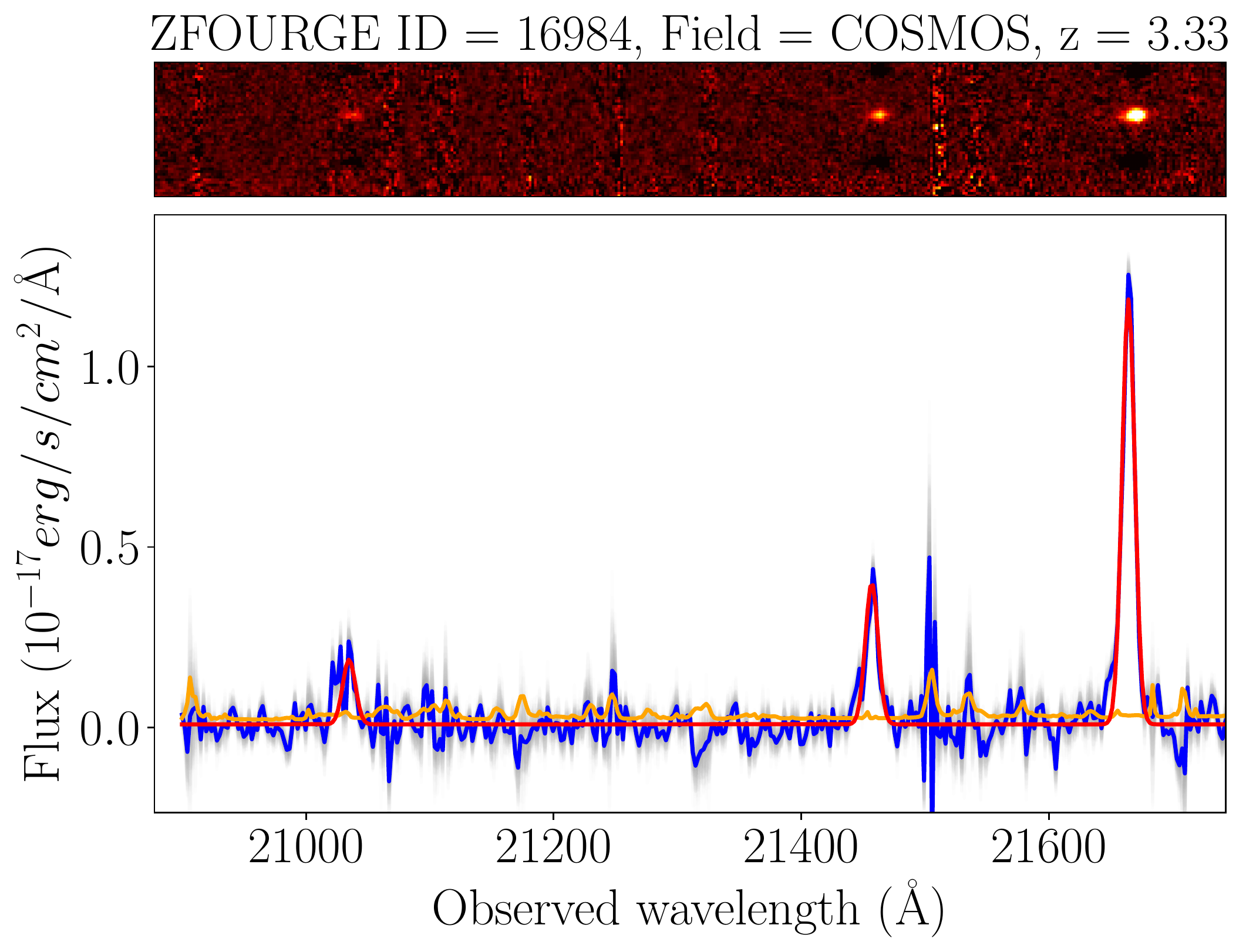}
	\includegraphics[scale=0.27, trim=0.2cm 0.0cm 0.0cm 0.0cm,clip=true]{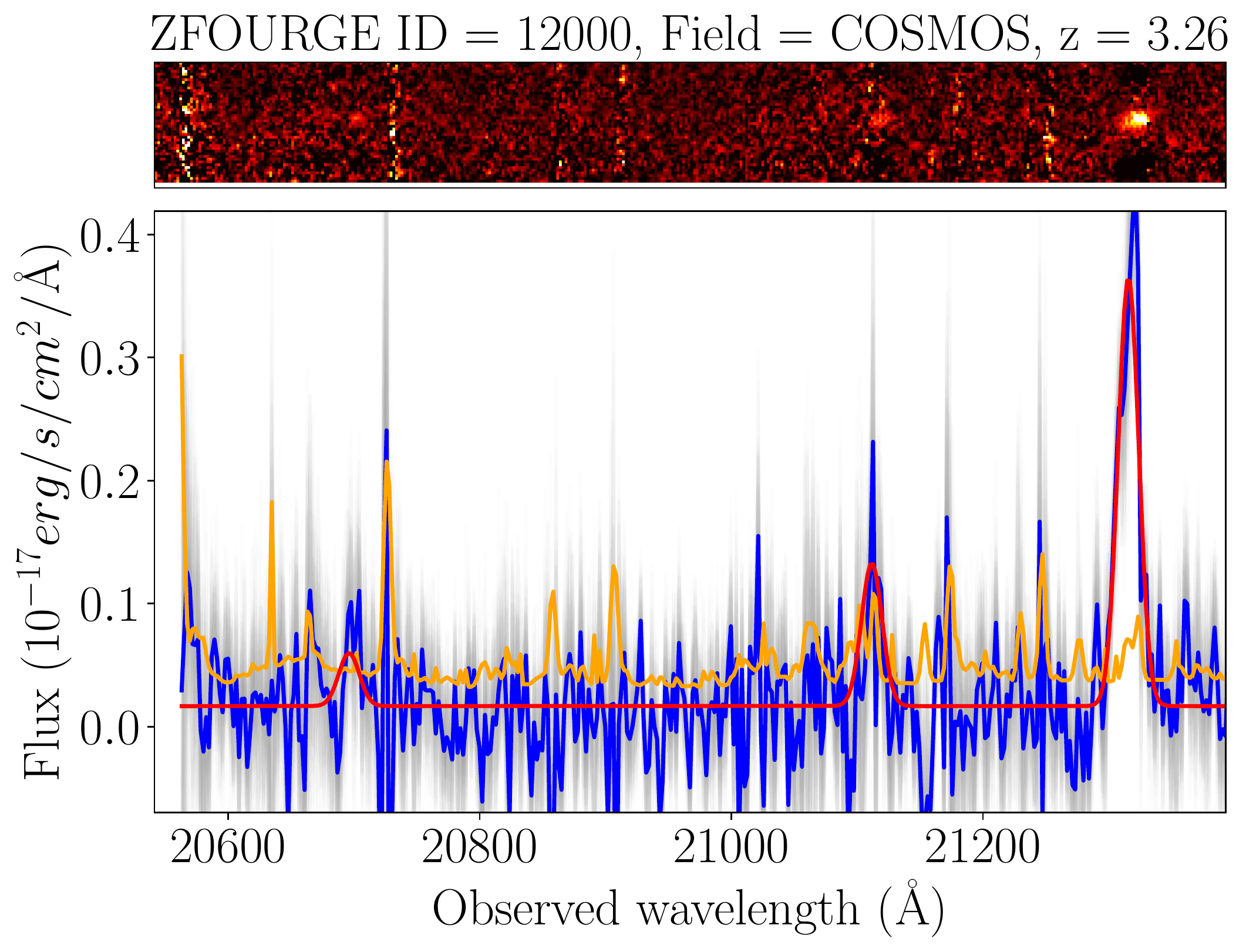}	
	\caption{Spectra of galaxies selected for the stacking analysis from the MOSFIRE observations. The blue curve is the 1D spectra in the observed frame, orange is the noise spectra, and the grey shaded region is the bootstrapped iterations. The red curves correspond to the best-fit curves to the \oiii${, 4959}$ and \hb\ emission lines. The image on top of each panel shows the 2D slit spectra.  }
	\label{fig:o3_fits_mosfire}
\end{figure*}

Figures \ref{fig:o3_fits_kmos} and \ref{fig:o3_fits_mosfire} shows the 1D spectra of the galaxies selected for the stacking analysis in this work. 

 \section{Stacked Sample - Imaging}\label{sec:images_appendix}

\begin{figure*}
	\centering
	\tiny
	\includegraphics[scale=0.5, trim=0.0cm 0.0cm 0.0cm 0.0cm,clip=true]{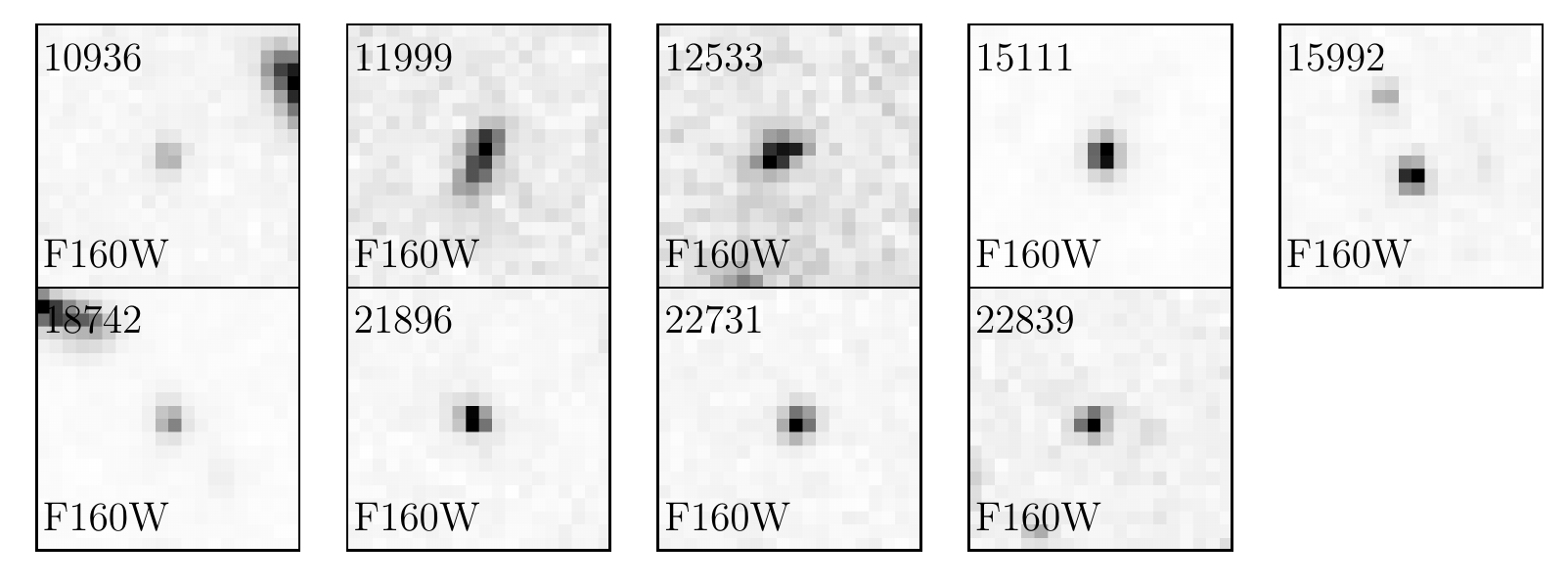}
	\caption{Cutouts of HST F160W mosaics by the \zfourge\ survey for the selected galaxies from the KMOS observations. The size of each image is $3''\times3''$.  }
	\label{fig:2d_img_kmos}
\end{figure*}

\begin{figure*}
	\centering
	\tiny
	\includegraphics[scale=0.5, trim=0.0cm 0.0cm 0.0cm 0.0cm,clip=true]{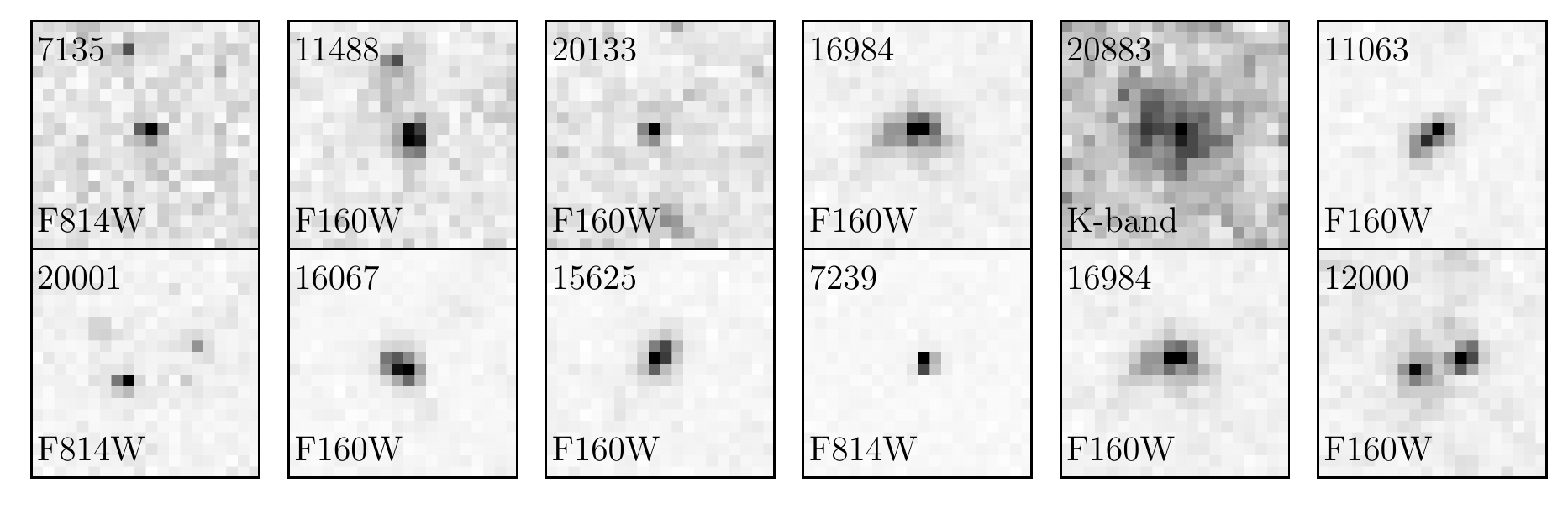}
	\caption{Cutouts from the mosaic created by the \zfourge\ survey for the selected galaxies from the MOSFIRE observations.  We show HST F160W or F814W filter if galaxy is not observed in F160W. The size of each image is $3''\times3''$. The galaxy 20883 is not observed with either HST filters, therefore we show its K-band image. }
	\label{fig:2d_img_mosfire}
\end{figure*}

Figures \ref{fig:2d_img_kmos} and \ref{fig:2d_img_mosfire} shows the HST images for  galaxies selected for the stacking analysis in this work. Most of our sample consist of compact and unresolved galaxies making the exclusion of mergers from the stacked spectra difficult.

\bsp	
\label{lastpage}
\end{document}